\documentclass[preprint,showpacs,pre,12pt]{revtex4-1}
\usepackage{amsfonts}
\usepackage{amssymb}
\usepackage{amsmath}
\usepackage{graphicx}
\usepackage{pgfplots}
\newcommand{\sgn}{\operatorname{sgn}}

\begin{document}
\title{Formation of superscar waves in plane polygonal billiards}
\author{Eugene Bogomolny}
\affiliation{LPTMS,  CNRS, Univ. Paris-Sud, Universit\'e Paris-Saclay, 91405 Orsay, France}
\date{\today}
\begin{abstract}
Polygonal billiards constitute a special class of models. Though they have zero Lyapunov exponent their classical and quantum properties are involved   due to scattering on singular vertices with angles $\neq \pi/n$ with integer $n$. It is demonstrated that in the semiclassical limit multiple singular scattering on such vertices when optical boundaries of many scatters overlap leads to vanishing of quantum wave functions along straight  lines built by these scatters.  This phenomenon has an especially important consequence for polygonal billiards where periodic orbits (when they exist)  form pencils  of parallel rays restricted from the both sides by singular vertices. Due to singular scattering on boundary vertices, waves propagated inside periodic orbit pencils in the semiclassical limit tend   to zero along pencil boundaries thus forming weakly interacting quasi-modes. Contrary to scars in chaotic systems the discussed quasi-modes in polygonal billiards become almost exact for high-excited states and for brevity they are  designated as superscars.   Many pictures  of eigenfunctions for a triangular billiard and a barrier billiard which have clear superscar structures are presented in the paper. Special attention is given to the development of  quantitative methods of detecting and analysing such superscars.  In particular, it is demonstrated that the overlap between superscar waves associated with a fixed periodic orbit and  eigenfunctions of a barrier billiard is distributed according to the Breit-Wigner distribution typical for weakly interacting quasi-modes (or doorway states).  For special sub-class of rational polygonal billiards called Veech polygons where all periodic orbits can be calculated analytically it is argued and checked numerically that their eigenfunctions are fractal in the Fourier space.        
\end{abstract}
\maketitle
\begin{flushright}
In memory of Charles Schmit
\end{flushright}


\section{Introduction}

The largest part of this  work has been prepared during  2003-2004 but an implacable malady of Charles Schmit  had permitted  to publish  uniquely its short account \cite{main}. It is only now that I collect different fragments of performed investigations and organise them in a readable  form. 

The paper examines  the structure of eigenfunctions for a special class of quantum models, namely two-dimensional  polygonal billiards whose boundaries are straight lines.  Classical mechanics of these problems corresponding to rays propagation with specular reflection from boundaries is intricate, surprisingly rich, and notorious difficult   (see e.g. \cite{gutkin} and references therein). 

The most  investigated is the case of rational or pseudo-integrable billiards where all (internal) billiard angles $\theta_j$ are rational fractions of $\pi$
\begin{equation}
\theta_j=\pi \frac{m_j}{n_j}
\label{angles}
\end{equation} 
with co-prime integers $m_j$ and $n_j$.   A characteristic property of such billiards is that their classical trajectories belong to  2-dimensional surfaces of finite genus $g$ related with angles \eqref{angles} as follows (see e.g. \cite{Richens})
\begin{equation}
g=1+\frac{N}{2}\sum_{j} \frac{m_j-1}{n_j}
\end{equation}
where $N$ is the least common multiple of all $n_j$. 

\begin{figure}[!]
\begin{minipage}{.3\linewidth}
\begin{center}
\includegraphics[ width=.9\linewidth]{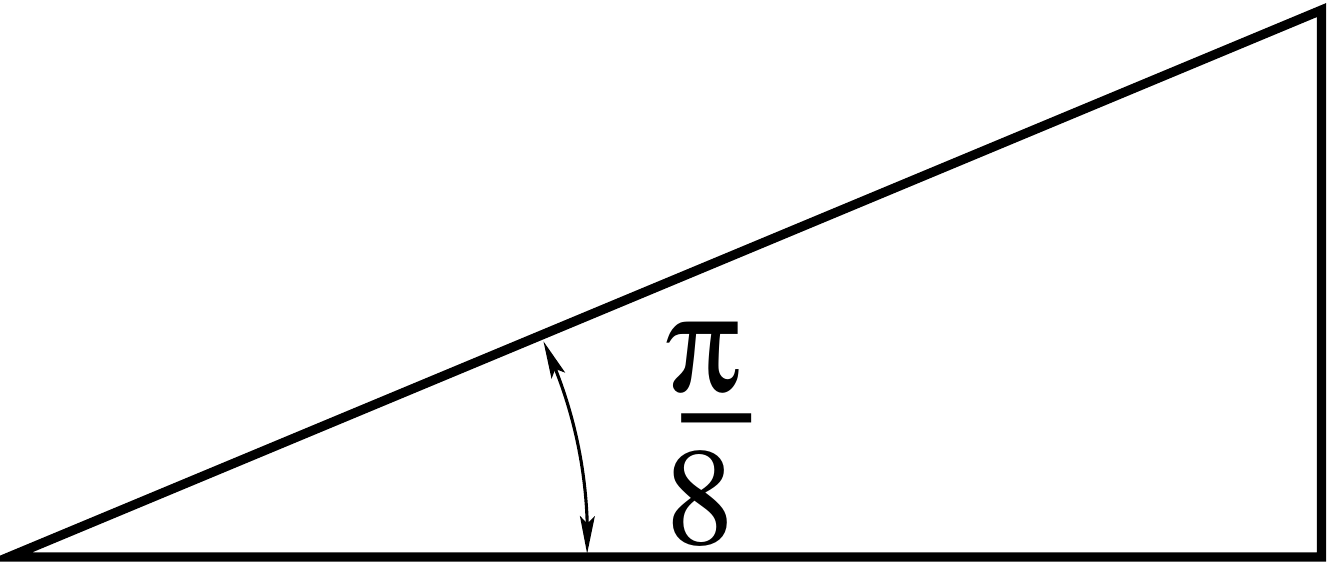}\\
a)
\end{center}
\end{minipage}
\begin{minipage}{.3\linewidth}
\begin{center}
\includegraphics[ angle=-90, width=.9\linewidth ]{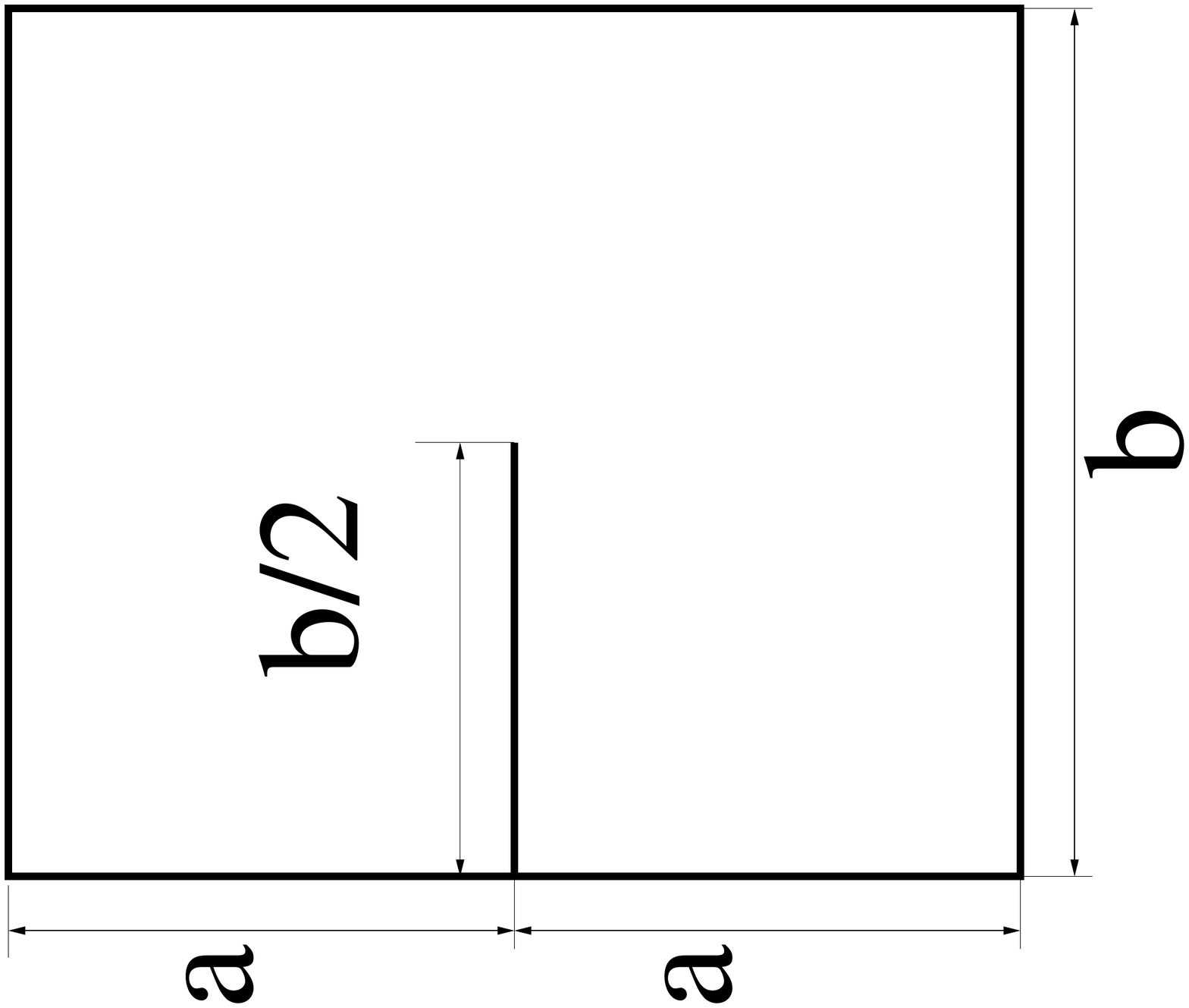}\\
b)
\end{center}
\end{minipage}
\begin{minipage}{.3\linewidth}
\begin{center}
\includegraphics[width=.9\linewidth ]{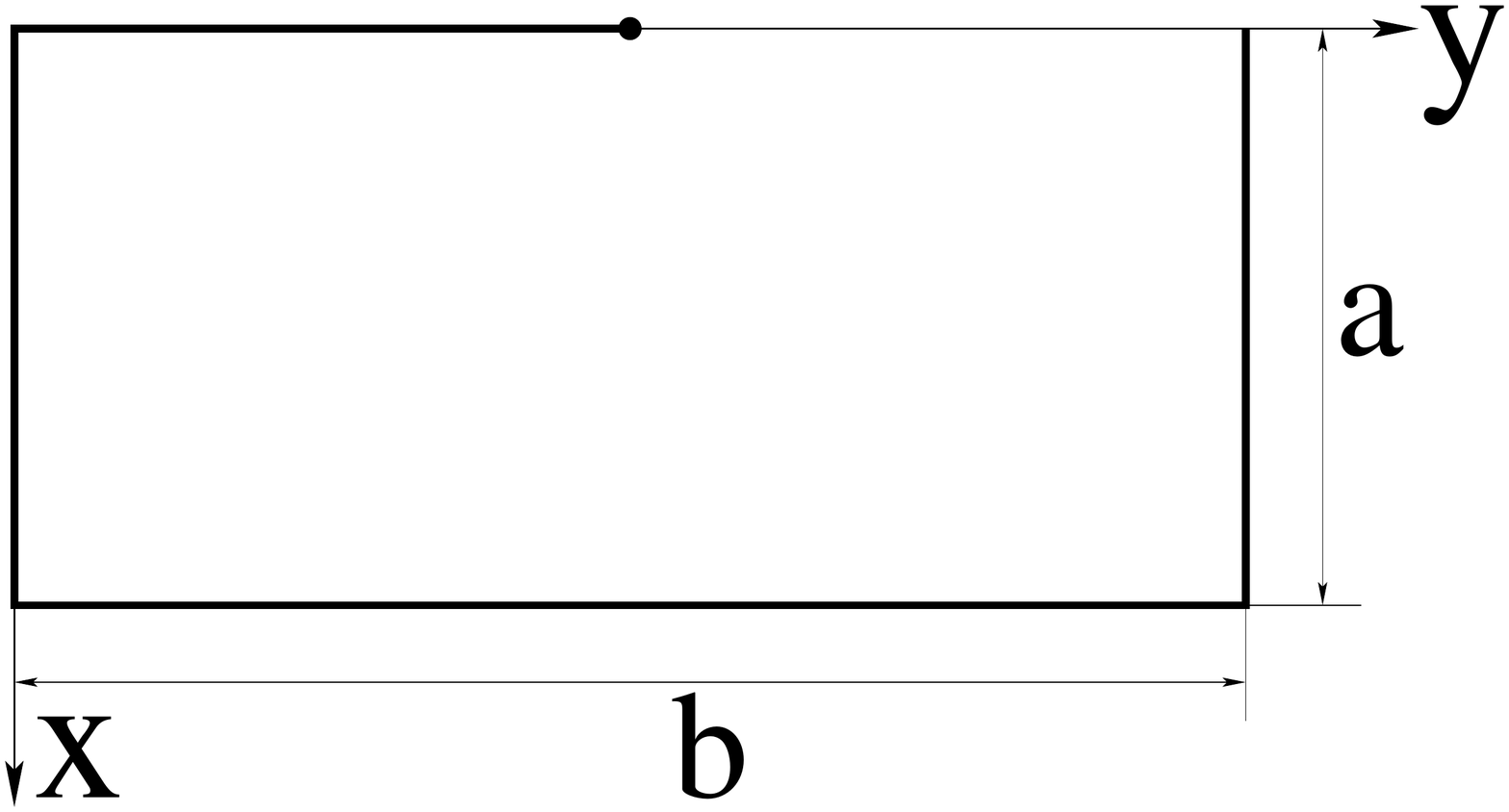}\\
c)
\end{center}
\end{minipage}
\caption{(a) Right triangular billiard with one angle $\pi/8$. (b) Barrier  billiard. (c) Desymmetrised barrier billiard. Eigenfunctions obey the Dirichlet boundary conditions at thick boundaries and the Neumann condition at thin part of the boundary. }
\label{examples}
\end{figure}
Two particular examples of such models  discussed in the paper are depicted in Fig.~\ref{examples}. The first model is a right triangular billiard with angles $\big [\frac{\pi}{8}, \frac{3\pi}{8}, \frac{\pi}{2}\big ]$. The second is a rectangular billiard of sides $2a$ and $b$ with a barrier of height $b/2$ in the middle. This model has $6$ corners with angles $\pi/2$ and one conner with angle $2\pi$. 

In a billiard where all angle  numerators $m_j=1$  trajectories belong to tori with $g=1$ and the model is  classically integrable.  The list of 2-dimensional integrable polygonal billiards is  limited. It includes rectangular (square) billiards and 3 triangular billiards with angles 
$\big [ \frac{\pi}{4}, \frac{\pi}{4},\frac{\pi}{2} \big ]$,  $\big [ \frac{\pi}{3}, \frac{\pi}{3},\frac{\pi}{3} \big ]$, and  $\big [\frac{\pi}{6}, \frac{\pi}{3},\frac{\pi}{2} \big ]$.

\begin{figure}
\begin{center}
\includegraphics[width=.9\linewidth ]{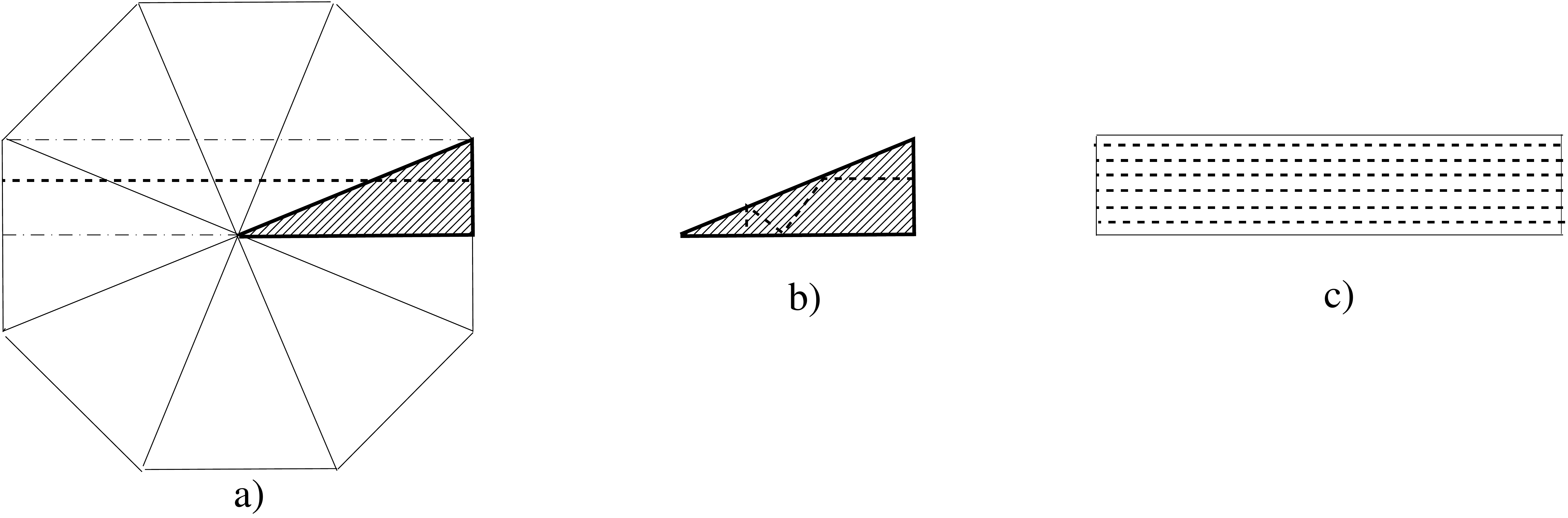}
\end{center}
\caption{(a) Unfolding of the $\pi/8$ right triangle (shaded) into the regular octagon. Dashed line is one periodic orbit for the motion inside the octagon. All trajectories parallel to  this line and lying inside the rectangle bounded by dot-dashed lines form  periodic orbit family.  (b) The same periodic orbit folded back to the original triangle. (c) Periodic orbit pencil formed by trajectories parallel to  the dashed one.}
\label{octagon}
\end{figure}

If at least one numerator is bigger than $1$, trajectories lie on surface of genus $g\geq  2$ and such  models are  genuine  pseudo-integrable models. The both billiards at  Fig.~\ref{examples}~a)  have genus $g=2$.  The values of genus can be obtained by explicit unfolding of the initial billiard table. For example, at  Fig.~\ref{octagon} the unfolding of the right triangular billiard with angle $\pi/8$ is performed. By reflections one gets a surface with shape of regular octagon whose opposite parallel sides are identified. Topologically the resulting surface is a sphere with 2 handles which is the canonical image of genus 2 surfaces. 

 The quantisation of billiards consists in finding eigenvalues  and eigenfunctions of the wave equation
\begin{equation}
(\Delta +E_{\alpha})\Psi_{\alpha}(x)=0
\label{equation}
\end{equation}
provided eigenfunctions obey certain boundary conditions along the billiard boundaries. In the paper  the Dirichlet boundary conditions are chosen
\begin{equation}
\Psi_{\alpha}(x)|_{\mathrm{boundaries}}=0. 
\end{equation}
For all integrable polygonal billiards cited above the solution of the quantum problem  is well known (see e.g. \cite{Richens}). Their eigenvalues and eigenfunctions depend on two integers. Eigenvalues are quadratic functions of these integers and eigenfunctions are finite combination of trigonometric functions. Even the inverse theorem is valid: the list of billiards whose all eigenfunctions are finite combinations of trigonometric functions is exhausted by the above integrable billiards \cite{triangles}.   

The structure of eigenvalues and eigenfunctions of polygonal billiards are much more complicated and only partial results are available.

 In quantum chaos there are two big conjectures concerning  spectral statistics of generic integrable and fully chaotic systems.  For integrable models spectral statistics (after unfolding) coincides with the Poisson statistics of independent random variables \cite{Berry} and  for chaotic systems it corresponds to eigenvalue statistics of random matrix ensembles depended only on symmetries \cite{BGS}.  The difference between these two types of universal statistics is clearly seen in the behaviour of the nearest-neighbour distribution, $p(s)$, which gives the probability that two nearest levels are separated by distance $s$ (see e.g. \cite{Mehta}, \cite{Bohigas}). For integrable systems $p(s)=\exp (-s)$ which implies the absence of level repulsion ($p(0)\neq 0$) and exponential decrease of correlations at large distances. For chaotic systems $p(s)$ is well approximated by the Wigner ansatz  $p(s)=a s^{\beta} \exp(-b s^2)$ where $\beta=1,2,4$ with  $a,b$ being constants determined from normalisation conditions. Contrary to integrable models  fully chaotic  systems are characterised by the level repulsion ($p(0)=0$) and quadratic  falloff of $p(s)$  at large distances  ($p(s)\underset{s\to\infty}{\longrightarrow} \exp( - s^2)$). 

Numerical calculations of pseudo-integrable billiards \cite{Richens}, \cite{Cheon}--\cite{GW} demonstrate that their spectral properties are in-between these two universal distributions. Namely their nearest-neighbour  distribution reveals  a linear level repulsion  $p(s)\underset{s\to 0}{\longrightarrow}  s$ as for random matrix ensemble with $\beta=1$ but at large distances $p(s)$ decreases exponentially as for the Poisson statistics. 

\begin{figure}
 \begin{minipage}{.49\linewidth}
 \begin{center}
\includegraphics[width=.6\linewidth]{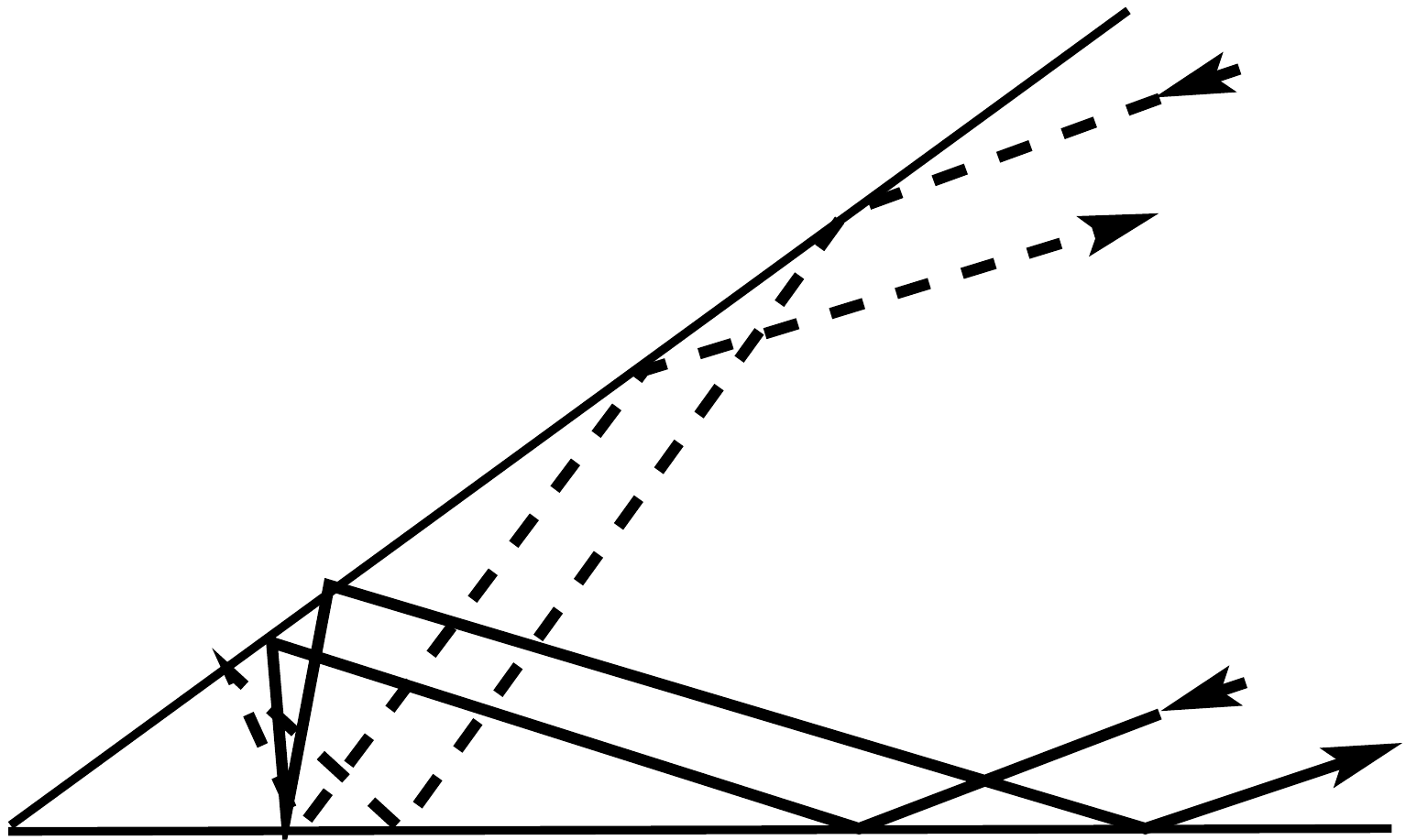}
\end{center}
\end{minipage}
\begin{minipage}{.49\linewidth}
\begin{center}
\includegraphics[width=.6\linewidth]{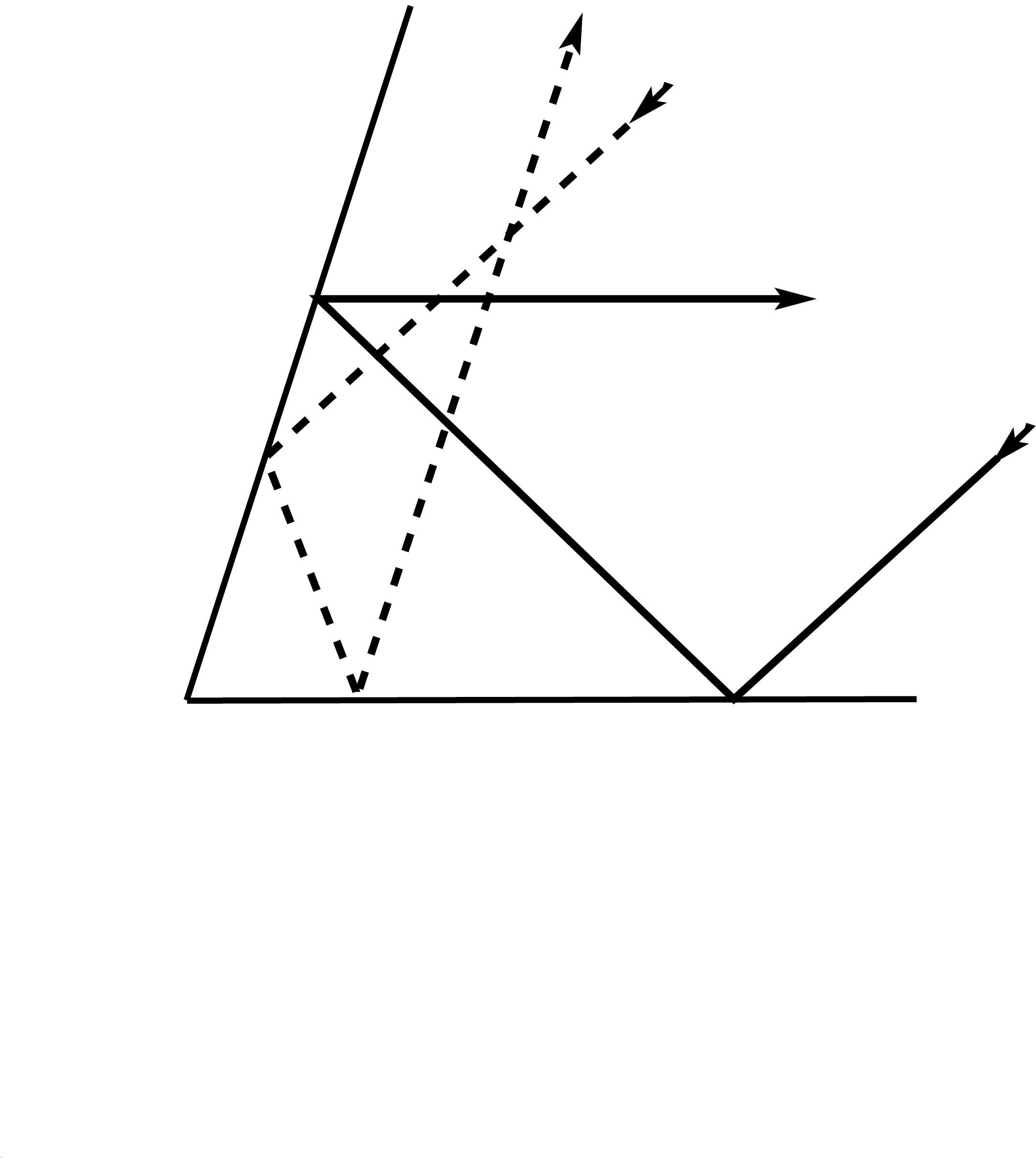}
\end{center}
\vspace{-2.5cm}
\end{minipage}
\caption{Reflection of two parallel rays (solid and dashed lines) from a vertex with (left) $\theta=\frac{\pi}{5}$  and (right) $\theta=\frac{2\pi}{5}$. }
\label{splitting}
\end{figure}

The purpose of this paper is to investigate properties of eigenfunctions for plane polygonal billiards. The main difficulty in treating such problems is the fact that in polygonal billiards  vertices with angles $\neq \pi/n$ with integer $n$ are singular points for classical motion. If a parallel pencil of rays hits such point it splits discontinuously into two different pencils (cf. Fig.~\ref{splitting}). 

\begin{figure}
\begin{minipage}{.38\linewidth}
\begin{center}
\includegraphics[width=.99\linewidth ]{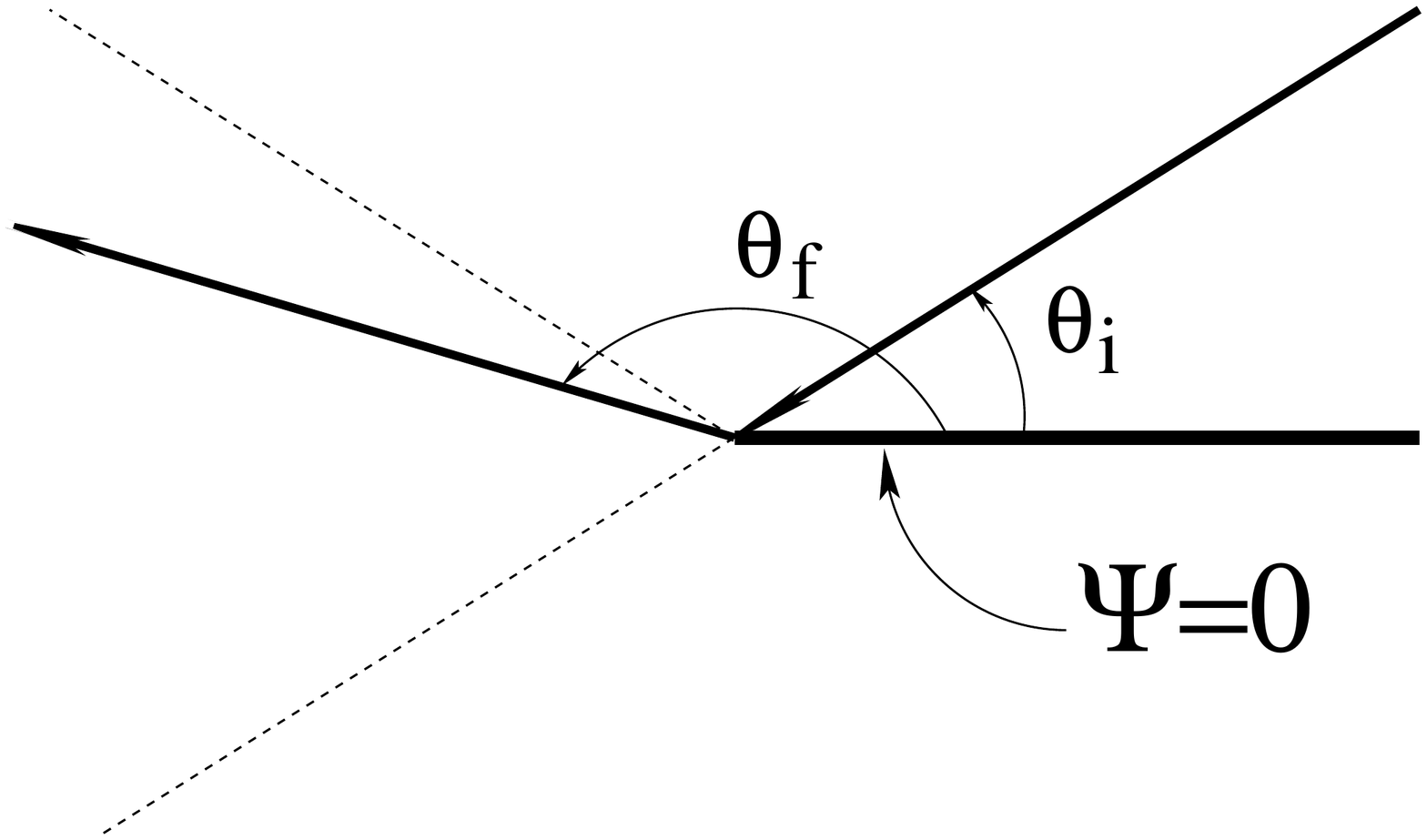}\\
a)
\end{center}
\end{minipage}
\begin{minipage}{.3\linewidth}
\begin{center}
\includegraphics[width=.8\linewidth ]{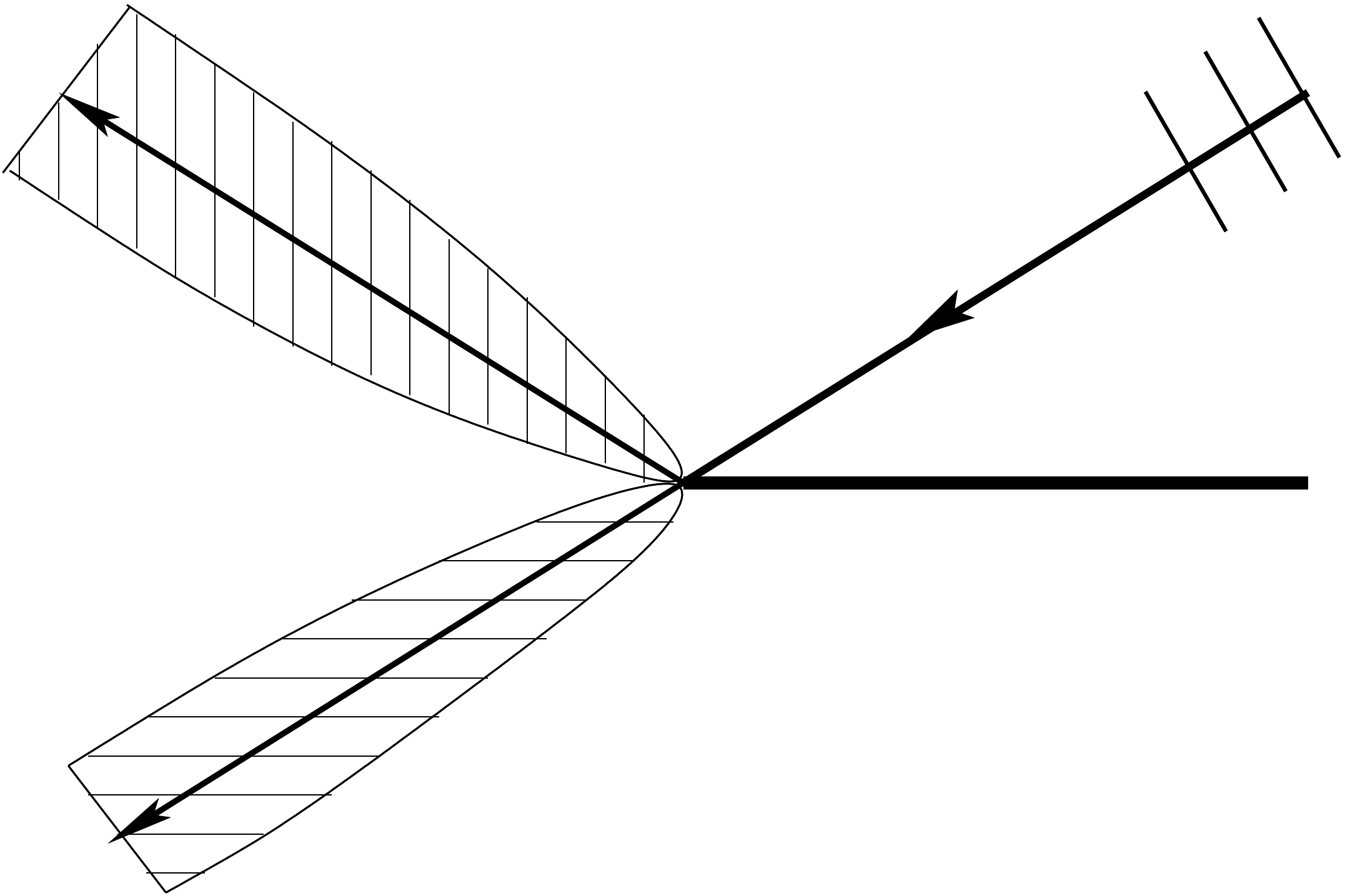}\\
b)
\end{center}
\end{minipage}
\begin{minipage}{.3\linewidth}
\begin{center}
\includegraphics[width=.86\linewidth]{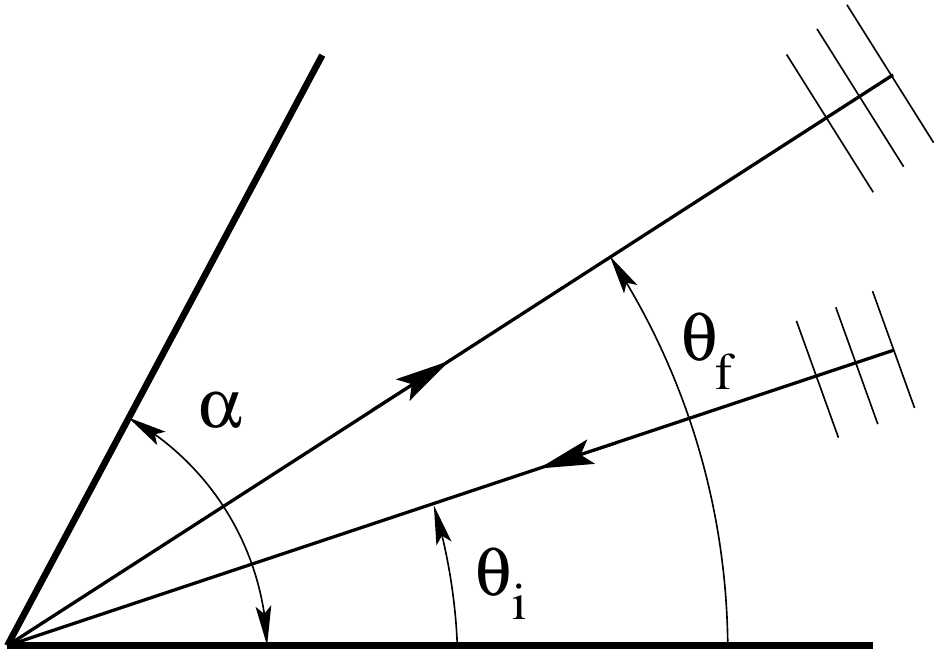}\\
c)
\end{center}
\end{minipage}
\caption{(a) Scattering on a half-plane. (b) Optical boundaries (dashed) for the scattering on a half-plane. (c) Scattering on a wedge with angle $\alpha$.}
\label{scattering} 
\end{figure}

Quantum mechanics has to smooth these singularities and leads to the notion of singular diffraction. The exact solution for the scattering on wedge has been obtained long time ago by Sommerfeld \cite{Sommerfeld} (cf. also \cite{Budaev}). The simplest case of such diffraction corresponds to the the scattering on a half-plane  with e.g. the Dirichlet boundary conditions, see Fig.~\ref{scattering}~a).  
The exact solution for this problem has been found by Sommerfeld in 1896 \cite{Sommerfeld} and it reads
\begin{equation}
\Psi(\vec{r}\ )=\mathrm{e}^{-\mathrm{i} kr \cos (\theta_f-\theta_i)}
F\left (-\sqrt{2kr}\cos\frac{\theta_f-\theta_i}{2}\right )-
\mathrm{e}^{-\mathrm{i} kr \cos (\theta_f+\theta_i)}
F\left (-\sqrt{2kr}\cos\frac{\theta_f+\theta_i}{2}\right )
\end{equation}
where $F(u)$ is the Fresnel integral
\begin{equation}
F(u)=\frac{\mathrm{e}^{-\mathrm{i}\pi/4}}{\sqrt{\pi}}\int_{u}^{\infty}\mathrm{e}^{\mathrm{i} t^2}\mathrm{d} t.
\end{equation}
From the expansion of $\Psi(\vec{r}\ )$ at large distances one finds that the total wave  splits into two contributions, the incident plane wave and the out-going cylindrical wave 
\begin{equation}
\Psi(\vec{r}\ )=\mathrm{e}^{\mathrm{i}\vec{k}\vec{r}}+
\frac{D(\theta_f,\theta_i)}{\sqrt{8\pi k r}}\mathrm{e}^{\mathrm{i}(kr-3\pi/4)}
\end{equation}
where $D(\theta_f,\theta_i)$ is  the diffraction coefficient
\begin{equation}
D(\theta_f,\theta_i)=  \dfrac{1}{\cos \frac{\theta_f-\theta_i}{2}}
-\dfrac{1}{\cos \frac{\theta_f+\theta_i}{2}}\ .
\end{equation}
Sommerfeld \cite{Sommerfeld} also found the exact solution for the scattering on arbitrary wedge with the Dirichlet boundary conditions as at Fig.~\ref{scattering}~c). In this case the diffraction coefficient has the following form
\begin{equation}
D(\theta_f, \theta_i)=\frac{2}{\gamma}\sin \frac{\pi}{\gamma}\left [
\dfrac{1}{\cos \frac{\pi}{\gamma} -\cos \frac{\theta_f+\theta_i}{\gamma}}-
\dfrac{1}{\cos \frac{\pi}{\gamma} -\cos \frac{\theta_f-\theta_i}{\gamma}} \right ]
\end{equation}
where  $\gamma=\alpha/\pi$ and $\alpha$ is the wedge angle.
 
The main feature of such diffraction coefficients is the existence of certain lines where diffraction coefficients formally blow-up.  These lines are called optical boundaries and they correspond to zeros of the denominators in the above formulas. For the scattering on a half-plan they appear when 
\begin{equation}
\theta_f=\pi \pm \theta_i.
\end{equation} 
Physically these lines separate regions with  different numbers of geometrical rays and are manifestation of discontinuous character of classical (rays) motion (see Fig.~\ref{scattering}~b)).  As in quantum mechanics wave fields are continuous, the separation of the exact field into a sum of  free motion (plane wave) plus  small reflected field is not possible in a vicinity of such optical boundaries which forces the diffraction coefficient to diverge. Consequently,  diffractive coefficient description cannot be applied in parabolic regions near optical boundaries where the dimensionless arguments of the $F$-functions are of the order of 1, $u=\sqrt{kr}\sin \frac{\delta \varphi}{2}\sim 1$, and  $\delta \varphi $  is the  angle of deviation  from optical boundaries (cf. Fig.~\ref{scattering}~b)).

Difficult problems appear when inside these intermediate regions there are other points of singular diffractions which is inevitable for plane polygonal billiards.  For finite number of singular diffraction vertices it is possible to develop uniform approximations which give good description of multiple singular diffraction in the semiclassical limit  $k\to\infty$ (see e.g. \cite{BPS} and references therein).  For infinite number of singular diffractions the situation is less clear. To understand the behaviour of waves scattered on many singular scatters where optical boundaries strongly overlap three interrelated  approaches  are discussed in Section~\ref{singular_diffraction}.  All these methods  demonstrate that multiple singular diffraction in the semiclassical limit of high energy scattering leads to a non-perturbative  effect of (almost) vanishing of eigenfunctions along straight lines passing through singular scatters (vertices with angles $\neq \pi/n$). Consequently, a wave scattered with a small incident angle from many singular scatters arranged along a line  will be reflected from them  as from a mirror with the Dirichlet boundary condition though the mirror itself does not exist.  The importance of this phenomenon for polygonal billiards is related with the fact that periodic orbits in such billiards (when they exist) form families of parallel trajectories (cf. Fig.~\ref{staggered}~c)).  When unfolded each family constitutes an infinite  pencil (or channel)  restricted from the both sides by singular scatters. Such configuration is exactly the one which permits the propagation of plane wave with (approximately) Dirichlet boundary conditions along two fictitious mirrors built by singular scatters.   The validity of such approximation becomes better in the semiclassical limit of high energy. Therefore we propose to call these waves  'superscars' to distinguish them from the scar phenomena in chaotic systems \cite{scar_heller}--\cite{kaplan}  where individual scar amplitudes decrease in the semiclassical limit.  

 Numerous examples of numerically computed  high-excited eigenfunctions  with  clear superscar structures for the triangular and the barrier billiards depicted at Fig.~\ref{examples} are presented in Section~\ref{superscar_examples}. Additional confirmation of applicability of superscar picture is the very good agreement of true eigenenergies of such states with  superscar energies computed analytically from the knowledge of periodic orbit parameters.   

To get quantitative information about the formation of superscar waves the overlaps between consecutive barrier billiard eigenfunctions and specific  folded superscar waves are investigated in Section~\ref{Fourier_expansion}. It is observed that in a small vicinity of  all superscar energies there exist true eigenstates having  large overlaps with the corresponding superscar waves. In a finite energy window  the values of the overlap fluctuate according to the Breit-Wigner distribution whose parameters agree with the ones calculated analytically  in Section~\ref{singular_diffraction}.  Another useful approach discussed in the same Section is  the Fourier-type expansion method.  It consists in the expansion of true eigenfunctions in a series of convenient basis functions. The existence of superscars manifests as anomalously large values of certain expansion coefficients.

If a periodic orbit family exits in a given polygonal billiard it may and will support superscar waves. But for generic polygonal billiards very little is known about periodic orbits. Only for a special sub-class of pseudo-integrable billiards called Veech polygons \cite{Veech} one can find all periodic orbits analytically. Billiards considered in the paper belong to this class. For Veech polygons it is possible to calculate analytically the level compressibility \cite{Giraud}, \cite{GW}  which is practically the only one spectral characteristic accessible to analytical calculations. It is believed (and confirmed numerically for many different models, see e.g. \cite{entropy}) that systems with non-trivial compressibility should have eigenfunctions with non-trivial fractal dimensions. For pseudo-integrable billiards the above mentioned strong fluctuations of Fourier coefficients mean that eigenfunctions in the momentum  space may have fractal dimensions. In Section~\ref{fractal_dimensions} it is numerically demonstrated  that indeed eigenfunctions of the barrier billiard do have non-trivial fractal dimensions.  Section~\ref{summary} contains a brief summary of obtained results.  Appendix~\ref{appendix_A} is devoted to investigations of periodic orbit pencils in the barrier billiards and the folding of corresponding superscar waves.  
 

\section{Singular multiple diffraction}\label{singular_diffraction}

The purpose of this Section is to present different approaches to multiple singular scattering on a periodic array of singular vertices (wedges with angles $\neq \pi/n$ with integer $n$) arranged along a straight line  as indicated at Fig.~\ref{Kirchhoff_diff}. The simplest method consists in the construction of the Kirchhoff-type approximation to this problem. It has been done in Ref.~\cite{BPS} and briefly reviewed  in Section~\ref{Kirchhoff_approximation}.  It is  known that  the condition of applicability of the Kirchhoff approximation is not easy to be rigorously established. To get more precise information of this process,  the exact solution for the scattering on staggered periodic set of half-planes as indicated at Fig.~\ref{staggered}~a) derived by  Carlson and Heins in 1947 \cite{Carlson} and analysed in the semiclassical limit in \cite{BS}  is discussed in  Section~\ref{Carlson_Heins}.  Section~\ref{periodic_slits} is devoted to numerical investigation of wave propagation inside periodic array of slits depicted at Fig.~\ref{staggered}~b). 

The main result established in  that Sections is the fact that  small-angle high-energy multiple scattering on singular wedges is equivalent to much simpler specular (i.e. mirror) reflection from a straight line passing through the apex of the wedges  though the line itself does not constitute a physical boundary. In Section~\ref{applications}  it is  demonstrated that this result  applied to   polygonal billiards proofs   the existence of special weekly interacting quasi-modes corresponding  to plane waves propagating inside periodic orbit channels (when they exist). These quasi-modes called in the paper superscars are a specific feature of polygonal billiards. They do not exit neither in integrable nor  in  chaotic systems and are a non-perturbative consequence of multiple singular diffraction inherent for polygonal billiards.  


\subsection{The Kirchhoff approximation}\label{Kirchhoff_approximation}

\begin{figure}
\begin{center}
\includegraphics[width=0.7\linewidth]{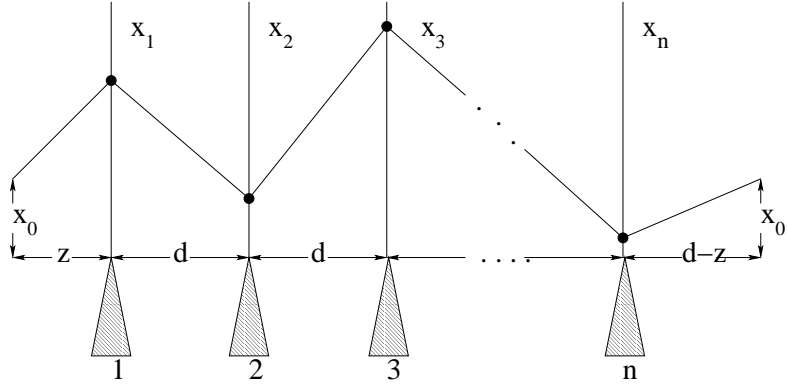}
\end{center}
\caption{Multiple diffraction near optical boundary of $n$ equally spaced wedges separated by distance $d$.}
\label{Kirchhoff_diff}
\end{figure}

A direct approach to  multiple singular scattering  consists in the construction of uniform approximation based on the Kirchhoff approach (see e.g. \cite{Sommerfeld}) which corresponds to the summation over all  trajectories  indicated at Fig.~\ref{Kirchhoff_diff}.  In this approximation the role of wedges is reduced to the restriction of integration domains to  half-lines $(0,\infty)$ (cf. Fig.~\ref{Kirchhoff_diff}).  This problem has been investigated in Ref.~\cite{BPS} where  it has been proved  that the contribution to the trace formula from such trajectories (i.e. $(n+1)$-fold integral over all $x_j$ at Fig.~\ref{Kirchhoff_diff}) can be calculated analytically even for a finite number ($n$) of wedges and the result is 
\begin{equation}
\rho^{(diff)}(E)=-\frac{d}{16\pi k} A_n \mathrm{e}^{\mathrm{i} k\,d\,n}+\mathrm{c.c.} , \qquad A_n=\frac{1}{\pi}\sum_{q=1}^{n-1}\frac{1}{\sqrt{q(n-q)}}\, .
\label{diff}
\end{equation}

For large number of scatters the sum over $q$ can be substituted by the integral and \cite{BS}
\begin{equation}
A_n\underset{n\to\infty}{\longrightarrow}1+\frac{2\zeta(1/2)}{\pi \sqrt{n}}
\label{limit}
\end{equation}
where $\zeta(s)$ is the Riemann zeta function  ($\zeta(1/2)=-1.460354$). 

It has been demonstrated in \cite{BS} that this result for multiple scattering on periodic set of wedges is equivalent to the specular reflection of  the incident wave from a straight (fictitious)  mirror which passes through the apex of all wedges. For small incident angle $\varphi$ the effective reflection coefficient for high-energy scattering determined from Eq.~\eqref{limit}  is the following 
\begin{equation}
R_0=-1-\sqrt{\frac{kd}{\pi}}(1-\mathrm{i})\zeta(1/2)\,\varphi, \qquad k=\sqrt{E} \, .
\label{specular_reflection}
\end{equation}


\subsection{Scattering on staggered periodic set of half-planes}\label{Carlson_Heins}

\begin{figure}
\begin{minipage}{.35\linewidth}
\begin{center}
\includegraphics[ width=.9\linewidth]{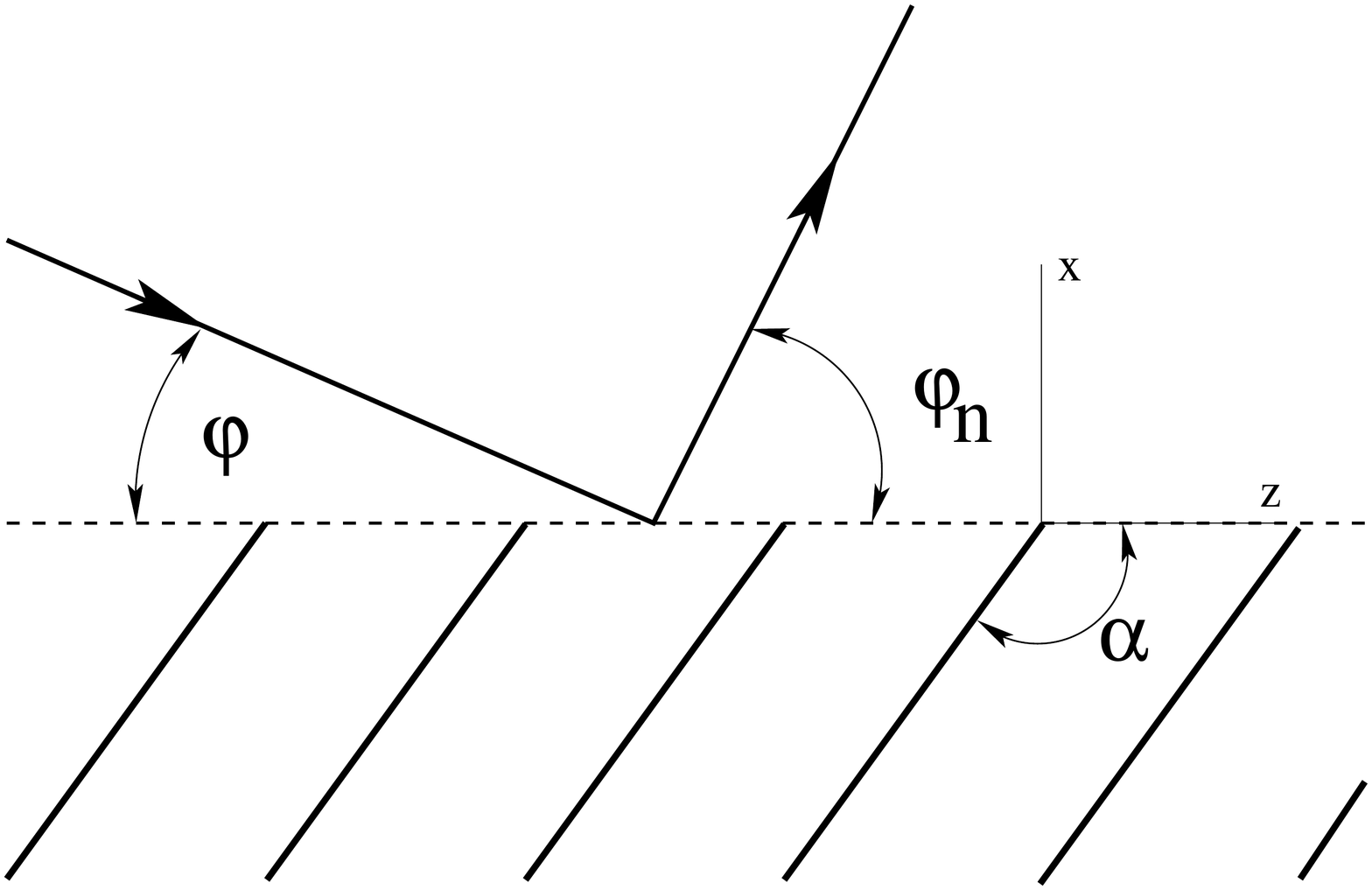}\\
a)
\end{center}
\end{minipage}
\begin{minipage}{.35\linewidth}
\begin{center}
\includegraphics[ width=.9\linewidth]{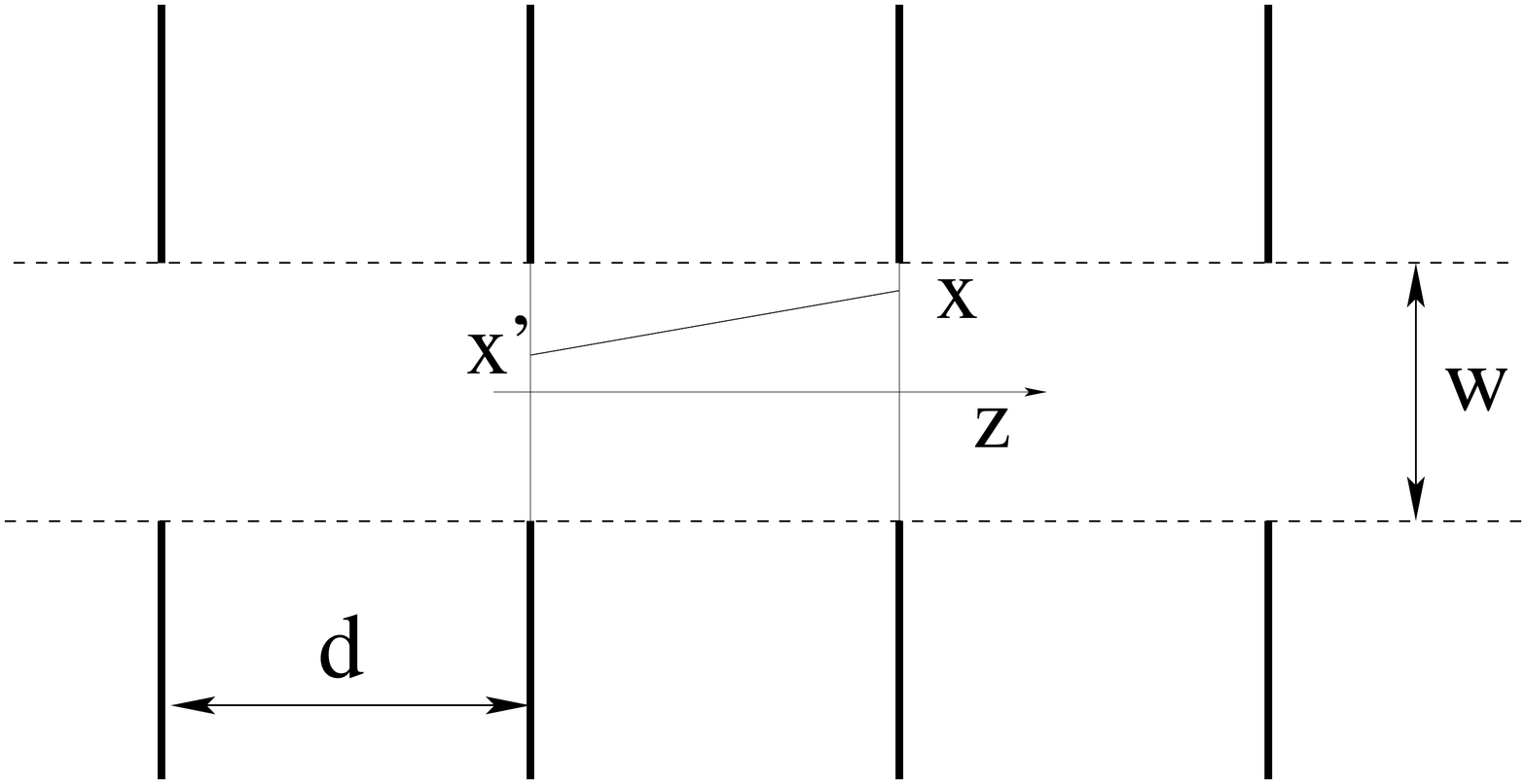}\\
b)
\end{center}
\end{minipage}
\begin{minipage}{.25\linewidth}
 \begin{center}
 \includegraphics[ width=.99\linewidth]{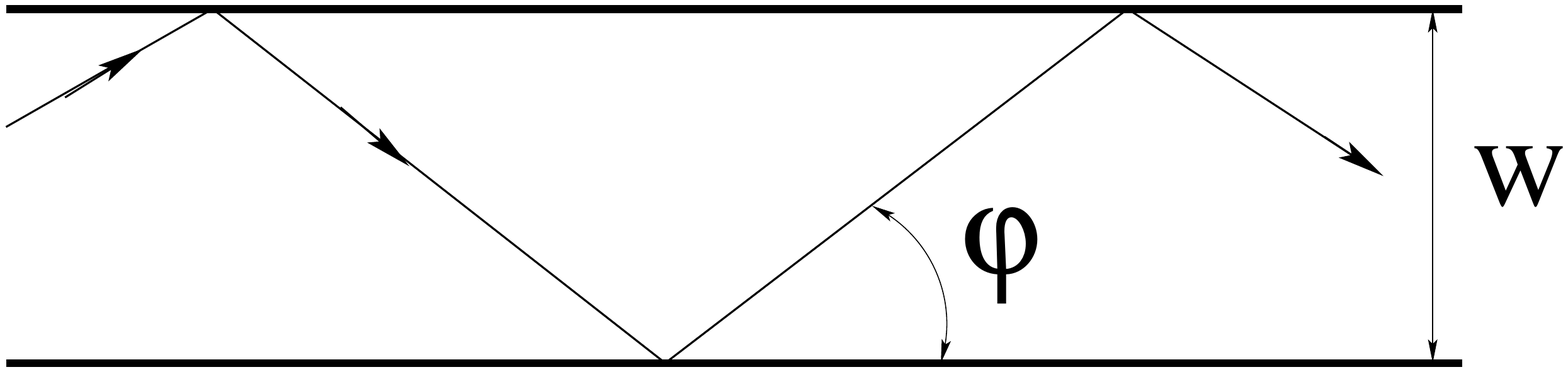}\\
 c)
 \end{center}
 \end{minipage}
\caption{(a) Scattering on periodic set of half-planes inclined at angle $\alpha$ with respect to a dashed straight line and separated by  perpendicular distance $a$. (b) Multiple diffraction of a periodic sequence of slits. (c) Wave propagating inside  a periodic orbit channel.}
\label{staggered}
\end{figure}

Though the Kirchhoff approximation discusses in the preceding Section does indicate that multiple singular diffraction leads to effective scattering from (fictitious) mirror formed by singular scatters  it is difficult, in general, to prove rigorously the applicability of this approximation. In this Section an exact solution of a similar problem of scattering  of a  plane wave with incident angle $\varphi$
\begin{equation}
\Psi^{(\mathrm{inc})}(z,x)=\mathrm{e}^{\mathrm{i} k(z\cos \varphi-x\sin \varphi)}
\end{equation}
on a  periodic set of half-planes separated by perpendicular distance $a$  is discussed. The apex of all half-planes belong to a straight line and planes are inclined with respect to this line by angle $\alpha$ (cf. Fig.~\ref{staggered}~a)).  

The field at large distances is the sum of the reflected (into the upper half-plane) and transmitted (into the lower half-plane) fields. 
The total reflected field is the sum of finite number of  reflected plane waves
\begin{equation}
\Psi^{(\mathrm{ref})}(z,x)=\sum_{n} R_n \mathrm{e}^{\mathrm{i}k\, (z\cos \varphi_n+x\sin  \varphi_n )}
\end{equation}
where $R_n$  are  reflection coefficients and   $\varphi_n$  are  reflected angles determined  due to the periodicity from the grating equation
\begin{equation}
 kd(\cos \varphi -\cos \varphi_n)=2\pi n,\quad n=\mathrm{integer},\quad 
 -Q\sin^2 \frac{\varphi}{2}\leq n \leq   Q\cos^2 \frac{\varphi}{2}.
 \label{grating}
\end{equation}
Here $d$ is the distance between the apex of half-planes, $d=a/\cos \alpha$, and $Q=kd/\pi$ is dimensionless momentum.

The total transmitted field is the same as  inside straight tubes built by half-planes with the Dirichlet boundary conditions 
\begin{equation}
\Psi_m^{(\mathrm{trans})}(z,x)=\sum_{m=1}^{m_{\mathrm{max}}}
T_m \mathrm{e}^{-\mathrm{i}\omega_m\, (x\sin \alpha+z\cos \alpha) } \sin \left (\frac{\pi m}{a}\, (x\cos \alpha-z\sin \alpha)\right ),\quad m_{\mathrm{max}}=\left [\frac{ka}{\pi}\right ]
\end{equation}
where   frequencies of transmitted waves $\omega_m=\sqrt{k^2-\left (\frac{\pi m}{a}\right )^2}$ and $T_m$ are  transmission coefficients.

It has been shown in \cite{Carlson} that the above problem is soluble  by the Wiener-Hopf method  but the calculations in that article  were performed only for low values of momenta. In \cite{BS} this problem has been reconsidered in the semiclassical limit $Q\to\infty$ 
and  it was demonstrated that in that limit infinite products inherent in the Wiener-Hopf method and, consequently,  reflection and transmission coefficients can be obtained analytically.  The most difficult (and the most interesting for us) is the case of  small-angle scattering when incident angle $\phi\to 0$  as within the optical boundary of one scatter there exist many other scatters.  

The main conclusions of Ref.~\cite{BS} for this problem in the limit  $\sqrt{Q}\varphi \ll1$ and $Q\to\infty$ are as  follows:
\begin{itemize}
\item The reflection coefficient with $n=0$ in Eq.~\eqref{grating} corresponding to the specular (mirror-like) reflection, $\varphi_0=\varphi$, is special  
\begin{equation}
R_0 =-1- \sqrt{Q}  \, (1-\mathrm{i})\zeta(1/2)\,\varphi .
\label{R_0_small}
\end{equation}
Notice that this expression coincides with Eq.~\eqref{specular_reflection} obtained in the Kirchhoff approximation. 
\item Reflection coefficients  when $n>0$ in Eq.~\eqref{grating} is kept fixed and $Q\to\infty$ corresponding   to  small reflection angle,   $\varphi_n\approx 2\sqrt{n/Q}$ (independent on incident angle $\varphi$ provided $\sqrt{Q}\varphi \ll1$) are small and proportional to 
$\sqrt{Q} \, \varphi$
\begin{equation}
R_n^{(\mathrm{small})}=\sqrt{Q} \, \varphi \, r_n, \qquad |r_n|^2=\frac{\mathrm{e}^{2\sqrt{n}\zeta (1/2)-2}}{n}
\prod_{\substack{m\neq n\\ m> 0} }^{\infty}\frac{1+\sqrt{n/m}}{1-\sqrt{n/m}}\mathrm{e}^{-2\sqrt{n/m}}\ .
\label{small_reflection}
\end{equation}
\item When $\pi/2<\alpha<\pi$  transmission is negligible and large-angle reflection coefficients  dominate  when $n$ is close to $n^{*}=Q\sin^2 \alpha$  and $\varphi_n$ is near to  $2\pi -2\alpha$ (as for the specular reflection from  the full inclined plane). For small $\varphi$ these coefficients are proportional to $\varphi$
\begin{equation}
 R_n^{(\mathrm{large})}=\frac{\varphi}{\sin 2\alpha} \, r(u_n(\alpha)),\quad |r(u)|^2= \mathrm{e}^{2\zeta(1/2)u} \prod_{m=1}^{\infty} 
\left (1+\frac{u}{\sqrt{m}}\right )^2
\left (1+\frac{u^2}{m}\right )\mathrm{e}^{-2u/\sqrt{m}}
\label{large_reflection}
 \end{equation}
where
\begin{equation} 
u_n(\alpha)=\frac{n-n^{*}}{\sqrt{Q}\sin 2\alpha}.
\label{u_n}
\end{equation} 
\item When $0<\alpha<\pi/2$ large-angle reflection coefficients are negligible and transmission coefficients are 
\begin{equation}
T_n=\varphi\, t(u_n(\pi-\alpha)),\qquad |t(u_n(\pi-\alpha))|^2=2|r(u_n(\pi-\alpha)|^2
\end{equation} 
with the same functions $r(u)$  and $u_n(\alpha)$ as in \eqref{large_reflection} and \eqref{u_n}.  
\end{itemize} 
 The main conclusion from the above expressions is that for the sliding-type multiple scattering when the incident angle is small, $\sqrt{Q}\,\phi\ll1$,   and $Q\to\infty$ the dominant contribution to the reflected field comes  only from one term with $n=0$ ($R_0\approx -1$ and $\varphi_n=\phi$). This term corresponds to the specular reflection  from fictitious mirror built from a straight line passing through singular scatters (indicated by  dashed lines at Figs.~\ref{staggered}) 
\begin{equation}
\Psi(z,x)\approx \mathrm{e}^{\mathrm{i}kz\cos \phi} \left [\mathrm{e}^{-\mathrm{i} kx\sin \phi}-\mathrm{e}^{\mathrm{i} kx\sin \phi}\right ]+\delta \Psi(z,x)
\end{equation}
where $\delta \Psi(z,x)$ is small when $\varphi \sqrt{Q}\ll 1$. For    $\pi/2<\alpha<\pi$
\begin{equation}
\delta \Psi(z,x)=\varphi \sqrt{Q}\Big [\sum_{n=0}r_n \mathrm{e}^{\mathrm{i}kz\cos \varphi_n+\mathrm{i}kx\sin \varphi_n}\Big ]
+ \frac{\varphi}{\sin 2\alpha}\Big [\sum_{u_n}r(u_n) \mathrm{e}^{\mathrm{i}kz\cos \varphi_n+\mathrm{i}kx\sin  \varphi_n}\Big ].
\end{equation}
The formation of quasi-mirror boundary where in the semiclassical limit the total field tends to zero  is a non-perturbative effect of small-angle multiple scattering on singular scatters (i.e. vertices with angle $\neq \pi/n$). 

The existence of the exact solution permits also to find a small leakage of the specular reflected wave after one scattering into other channels. The modulus of the amplitude of that wave  deviates from unity by a small amount (when $\sqrt{Q}\,\varphi\ll 1$) as follows
\begin{equation}
|R_0|^2=1-C\sqrt{kd}\,\varphi, \qquad C=-\frac{2\zeta(1/2)}{\sqrt{\pi} }\approx 1.65.
\label{leakage}
\end{equation}

\subsection{Scattering on periodic array of slits}\label{periodic_slits}

To investigate this phenomenon  further it is instructive  to investigate  the propagation of waves inside periodic array of slits with Dirichlet boundary conditions as indicated at Fig.~\ref{staggered}~b). The problem corresponds to find the solution of  the Helmholtz equation $(\Delta+k^2)\Phi(z,x)=0$  which vanishes at indicated slits and is generated by the plane wave along z-axis. 

In the Kirchhoff approximation (see e.g. \cite{Sommerfeld}) the wave $\Phi(d,x)$ at distance $d$ from the origin  is related with the wave  
$\Phi(0, x^{\prime})$ by the relation valid provided the width $w$ is much smaller than the distance between slits $d$
\begin{equation}
\Phi(d, x)= \mathrm{e}^{\mathrm{i}kd-\mathrm{i}\pi/4}\sqrt{\frac{k}{2\pi d}}\int_{-w/2}^{w/2}\mathrm{e}^{\mathrm{i}k(x-x^{\prime})^2/2d}\,
\Phi(0, x^{\prime})\mathrm{d} x^{\prime}.
\end{equation}
Periodicity of the slits requires that $\Phi(d, x)=\lambda  \Phi(0, x)$ where $\lambda$ determines the propagation and attenuation due to scattering on slits.  Therefore the considered problem is reduced to the following equation 
\begin{equation}
\mathrm{e}^{\mathrm{i}kd -\mathrm{i}\pi/4}\sqrt{\frac{k}{2\pi d }}\int_{-w/2}^{w/2}\mathrm{e}^{\mathrm{i}k (x-y)^2/2d }\Psi(y)\mathrm{d}y=
\lambda \Psi(x). 
\label{Kirchhoff}
\end{equation}
Here $\Psi(x)\equiv \Phi(0,x)$ is the value of the wave inside a slit.
  
\begin{figure}
\begin{minipage}{.49\linewidth}
\begin{center}
\includegraphics[width=.9\linewidth]{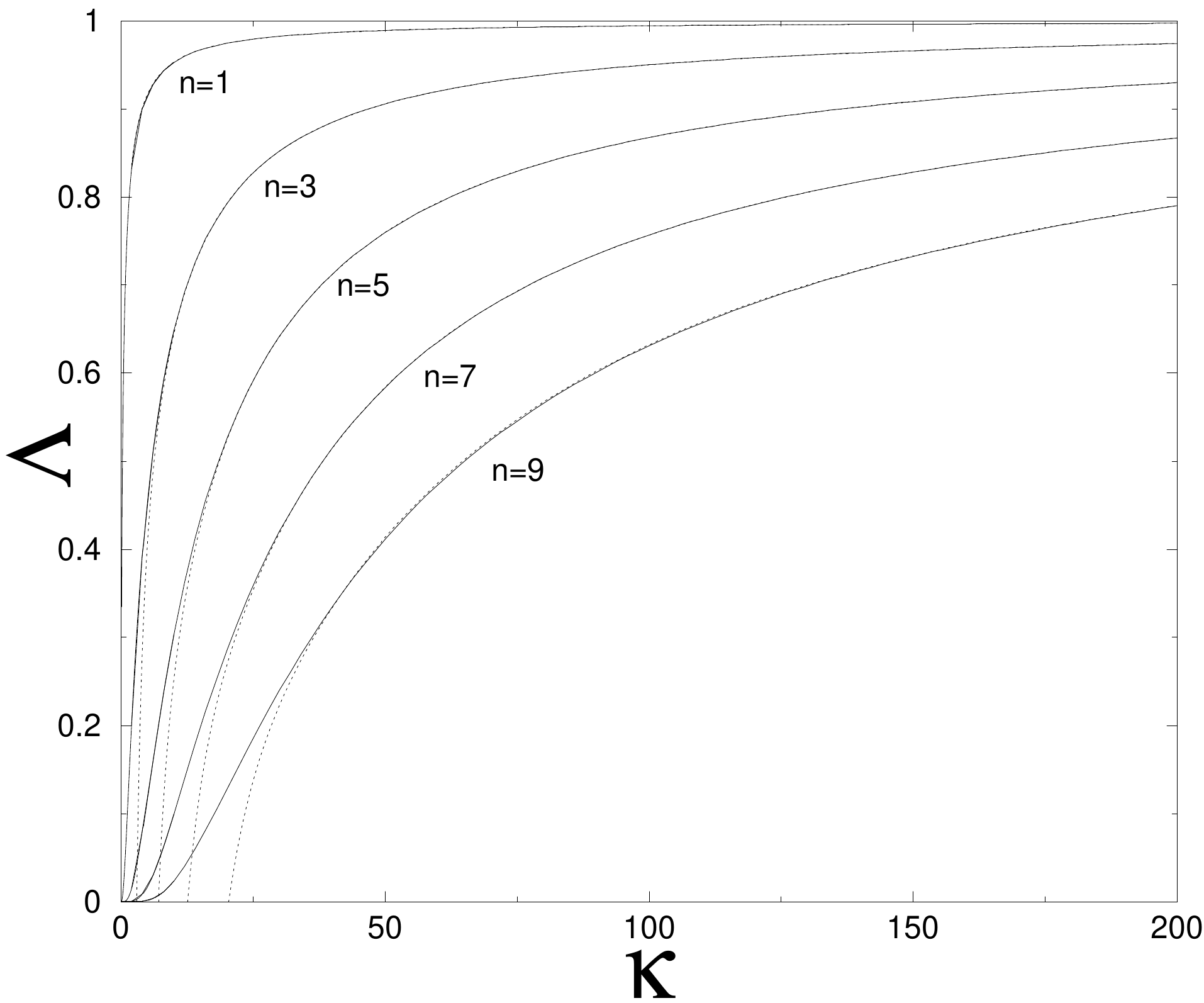}\\
a)
\end{center}
\end{minipage}
\begin{minipage}{.49\linewidth}
\begin{center}
\includegraphics[width=.9\linewidth]{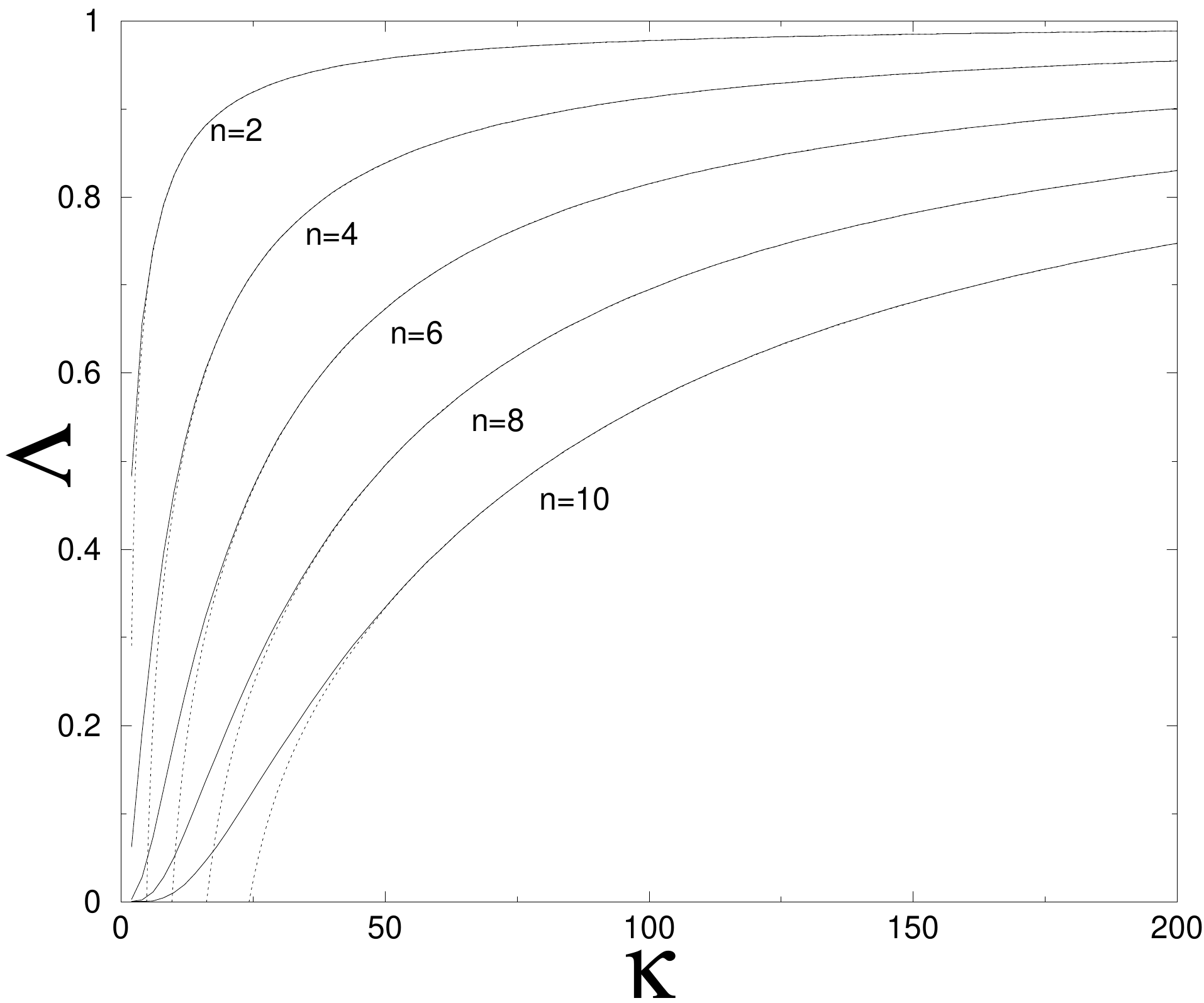}\\
b)
\end{center}
\end{minipage}
\caption{Ten largest eigenvalues of \eqref{real_eq} as functions of $\kappa$. (a) Even solutions. (b) Odd solutions. Dotted lines indicate approximate asymptotic expression \eqref{lambda_value}. }
\label{eigenvalues_scars}
\end{figure}

\begin{figure}
\begin{minipage}{.49\linewidth}
\begin{center}
\includegraphics[ width=.9\linewidth]{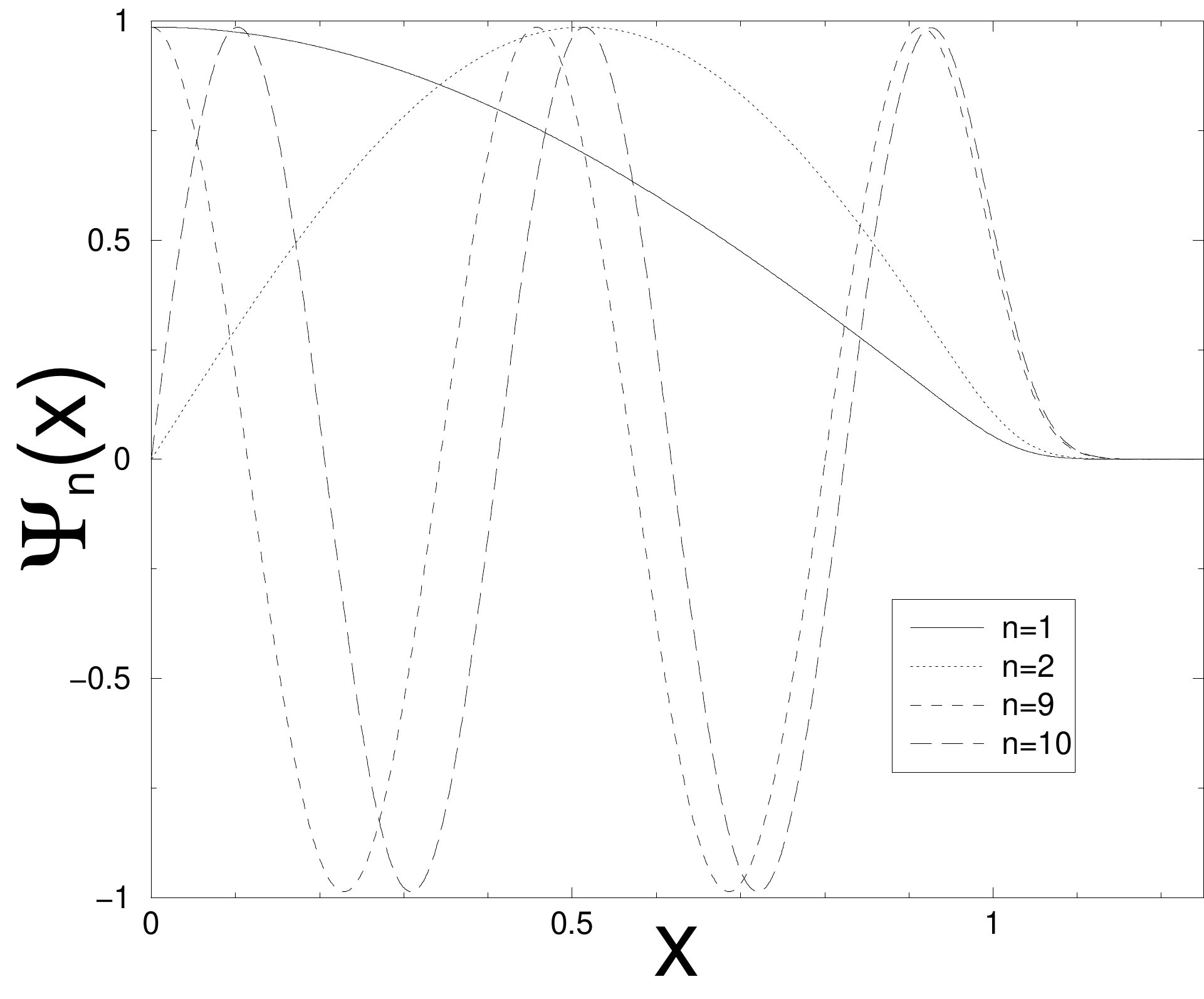}\\
a)
\end{center}
\end{minipage}
\begin{minipage}{.49\linewidth}
\begin{center}
\includegraphics[width=.9\linewidth]{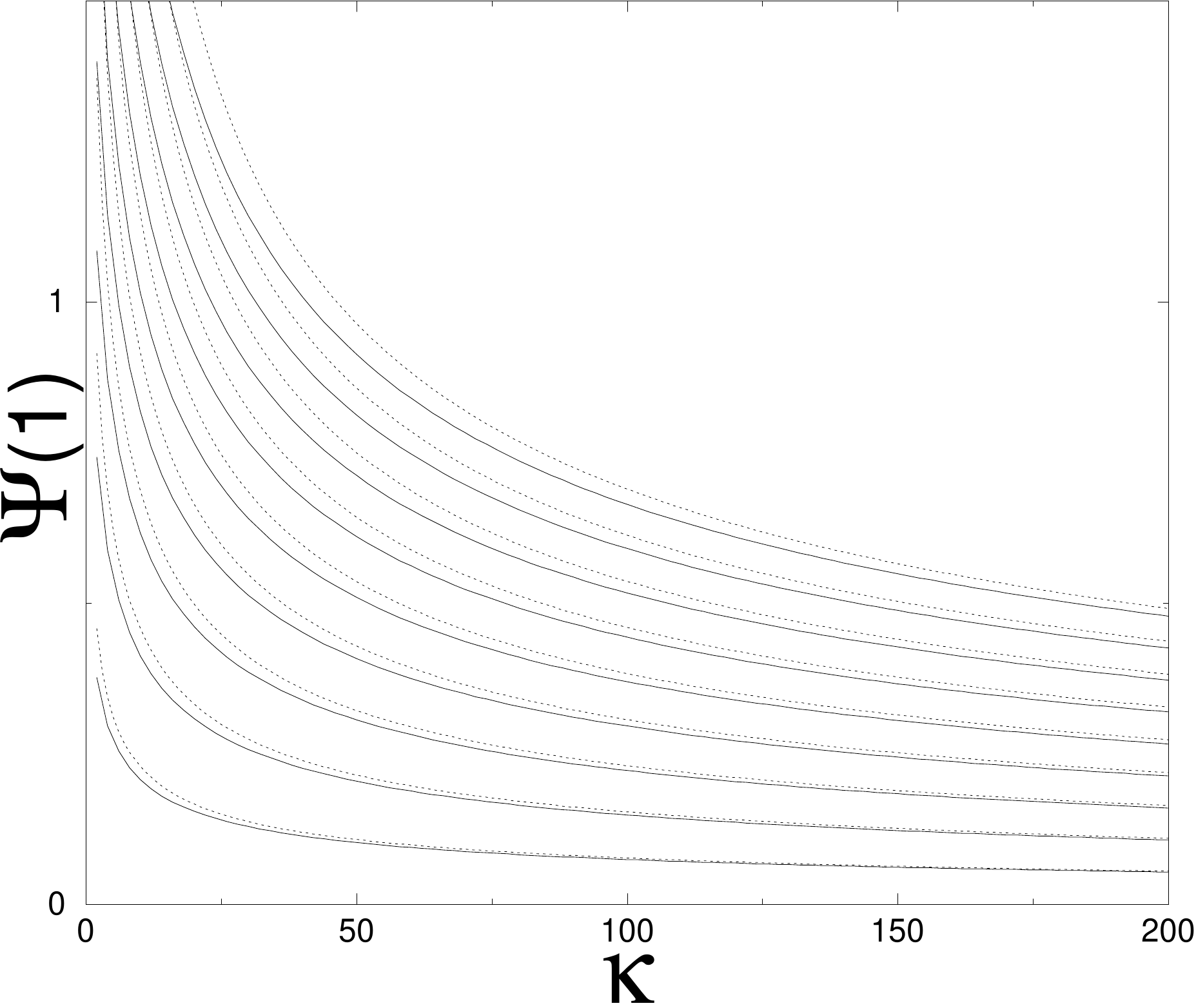}\\
b)
\end{center}
\end{minipage}
\caption{(a) Eigenfunctions corresponding to a few largest eigenvalues of  \eqref{real_eq} for $k=200$. (b) The value of eigenfunctions at the 	boundary of effective tube for 10 largest eigenvalues as at Fig.~\ref{eigenvalues_scars} for different values of $k$. From bottom to top $n=1,2,\ldots, 10$. Dashed lines indicate an approximate asymptotic formula \eqref{bondary_value}.}
\label{eigenfunctions_scars}
\end{figure}

No analytical  solution of this (simplified) equation is known. Nevertheless, as it has been discussed above,  in semiclassical limit $k\to\infty$ its solution should be close to waves propagating inside a rectangular slab restricted by straight lines passing through corners of the slits (denoted by dashed lines at Fig.~\ref{staggered}~b))
\begin{equation}
\Psi_n^{(\mathrm{approx})}(z,x)\sim \sin \big  (p_n(x+w/2)\big  )\,\mathrm{e}^{\mathrm{i}\sqrt{k^2-p_n^2}\,z },\qquad p_n=\frac{\pi}{w}n,\quad n=1,2,\ldots 
\label{slab_eigenfunctions}
\end{equation} 
To check this statement numerical calculation of Eq.~\eqref{Kirchhoff} was performed. To simplify numerics  all space variables were rescaled in units of  $w/2$ and  the Wick rotation has been performed. It leads to a simpler equation
\begin{equation}
\sqrt{\frac{\kappa }{\pi}}\int_{-1}^{1}\mathrm{e}^{-\kappa(x-y)^2}\Psi_n(y)\mathrm{d}y=\Lambda_n \Psi_n(x)
\label{real_eq}
\end{equation}
where dimensionless variable  $\kappa=-\mathrm{i} k w^2/8d$ and
\begin{equation}
\Lambda_n=\lambda\,  \mathrm{e}^{\mathrm{i}\left (\sqrt{k^2-p_n^2}-k\right )\,d}.
\label{slab_lambda}
\end{equation}
Eq.~\eqref{real_eq} is the Fredholm integral equation of the first kind with symmetric kernel and it has a discrete set of eigenvalues $\Lambda_n$  ($\Lambda_1\geq \Lambda_2,\geq \ldots$) and  eigenfunctions $\Psi_n(x)$  which were determined numerically. Due to the symmetry solutions are either even or odd with respect to coordinate inversion:  $\Psi_n(-x)=(-1)^{n+1} \Psi_n(x)$. At Fig.~\ref{eigenvalues_scars}  ten largest eigenvalues of this equation are represented for different values of parameter $\kappa$.  At Fig.~\ref{eigenfunctions_scars} a few corresponding  eigenfunctions are plotted for $\kappa=200$. 

The main conclusion of these  (and others) calculations is  that in the semiclassical limit $k\to\infty$ eigenfunctions of \eqref{real_eq} are indeed well described by simple waves  as in Eq.~\eqref{slab_eigenfunctions}.  An important characteristics of such waves is the requirement that they vanish as $\kappa \to\infty$ at boundaries of effective slab which implies the quantisation of transverse momentum. At Fig.~\ref{eigenfunctions_scars}~b)  the values of true eigenfunctions at the boundary, $\Psi_n(1)$, are plotted. It is clearly seen that with fixed $p_n$ and increasing of $\kappa$ this value indeed tends to zero. It was observed that these values are well described by the following asymptotic formula
\begin{equation}
\Psi_n(1)\underset{\kappa \to\infty}{\longrightarrow} |\tilde{p}_n|\left (\frac{1}{2\sqrt{\kappa}}-\frac{1}{8\kappa}\right )
\label{bondary_value}
\end{equation}
where dimensionless transverse momentum $\tilde{p}_n=\pi n /2$. 

Similar asymptotic formula are also established for eigenvalues $\Lambda_n$ of Eq.~\ref{real_eq}
\begin{equation}
\Lambda_n\underset{\kappa \to\infty}{\longrightarrow} 1-\frac{\tilde{p}_n^2}{4\kappa}+.206 \frac{\tilde{p}_n^2}{\kappa^{3/2}}
\label{lambda_value}
\end{equation}
At Fig.~\ref{eigenvalues_scars} the comparison of this formula with numerical calculations is performed and good agreement have been found for large $\kappa$ and fixed momenta $\tilde{p}_n$. 

The first term of the series \eqref{lambda_value} corresponds to large $k$ expansion of Eq.~\eqref{slab_lambda}.  The second term can be interpreted as a complex shift  of wave energy when propagating inside the slits. Expanding longitudinal momenta $\sqrt{k^2+\delta E-p_n^2}$ in the exponent of  Eq.~\eqref{slab_lambda} and comparing coefficients with  \eqref{lambda_value}  one gets (in the original units)
\begin{equation}
 \delta E=C(1+\mathrm{i}) p_n^2 \, \sqrt{\frac{d}{kw^2}}, \qquad p_n=\frac{\pi}{w}n
\label{imaginary_part}
\end{equation}
where constant $C\approx 1.65$ is numerically the same as in Eq.~\eqref{leakage}, $C=-2\zeta(1/2)/\sqrt{\pi}$.  The equality of these two constants can be explained as follows. The appearance of the imaginary part of  wave energy physically means that  propagating wave  escapes into other channels. The modulus squared of this wave after passing the distance $L$ decreases by $\mathrm{Im}\, \delta E \,L/k$. According to Eq.~\eqref{leakage}  after each collision with slits this quantity decreases by $C\sqrt{kd} \varphi$. When a wave propagates with incident angle $\varphi$ along distance $L$ it has  approximately $L/(w/\varphi)$ collisions.  Therefore the total leakage is 
\begin{equation}
  \frac{L}{k}\, \mathrm{Im}\, \delta E=C\sqrt{kd} \, \varphi^2 \, \frac{L}{w} .
\end{equation}
As $\varphi\approx p_n/k$ one reproduces the imaginary part of Eq.~\eqref{imaginary_part}.


\subsection{Application to polygonal billiards}\label{applications}

 The multiple scattering on singular wedges with $\alpha\neq \pi/n$ is in general a complicated problem, especially when optical boundaries of different scatters overlap. The above discussion  proofs that in semiclassical limit when singular scatters are arranged along a straight line and the incident wave is inclined with a small angle with respect to this line  the reflected wave dominates by a specular reflection from that line though the line itself does not constitute a physical boundary.  The Kirchhoff approximation discussed in Section~\ref{Kirchhoff_approximation} clearly demonstrates that this result is independent of  wedge shapes.
 
 Such non-perturbative effect is especially important for polygonal billiards where classical periodic orbits appear in families which after unfolding form infinite periodic pencils (or channels) limited from the both sides by singular vertices (cf. Fig.~\ref{octagon}).  Consider one pencil corresponding to a primitive periodic orbit with period $L_p$ and let $w$ be its  width (see Fig.~\ref{staggered}~c)).
Of course, the horizontal pencil boundaries do not exist but they are constituted by singular scatters.  Due to multiple singular diffraction a wave propagating inside such pencil approximately vanishes at effective  horizontal boundaries and therefore will take the form of a plane wave as in Eq.~\eqref{slab_eigenfunctions}
\begin{equation}
\Psi(z,x)=\sin\left (p(x+w/2)\right )\mathrm{e}^{\mathrm{i}q z}, \qquad p=\frac{\pi}{w}n,\quad n=1,2,\ldots 
\label{plane_wave}
\end{equation}
where due to periodicity, $\Psi(z+L_p,x)=\pm \Psi(z,x)$, longitudinal momenta $q$ is also quantised  
\begin{equation}
q=\frac{\pi}{L_p}m, \qquad m=\mathrm{integer}.
\end{equation}
The energy of such wave is 
\begin{equation}
\mathcal{E}_{m,n}=\frac{\pi^2\, m^2}{L_p^2 } +\frac{\pi^2\, n^2}{w^2} .
\end{equation}
It is plain that such wave is only an approximation to (a much more complicated) exact solution. The validity of this approximation is governed by parameter $\varphi \sqrt{k L_p/\pi } \ll1 $ where $\varphi$ is the angle  between the wave direction and the horizontal boundaries. For the plane wave  \eqref{plane_wave}  $\varphi\approx p/k$. Therefore the wave \eqref{plane_wave} will be good approximation provided the following condition is fulfilled
\begin{equation}
p\sqrt{\frac{L_p}{k\pi }}\leq \lambda_0\sim 1.
\label{parameter}
\end{equation}
As $p=\pi n/w$ the values of integer $n$ are restricted
\begin{equation}
1\leq n\leq \lambda_0 w \sqrt{\frac{k}{\pi L_p}}.
\label{restriction_n}
\end{equation}
The requirement that $n\geq 1$ leads to the conclusion that at fixed energy not all periodic orbit  pencils can support propagating waves.  
As  $w\,L_p=\gamma A$ where $A$ is the billiard area and $\gamma=\mathcal{O}(1)$, the length of propagating channel is restricted as follows
\begin{equation}
L_p\leq L_{\mathrm{max}}=\delta k^{1/3},\qquad \delta=(A\lambda_0\gamma/\sqrt{\pi})^{2/3}.
\end{equation} 
Long-period channels with $L_p>L_{\mathrm{max}}$ are closed and cannot support propagating waves.

An important property of discussed propagating waves is that they become more visible (i.e. more isolated from other states) when the parameter \eqref{parameter} is decreasing. But for a given periodic orbit when transverse momentum $p$ is kept fixed but energy increases this parameter goes to zero. Consequently  in the semiclassical limit any periodic orbit pencil may and will support such propagating quasi-modes. That phenomenon resembles the formation of scars around of periodic orbits in chaotic systems \cite{scar_heller}--\cite{scar_berry} but contrary to the usual scars the discussed quasi-modes  become  practically exact in the semiclassical limit. It explains the name superscars proposed for these quasi-modes. In the next Section many examples of such superscars are presented for the billiards  depicted at Fig~\ref{examples}. 
 

\section{Examples of superscars in triangular and barrier billiards}\label{superscar_examples}

\begin{figure}
\begin{minipage}{.49\linewidth}
\begin{center}
\includegraphics[width=1.2\linewidth]{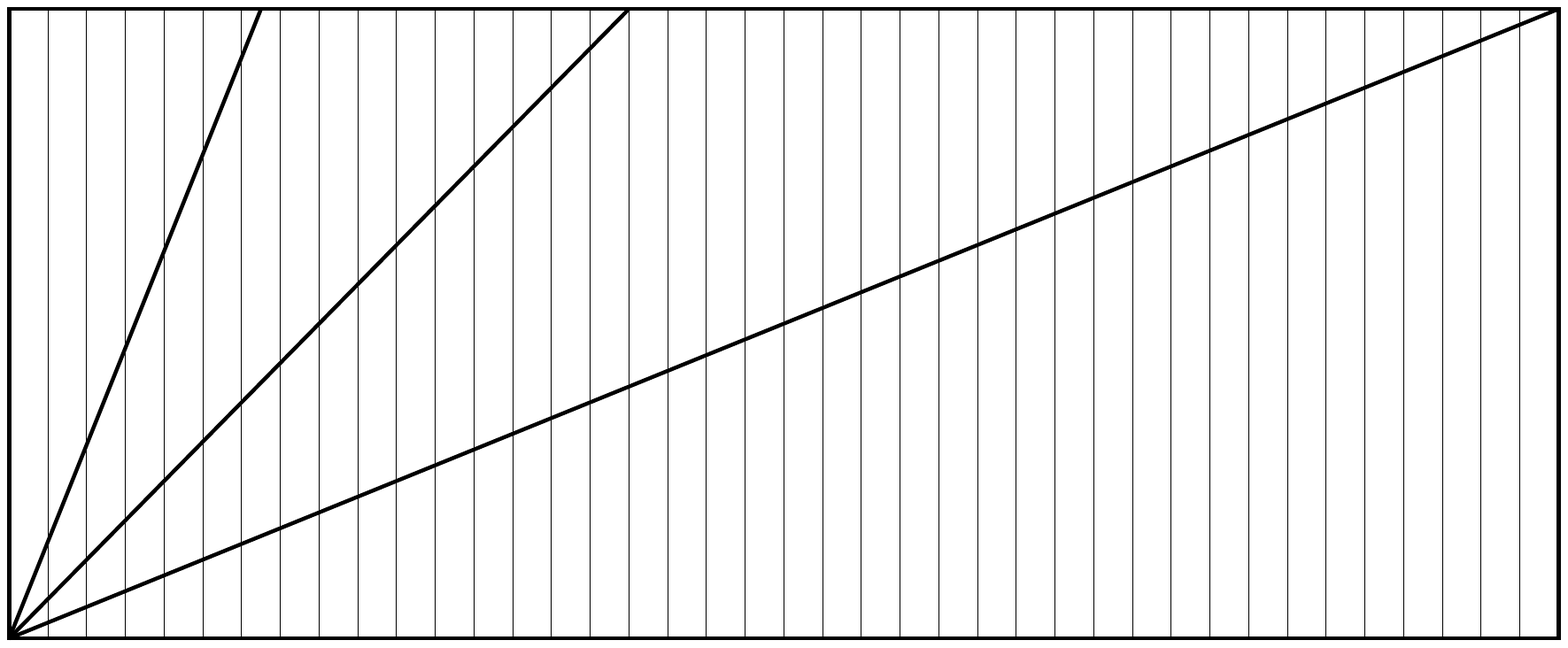}\\
a)
\end{center}
\end{minipage}
\begin{minipage}{.49\linewidth}
\begin{center}
\includegraphics[width=1.2\linewidth]{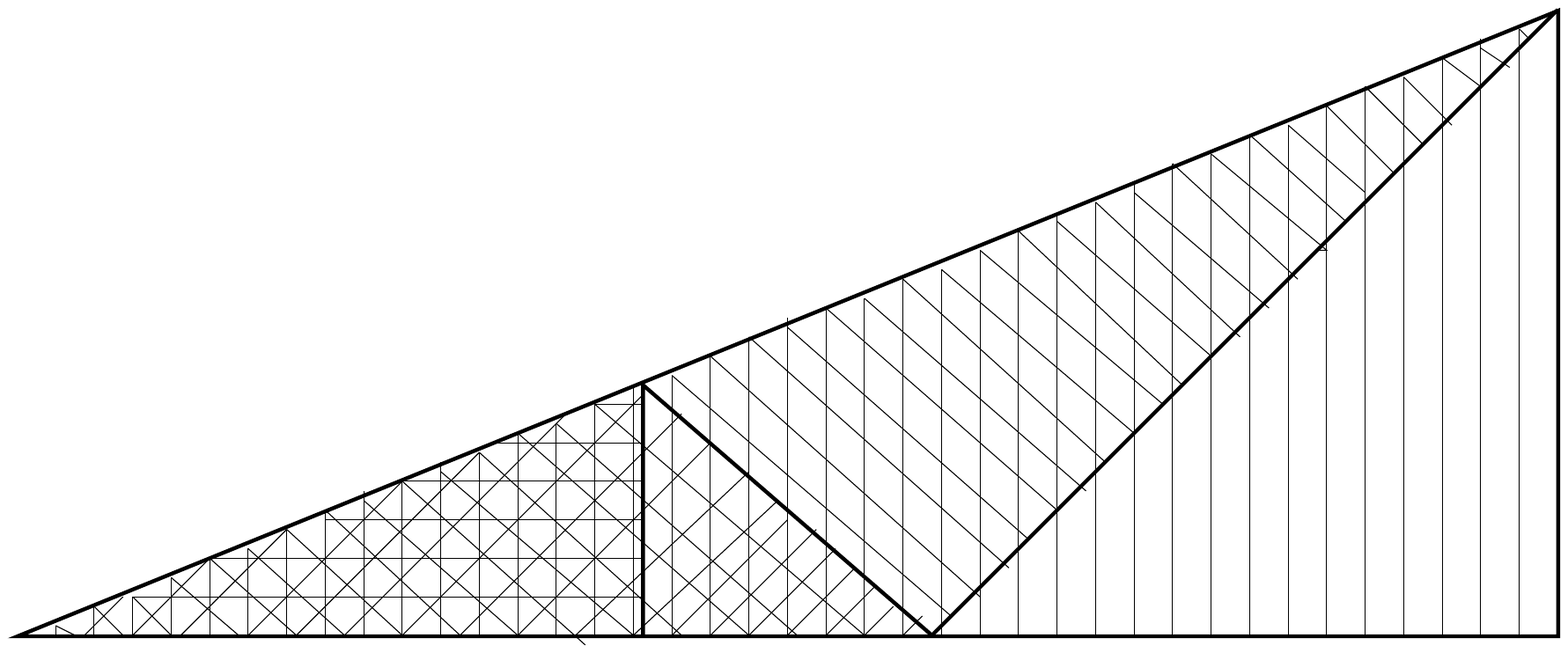}\\
b)
\end{center}
\end{minipage}
\caption{(a) Unfolded pencil of the simplest periodic orbit. Vertical lines  schematically indicate propagating wavefronts.  (b) The same but folded back into the original triangle. }
\label{simplest}
\end{figure}

Consider the billiard in the shape of the right triangle with one angle equals $\pi/8$ as at Fig~\ref{examples}~a).  One of the simplest periodic orbit family of this billiard corresponds to trajectories  perpendicular to the both  catheti as indicated at Fig.~\ref{octagon} b). When unfolded this family fills the rectangular pencil shown at Fig.~\ref{octagon} c). The length of this rectangle (i.e. the periodic orbit length) equals twice the length of the largest cathetus and its width is the length of the smallest cathetus. According to the above discussed multiple scattering on singular points the  superscar wave should propagate inside this rectangle with the Dirichlet boundary conditions on horizontal boundaries. As vertical boundaries are a part of the triangle boundaries the wave have to vanish on these boundaries as well.  Taking into account symmetry of the problem one concludes that the unfolded superscar wave in the semiclassical  limit obeys the Dirichlet boundary conditions on all sides of the rectangle indicated at Fig.~\ref{simplest} and has the form
\begin{equation}
\Psi_{m,n} (z,x)= \frac{2}{\sqrt{ab}}\sin \Big (\frac{\pi}{a}m z\Big ) \sin \Big (\frac{\pi}{b}n z\Big )\Theta(z)\Theta(b-z)
\label{triangle_wave}
\end{equation}
where $a$, $b$ are lengths of respectively the largest and the smallest cathetus ($b=\tan(\pi/8) a$). Two Heaviside $\Theta$-functions 
($\Theta(x)=1$ if $x>0$ and $\Theta(x)=0$ if $x<0$) are introduced to stress that this expression exists only inside the rectangle.  The energy of such state is
 \begin{equation}
 \mathcal{E}_{m,n}=\frac{\pi^2\,m^2 }{a^2}+\frac{\pi^2\, n^2 }{b^2}.
 \label{triangle_energy}
 \end{equation}
 
The superscar wave is simple (cf. Eq.~\eqref{triangle_wave}) only after unfolding. When folding inside the original triangle it takes the following form
\begin{equation}
\Psi^{(\mathrm{superscar})}(z,x)=\Psi_{m,n}(z,x)-\Psi_{m,n}\left (\frac{z+x}{\sqrt{2}},\frac{z-x}{\sqrt{2}}\right )
+\Psi_{m,n}\left (\frac{z-x}{\sqrt{2}},\frac{z+x}{\sqrt{2}}\right )-\Psi_{m,n}(x,z)
\label{exact_folding}
\end{equation}
where $\Psi_{m,n}(z,x)$ is given by Eq.~\eqref{triangle_wave} (with $\Theta$-functions included).

To examine the correspondence between (approximate) superscar waves  and true quantum eigenfunctions numerical calculations of high-excited states  were performed. The area of the billiard is normalised to $4\pi$ in order that the mean distance between  consecutive high-energy levels equals 1.  To find what numerically calculated (true) eigenfunctions resemble  to superscar waves the following procedure has been used.  First, values of integers $m$ and $n$ with $n\ll m$ were chosen and the superscar energy was calculated from Eq.~\eqref{triangle_energy}. Then from numerically calculated eigenvalues the one closest to the superscar energy has been selected. In  all investigated cases the corresponding eigenfunction  reveals clear picture very similar to the folded superscar wave \eqref{exact_folding}. 

A few examples of such comparison are presented below. At Fig.~\ref{50_1}~a) the folded superscar wave \eqref{exact_folding} with $m=50$ and $n=1$ is plotted. Notice the characteristic  picture of propagating wave fronts. At  Fig.~\ref{50_1}~b) the exact eigenfunction with energy $E_{\mathrm{exact}}=407.4$ which differs from the superscar energy by $0.2$ is shown. Though the energy is not too big this eigenfunction resembles well the superscar wave. At Figs.~\ref{79_1_110_1} and \ref{148_1_201_1} the exact eigenfunctions corresponding to superscar waves with $(m,n)$ equal respectively $(79,1)$,  $(110,1)$, $(148,1)$, and $(201,1)$ are presented. These eigenfunctions clearly have the same structure that  superscar waves  and the indicated exact energies agree well with superstar energies calculated from Eq.~\eqref{triangle_energy}.   

\begin{figure}
\begin{minipage}{.49\linewidth}
\begin{center}
\includegraphics[angle=90, width=.99\linewidth,clip]{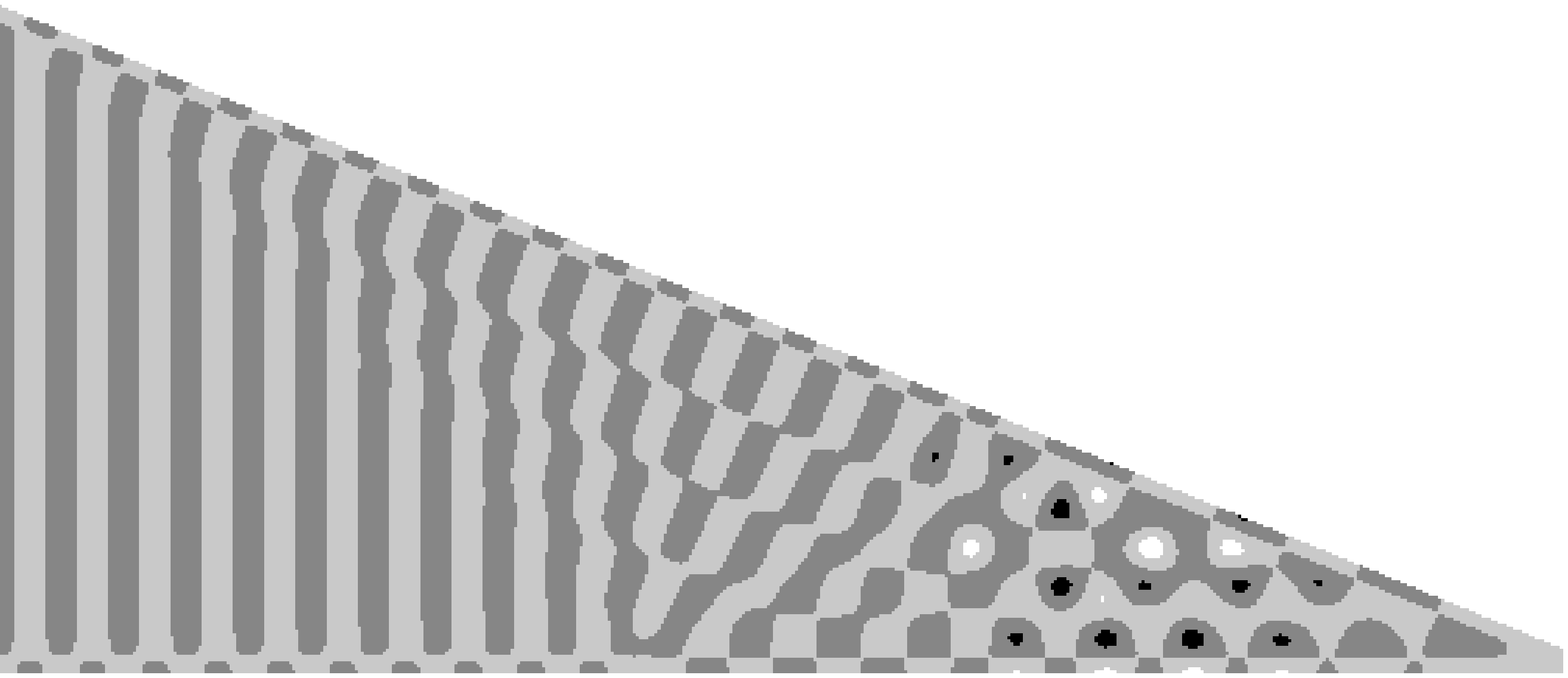}\\

   a)
\end{center}
\end{minipage}
 \begin{minipage}{.48\linewidth}
\begin{center}
\includegraphics[angle=90, width=.99\linewidth,clip]{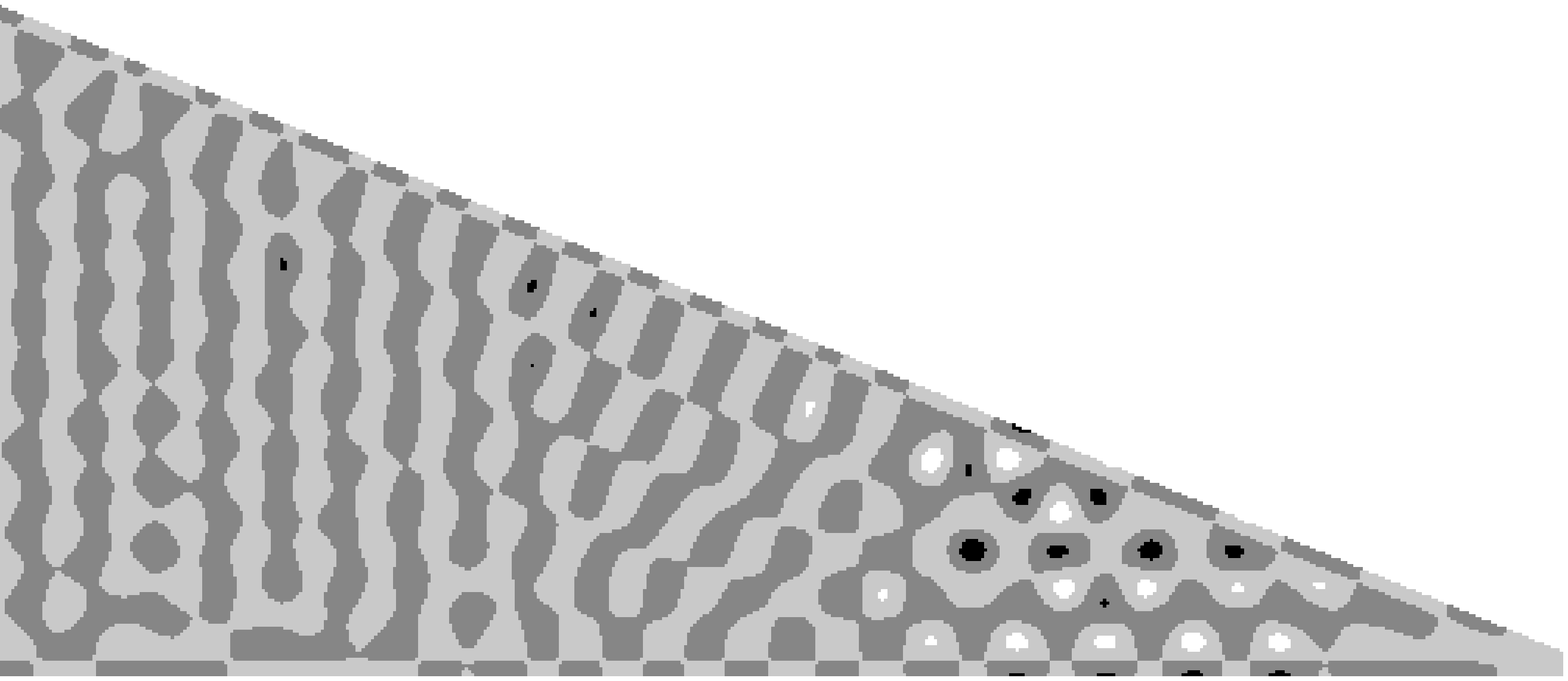}\\

   b)
\end{center}
\end{minipage}
\caption{(a) The folded superscar function given by Eq.~\eqref{exact_folding} for $m=50$ and $n=1$ with energy $\mathcal{E}_{50, 1}=407.6$. (b) Numerically calculated eigenfunction with energy $E_{\mathrm{exact}}=407.4$.}
\label{50_1}
\end{figure}

\begin{figure}
\begin{minipage}{.49\linewidth}
\begin{center}
\includegraphics[angle=90, width=.99\linewidth]{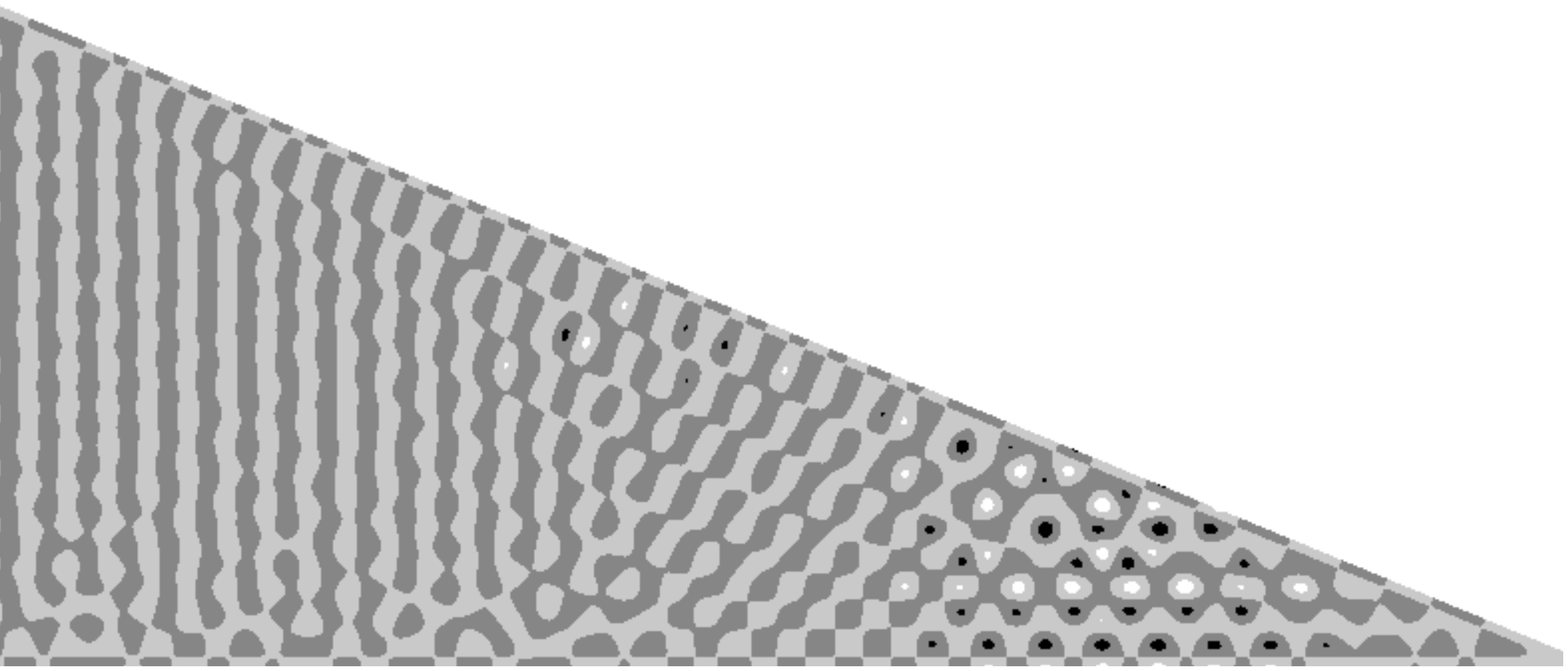} \\
a)
\end{center}
\end{minipage}
 \begin{minipage}{.49\linewidth}
\begin{center}
\includegraphics[angle=90, width=.99\linewidth]{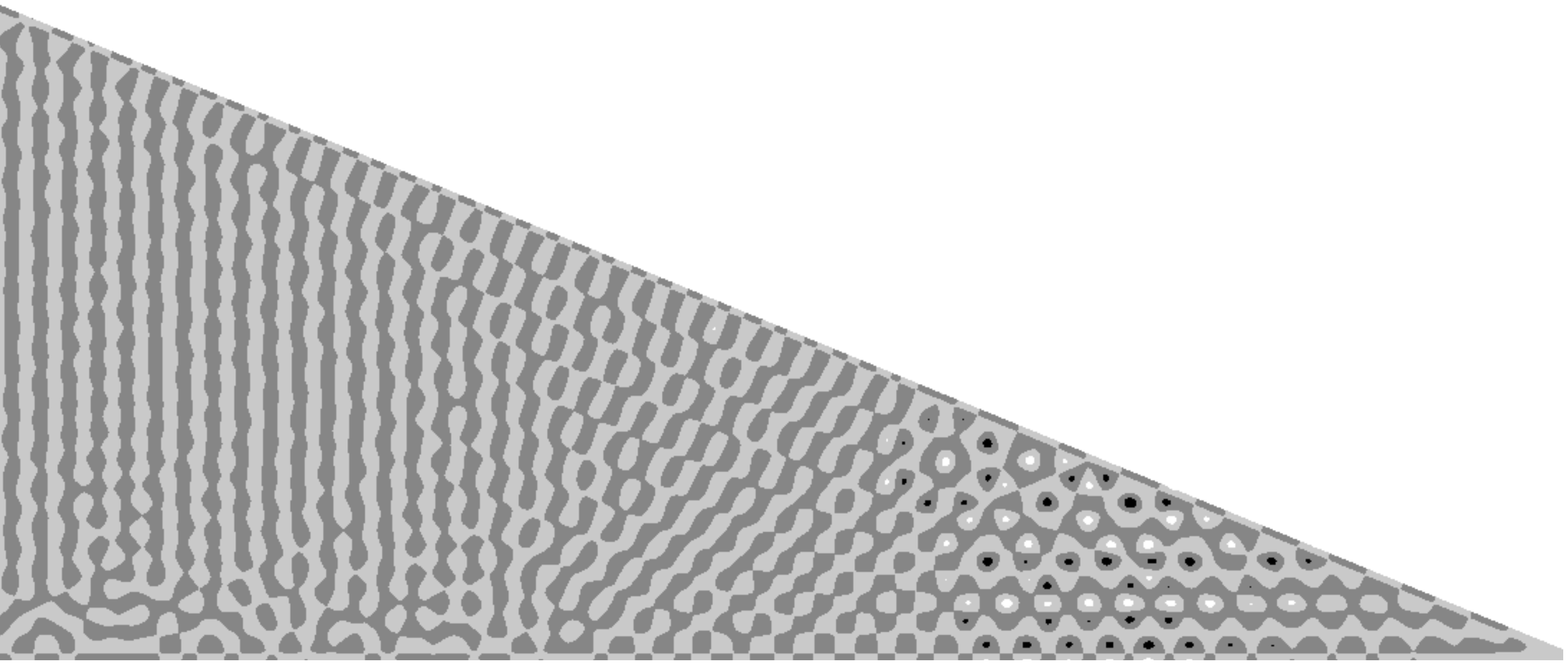} \\
b)
\end{center}
\end{minipage}
\caption{(a) The same as at Fig.~\ref{50_1} b) but with energy $E_{\mathrm{exact}}=1015.9$. The corresponding superscar energy $\mathcal{E}_{79,1}=1016.12$. (b) The same but with energy $E_{\mathrm{exact}}=1968.97$. The superscar energy $\mathcal{E}_{110, 1}=1969.15$.}
\label{79_1_110_1}
\end{figure}

\begin{figure}
\begin{minipage}{.49\linewidth}
\begin{center}
\includegraphics[angle=90, width=.99\linewidth]{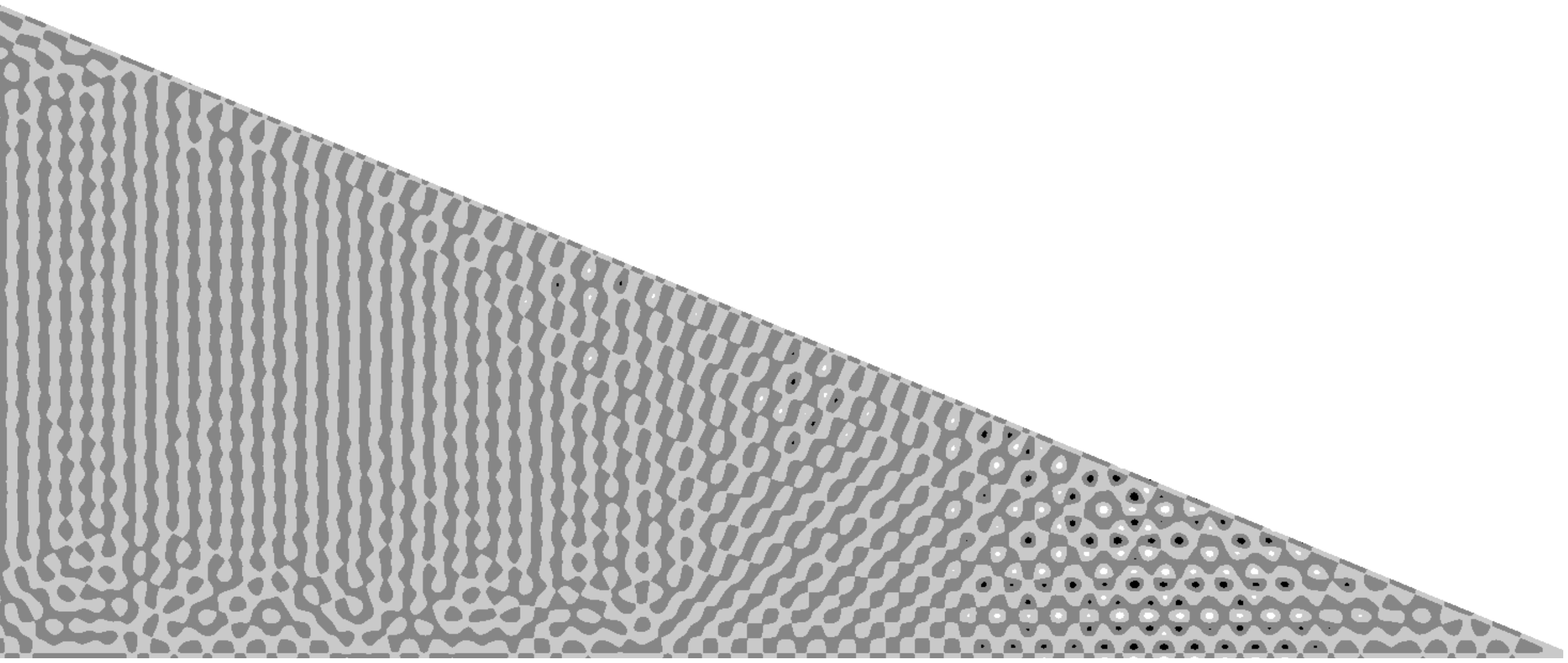} \\
a)
\end{center}
\end{minipage}
 \begin{minipage}{.49\linewidth}
\begin{center}
\includegraphics[angle=90, width=.99\linewidth]{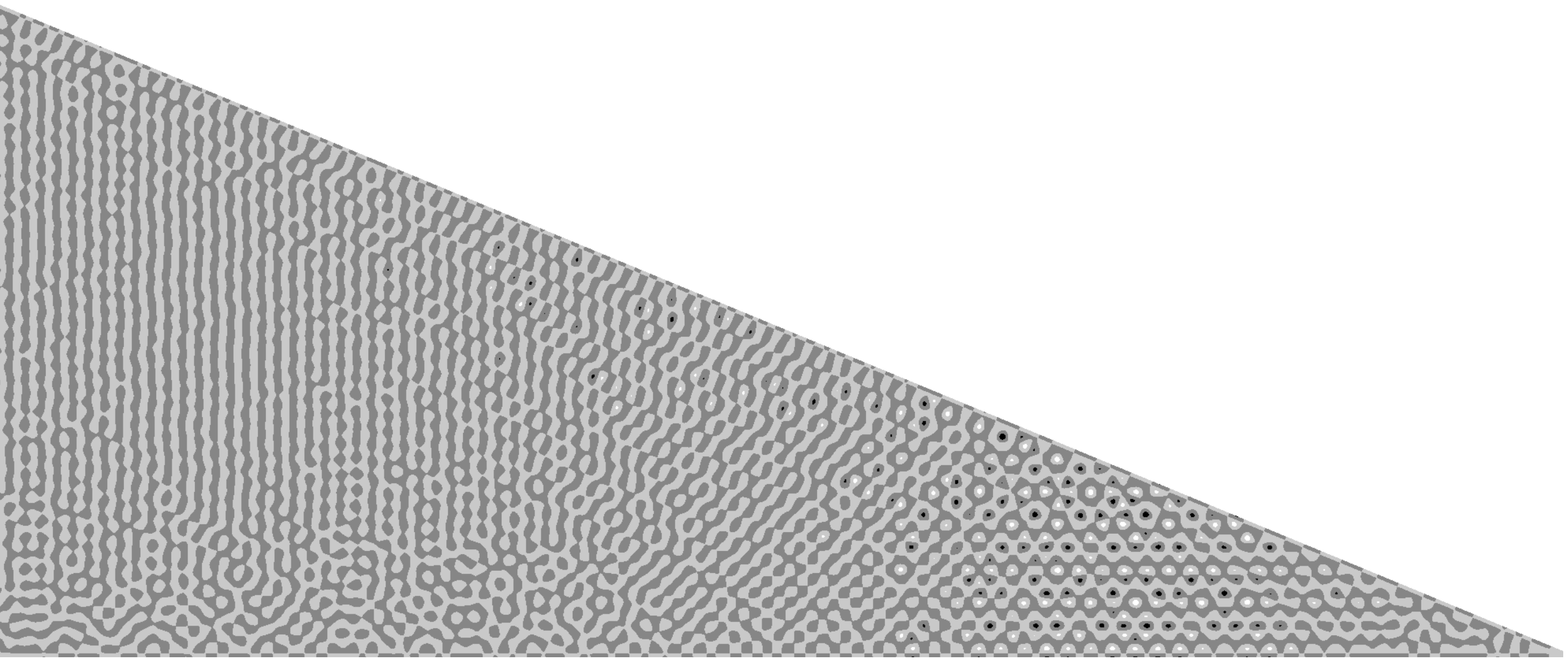} \\
b)
\end{center}
\end{minipage}
\caption{(a) The same as at Fig.~\ref{50_1} b) but with energy $E_{\mathrm{exact}}=3563.91$. The superscar energy $\mathcal{E}_{148, 1}=3563.88$. (b) The same but with energy $E_{\mathrm{exact}}=6572.47$. The superscar energy $\mathcal{E}_{201, 1}=6572.63$.}
\label{148_1_201_1}
\end{figure}

Figs.~\ref{50_1}-\ref{148_1_201_1}  clearly validate  the formation of superscar states around the simplest periodic orbit pencil in the triangular billiard.  But even that orbit requires 5 scatterings from the boundary (cf. Fig.~\ref{octagon} b))  and folded superscar function is complicated (cf. Fig.~\ref{50_1}~a)). Longer periodic orbit pencils   will necessarily be more elaborate and, consequently,  the structure of corresponding superscar functions would be less clear. 

To visualise better superscar structures, it is convenient to investigate the barrier billiard as at Fig.~\ref{examples}~b) where short-period orbits are simpler (see below). In numerical calculations only symmetric modes of this billiard were considered. Now the problem is reduced to solving the Helmholtz equation $(\Delta+k^2)\Psi(x,y)=0$ where  $\Psi(x,y)=0$ at all boundaries of the 
desymmetrised rectangle indicated at Fig.~\ref{examples}~c) except the segment $x=0, b/2<y<b$ where $\partial \Psi(x,y)/\partial x=0$. 

In calculations the aspect ratio of the barrier billiard, $b/a$, is chosen equal to $\sqrt{\sqrt{5}+1}\approx 1.8$ and the area of the billiard is normalised to $4\pi$.   A bunch of high excited eigenfunctions  around the $10000^{\mathrm{th}}$ level for this billiard was obtained numerically and eigenfunctions corresponding to a few superscar waves were selected  as it has been discussed above.  For clarity at certain figures below nodal domains of these eigenfunctions were plotted. Black (white) regions correspond to points where $\Psi(y,x)>0$ and $\Psi(y,x)<0$ respectively.  At other figures it was more convenient to show  grey images of the eigenfunction modulus. 

The structure of periodic orbit pencils in the barrier billiard is discussed in detail in Appendix~\ref{appendix_A}.  Any primitive periodic orbit in such billiard is characterised by 2 co-prime integers $n_a$ and $n_b$ which count the shifts by $2a$ and $2b$ in horizontal and vertical directions respectively on the unfolded rectangular lattice. Below such orbit is denoted by $(n_a-n_b)$. The length of such orbit is $L_p=\sqrt{(2a n_a)^2+(2b n_b)^2}$. In the barrier billiard periodic orbit  pencil with even $n_a$  has the width $w=4ab/L_p$ and such pencil fills the whole rectangle.  For odd $n_a$ there exit two different pencils of width $w=2ab/L_p$. Both pencils  may support superscar waves but the one with odd longitudinal quantum number $m$ and the other one with even $m$. The difference is due to different phases of reflection on boundaries with the Dirichlet boundary conditions. 

\begin{figure}
\begin{minipage}{.49\linewidth}
\begin{center}
\includegraphics[width=.99\linewidth]{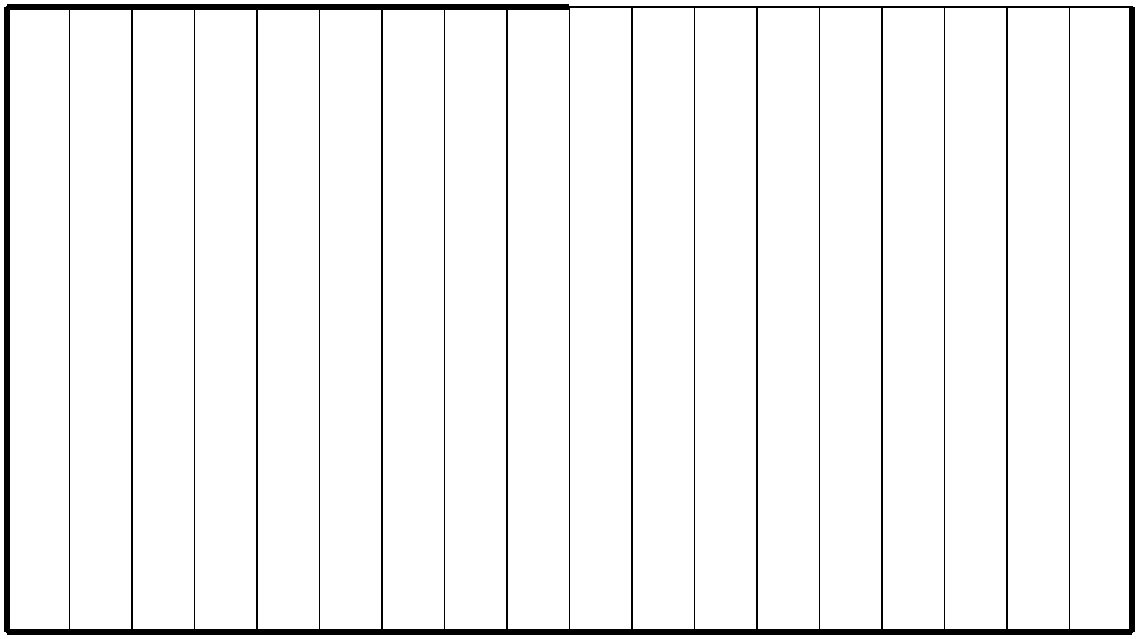} \\
a)
 \end{center}
 \end{minipage}
 \begin{minipage}{.49\linewidth}
 \vspace{-15pt}
\begin{center}
\includegraphics[angle=-90, width=.99\linewidth]{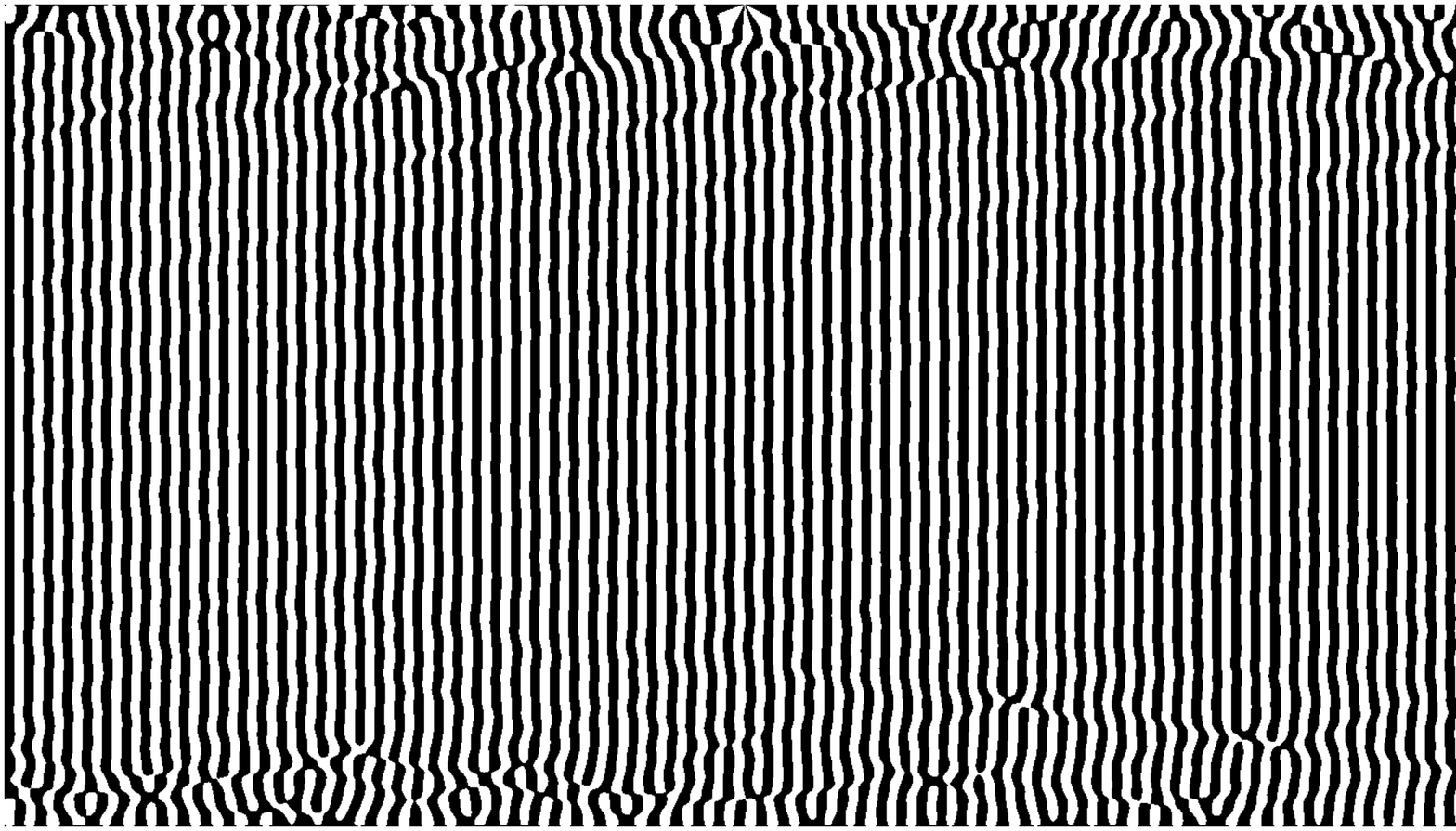} \\
b)
 \end{center}
 \end{minipage}
 \caption{(a): Wavefronts of the horizontal bouncing ball (the $(0-1) orbit$). (b) Nodal domains of numerically calculated  eigenfunction for the barrier billiard with $E_{\mathrm{exact}}=10088.61$. The superscar energy calculated from Eq.~\ref{horizontal_energy} $\mathcal{E}_{152, 1}^{(0-1)}=10088.56$.}
 \label{horizontal}
 \end{figure}

One of the simplest periodic orbit of the barrier billiard  corresponds to the horizontal motion inside the rectangle (i.e. the $(0-1)$ periodic orbit).  The superscar wave associated with this motion should have  the form (the axes are indicated at Fig.~\ref{examples}~c))
\begin{equation}
\Psi^{(\mathrm{superscar})}(y,x)=\frac{2}{\sqrt{ab}} \sin \left ( \frac{\pi}{b}m\, y \right )\sin \left (\frac{\pi}{a}n\,x\right ).
\label{horizontal_wave}
\end{equation}
When $n=1$ it has vertical wavefronts as at Fig.~\ref{horizontal} a). For larger $n$ the wavefronts have same form but with additional $n-1$ equidistant horizontal  lines where the function vanishes. 
 
\begin{figure}
\begin{minipage}{.49\linewidth}
\begin{center}
\includegraphics[angle=-90,width=.99\linewidth]{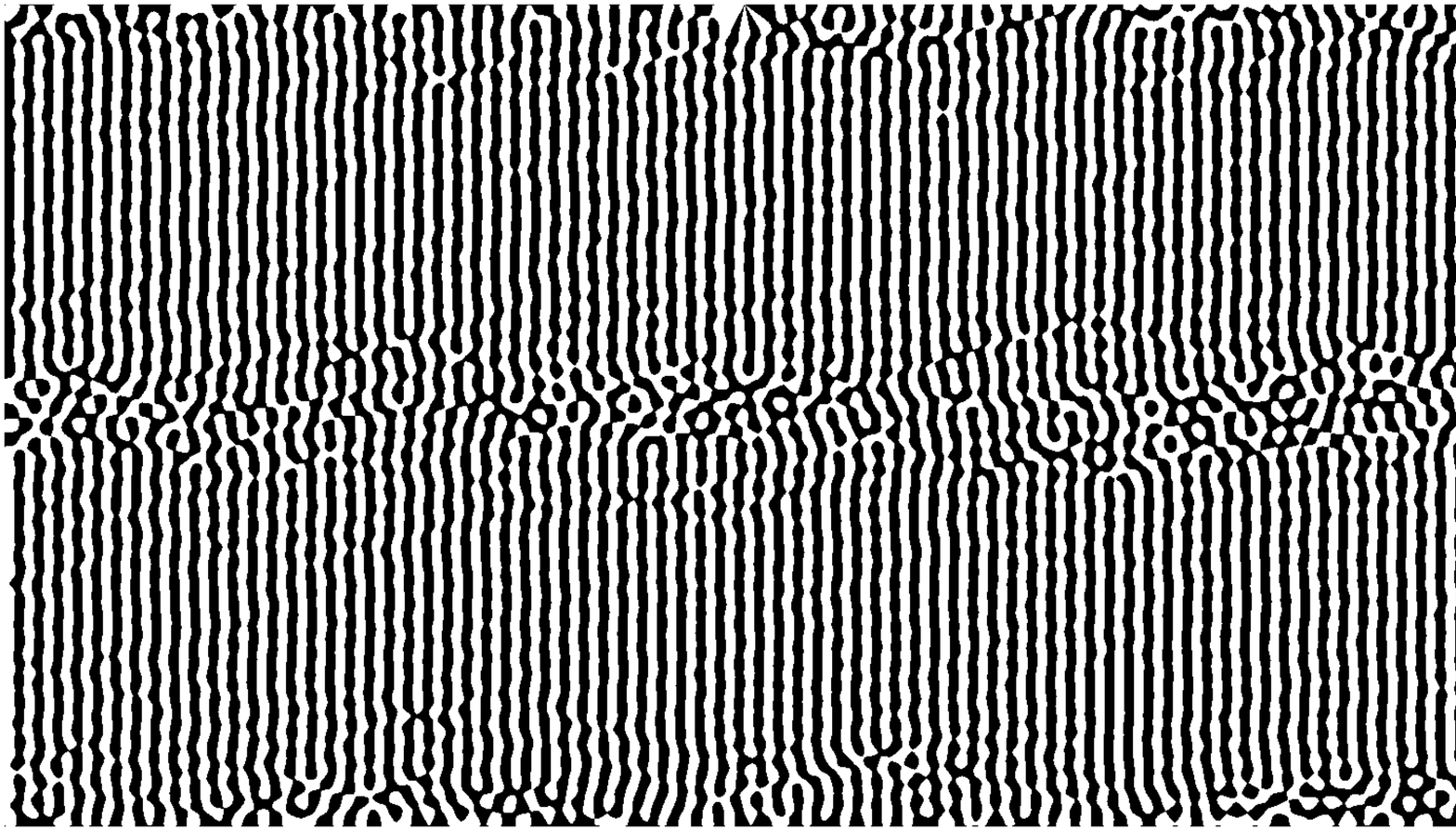} \\
a)
 \end{center}
 \end{minipage}
 \begin{minipage}{.49\linewidth}
\begin{center}
\includegraphics[angle=-90,width=.99\linewidth]{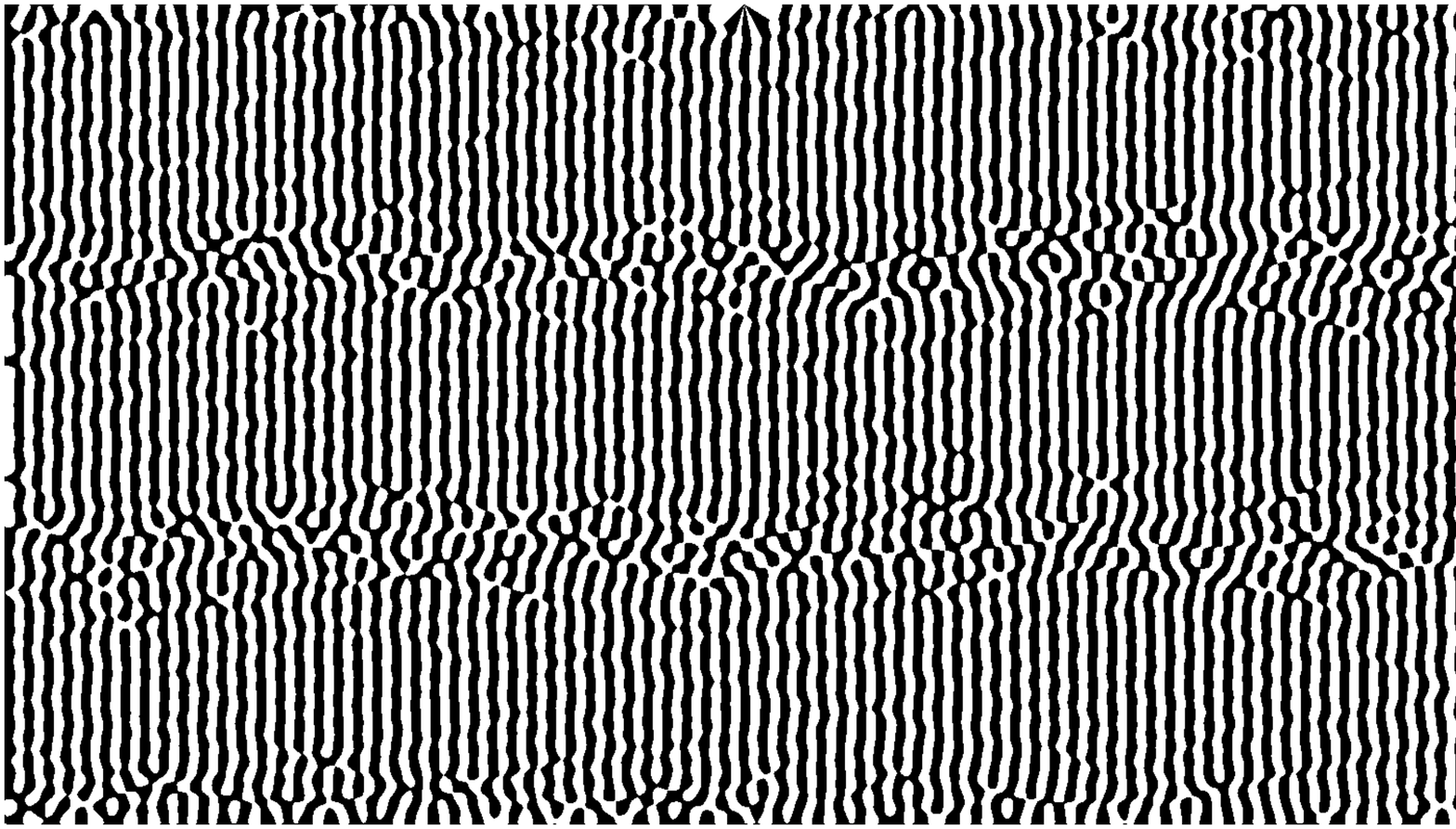} \\
b)
 \end{center}
 \end{minipage}
 \caption{The same as at Fig.~\ref{horizontal}~b) but  (a): $E_{\mathrm{exact}}=10092.63$. The superscar energy 
 $\mathcal{E}_{152, 2}^{(0-1)}=10092.80$, (b):  $E_{\mathrm{exact}}=10232.53$. The superscar energy 
 $\mathcal{E}_{153, 3}^{(0-1)}=10233.02$.}
 \label{horizontal_2}
 \end{figure}

The energy of such horizontal bouncing ball  is $(1\leq n\ll m)$
\begin{equation}
\mathcal{E}_{m,n}^{(0-1)} =\frac{\pi^2\,m^2}{b^2}+\frac{\pi^2\,n^2}{a^2}.
\label{horizontal_energy}
\end{equation}
At Figs.~\ref{horizontal}~b), \ref{horizontal_2}~a) and b) eigenfunctions related with such superscar waves with $m=152$ and  $n=1,2$ with $m=153$ and $n=3$  are shown.  The superscar structures are clear visible on these figures. The superscar energies (obtained from Eq.~\eqref{horizontal_energy}) indicated in figure captions are also very close to numerically calculated  energies for that states.

\begin{figure}
 \begin{minipage}{.49\linewidth}
 \begin{center}
 \includegraphics[width=.99\linewidth]{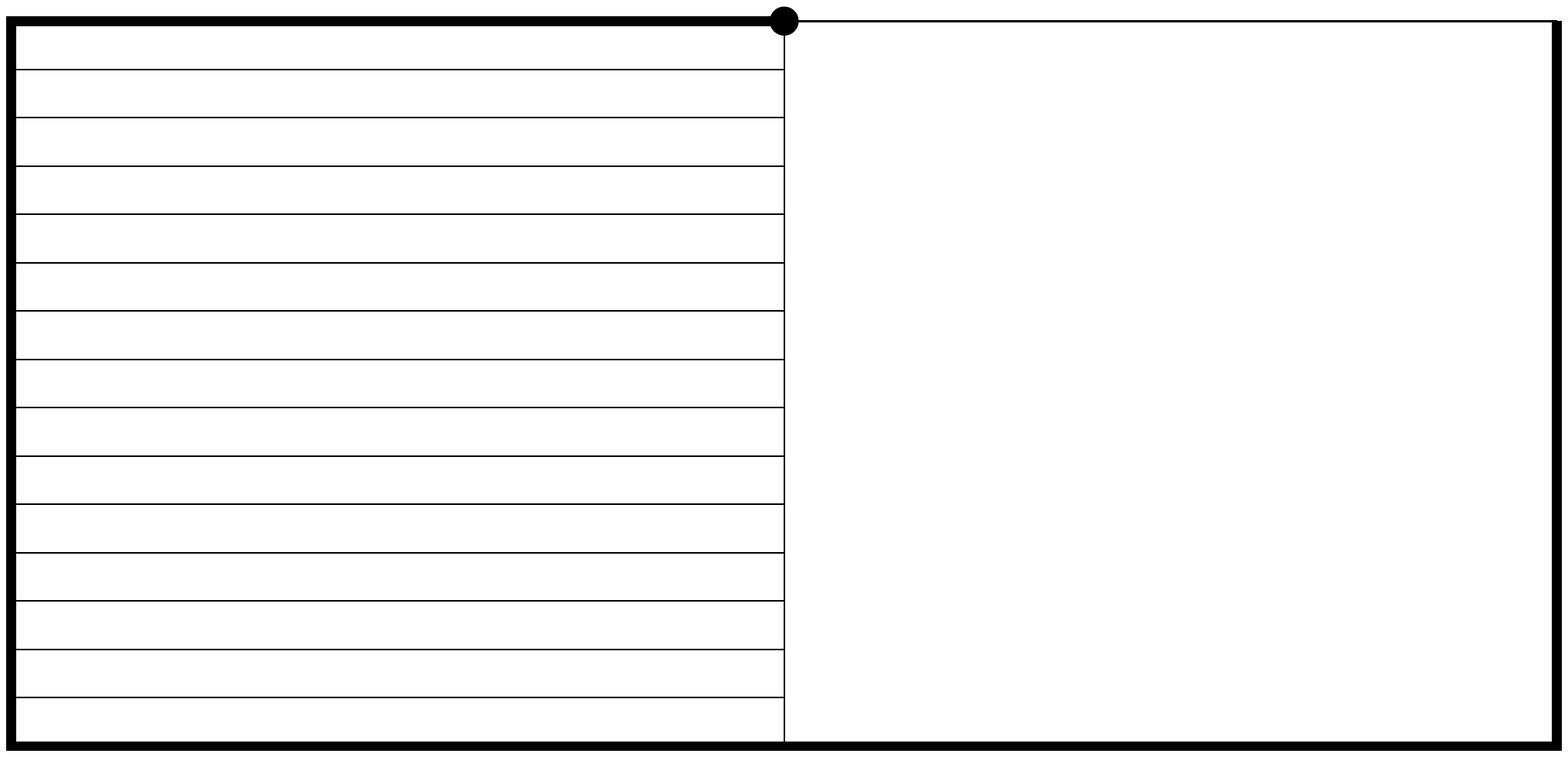}\\
 a)
 \end{center}
 \end{minipage}
 \begin{minipage}{.49\linewidth}
 \begin{center}
 \includegraphics[width=.99\linewidth]{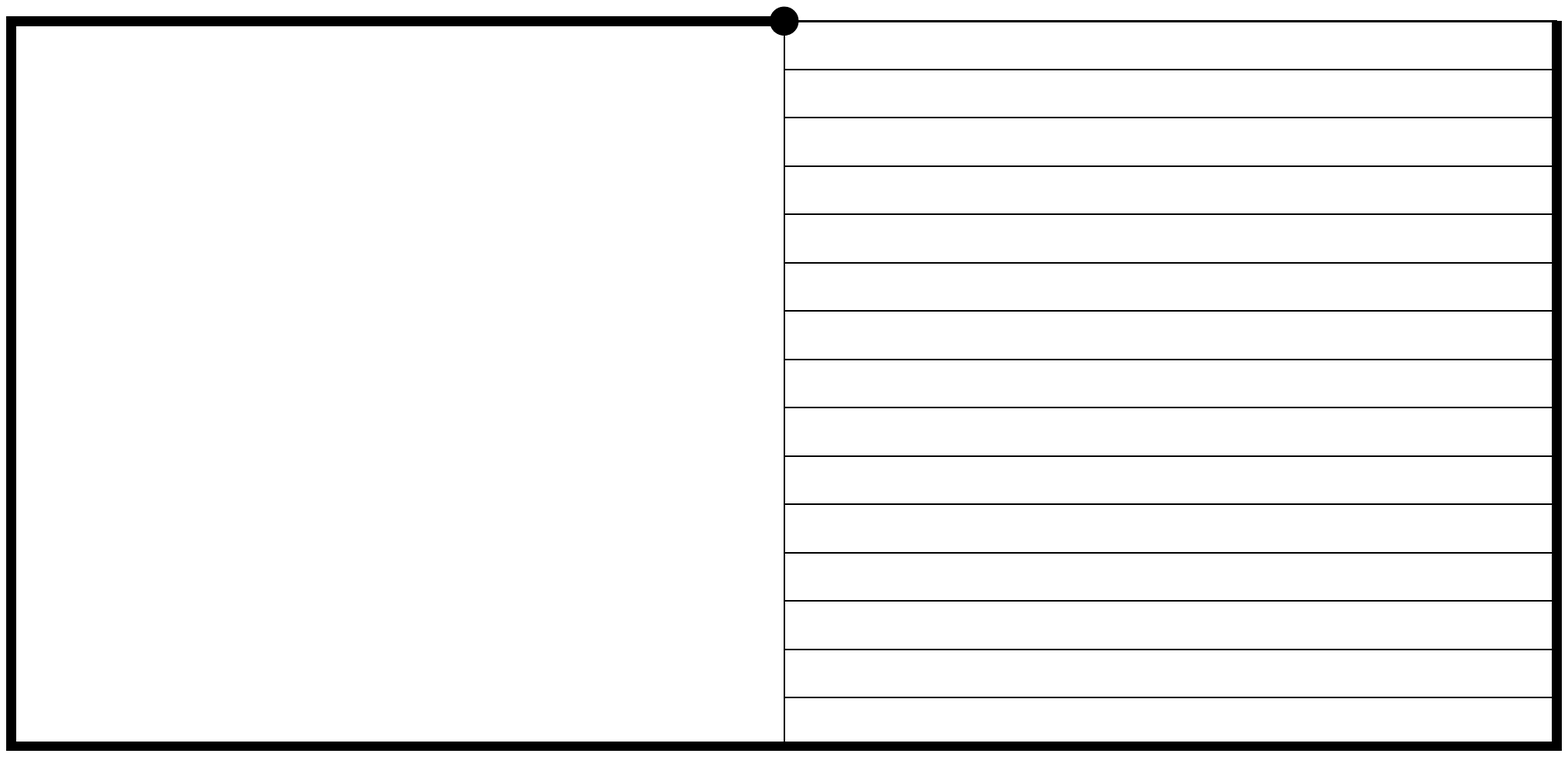}\\
 b)
 \end{center}
 \end{minipage}
 \caption{Wavefronts of two types of vertical bouncing balls. (a) Motion between two Dirichlet boundaries. (b) Motion between the Dirichlet and the Neumann boundaries. }
 \label{vertical}
 \end{figure}

 Another simple periodic orbit of the barrier billiard is the $(1-0)$ orbit  i.e. the  vertical bouncing ball. There are two types of such orbits. The first is related with the motion between two Dirichlet boundaries ($0<y<b/2$) and the second is associated with the motion between the Dirichlet and Neumann boundaries ($b/2<y<b$), cf. Fig.~\ref{vertical}.   
 
\begin{figure}
\begin{minipage}{.49\linewidth}
\begin{center}
\includegraphics[angle=-90,width=.99\linewidth]{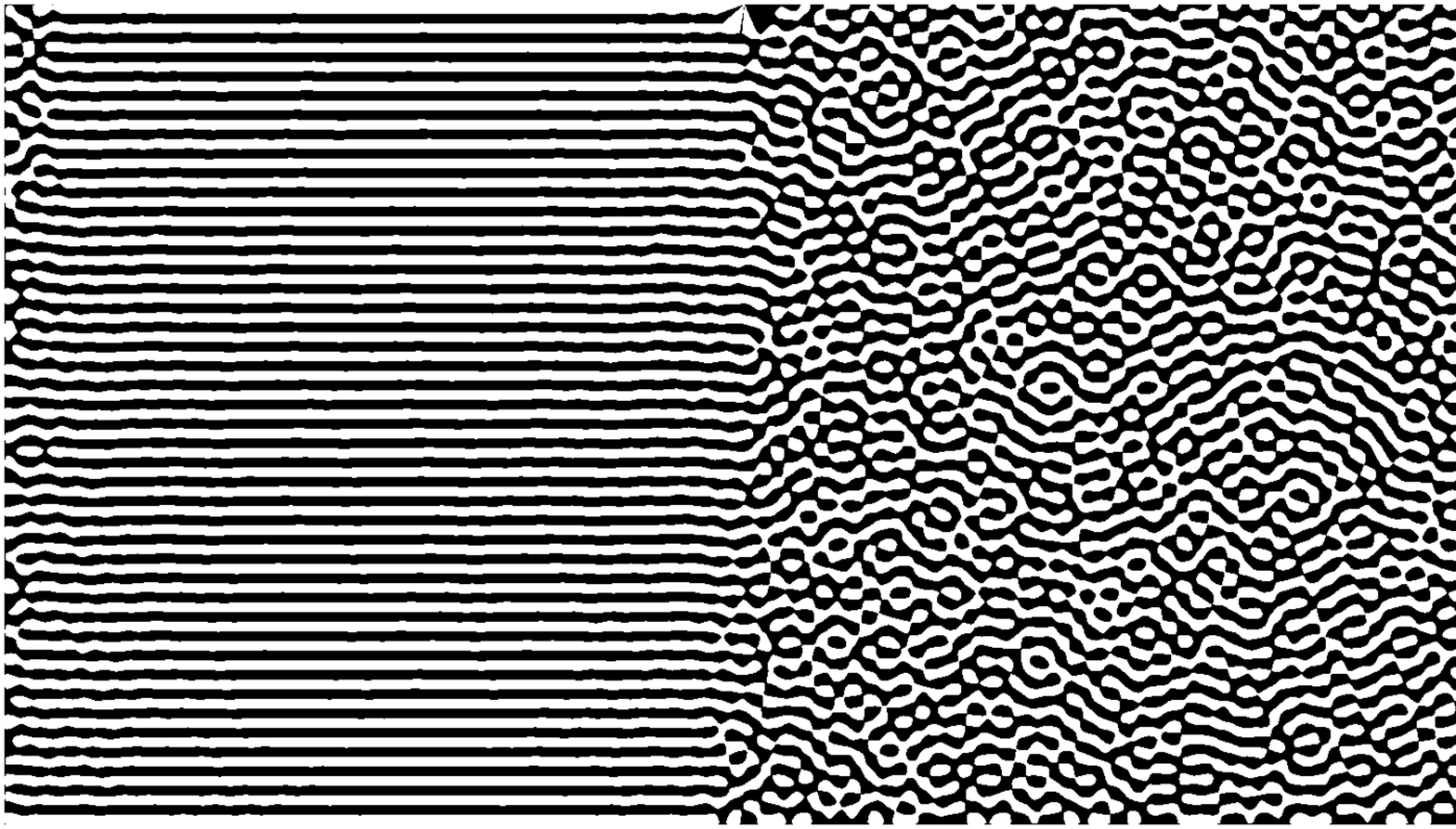} \\
a)
 \end{center}
 \end{minipage}
 \begin{minipage}{.49\linewidth}
\begin{center}
\includegraphics[angle=-90,width=.99\linewidth]{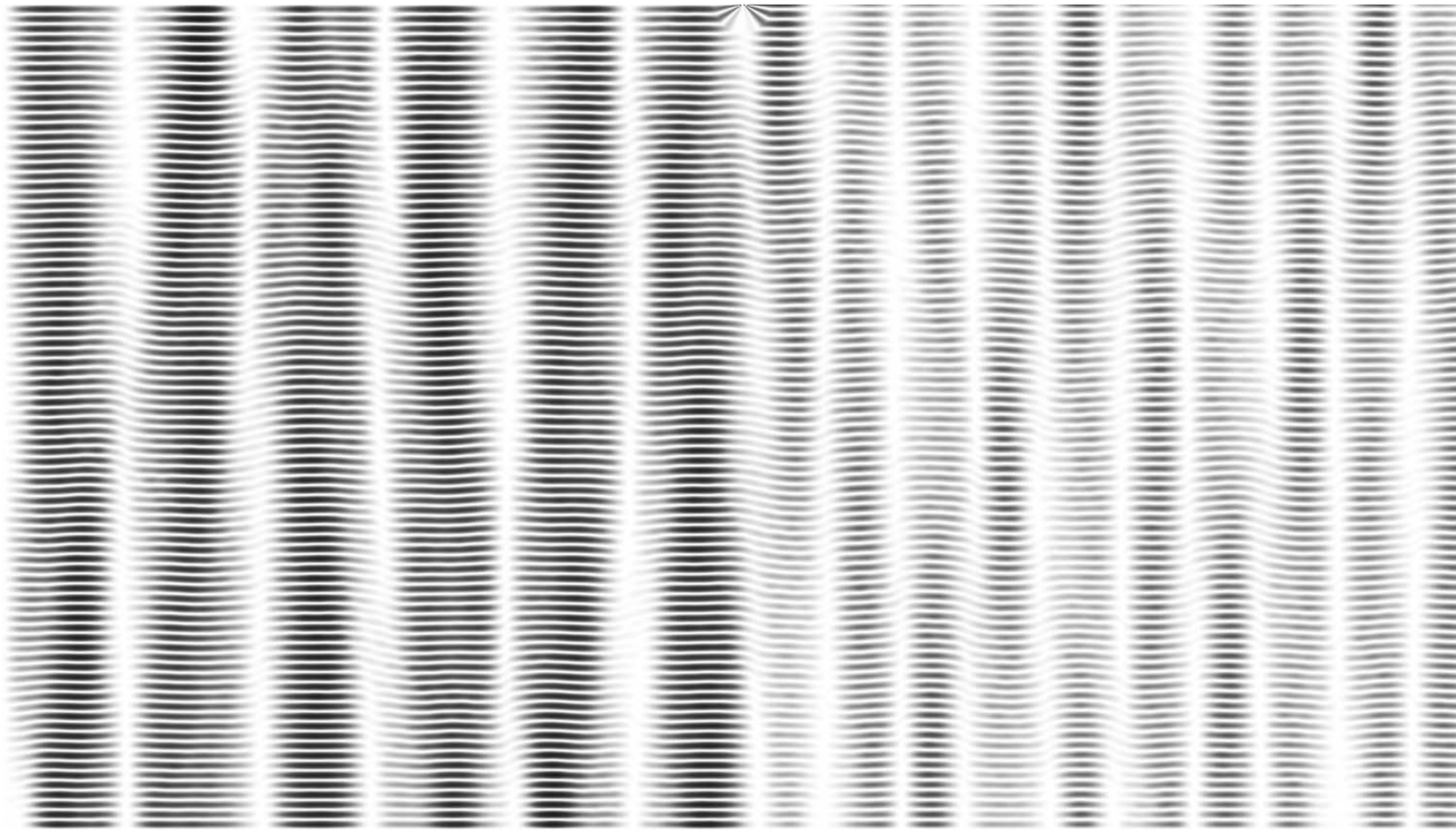} \\
b)
 \end{center}
 \end{minipage}
 \caption{(a):  Nodal domain of numerically calculated eigenfunction of the barrier billiard with $E_{\mathrm{exact}}=10209.55$. The superscar energy $\mathcal{E}_{85,1}^{\mathrm{DD}}=10209.65$. 
 (b) The  grey image of the modulus of an exact eigenfunction with $E_{\mathrm{exact}}=10029.45$. The superscar energy $\mathcal{E}_{84,6}^{\mathrm{DD}}=10032.00$. }
 \label{vertical_DD}
 \end{figure}

 The superscar waves propagating inside  these two pencils are 
 \begin{eqnarray}
 \Psi_{\mathrm{DD}}^{(\mathrm{superscar})}(y,x)&=&\frac{2}{\sqrt{ab}} \sin \left (\frac{\pi}{a}m\,x\right ) \sin \left ( \frac{2\pi}{b}n\, y \right )\Theta(b/2-y),\label{DD}\\
 \Psi_{\mathrm{DN}}^{(\mathrm{superscar})}(y,x)&=&\frac{2}{\sqrt{ab}} \cos \left (\frac{\pi}{a}(m-\tfrac{1}{2})\,x\right ) \sin \left ( \frac{2\pi}{b}n\, y \right )\Theta(y-b/2)\label{DN}
 \end{eqnarray}
 and their energies are as follows $(1\leq n\ll m)$
 \begin{eqnarray}
 \mathcal{E}_{m,n}^{\mathrm{DD}}&=&\frac{\pi^2 \,m^2}{a^2}+\frac{4\pi^2\, n^2}{b^2} , \label{EDD}\\
 \mathcal{E}_{m,n}^{\mathrm{DN}}&=&\frac{\pi^2\, (m-1/2)^2}{a^2}+\frac{4\pi^2 \,n^2}{b^2}.\label{EDN}
 \end{eqnarray}
 At Fig.~\ref{vertical_DD} a) the nodal domain of a numerically calculated eigenfunction of the barrier billiard with energy $E_{\mathrm{exact}}=10209.55$ is presented. Its structure consists of two different parts. The one corresponds to regular  waves propagating between the left part of the billiard as it should be for the Dirichlet-Dirichlet vertical bouncing ball 
 (cf. Fig.~\ref{vertical}~a)) and  \eqref{DD}). The second part is built from irregular waves with much smaller amplitudes. If superscar picture would be exact, this part should be exactly zero but as superscar wave is only an approximation such regions have to be constituted  of small-amplitude waves with irregular nodal domains. Such co-existence of two different parts of eigenfunctions is typical for superscar waves propagating in pencils with odd $n_a$ (see bellow). The calculated energy of the corresponding superscar with $m=85$ and $n=1$  $\mathcal{E}_{85,1}^{\mathrm{DD}}=10209.65$ is very close to the exact energy.   
 
With the increasing of perpendicular momentum $n$ superscar waves become less pronounced  as the parameter  \eqref{parameter} which controls the validity of superscar approximation grows. Nevertheless the vertical bouncing ball structure remains visible even for $n=6$ and $m=84$  as shown at Fig.~\ref{vertical_DD}~b). Notice that the second part of the
this picture is not irregular as at Fig.~\ref{vertical_DD}~a) but contains a (deformed) wave corresponding to Dirichlet-Neumann vertical bouncing ball structure. It can be explained by the fact that  $\mathcal{E}_{84,10}^{\mathrm{DN}}=10025.44$ which is also close to the exact energy of this state. 

\begin{figure}
\begin{minipage}{.49\linewidth}
\begin{center}
\includegraphics[angle=-90,width=.99\linewidth]{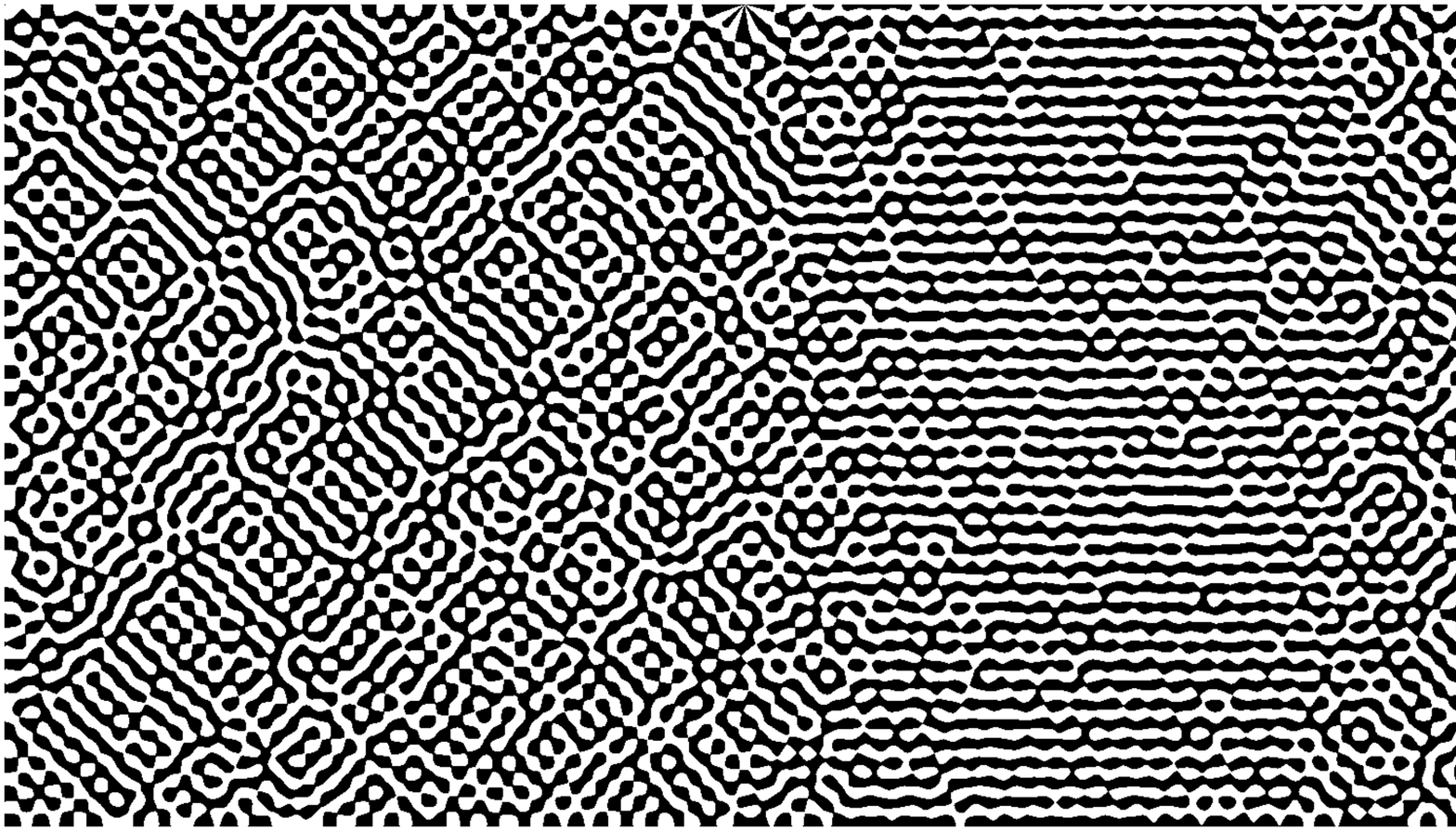} \\
a)
 \end{center}
 \end{minipage}
 \begin{minipage}{.49\linewidth}
\begin{center}
\includegraphics[angle=-90,width=.99\linewidth]{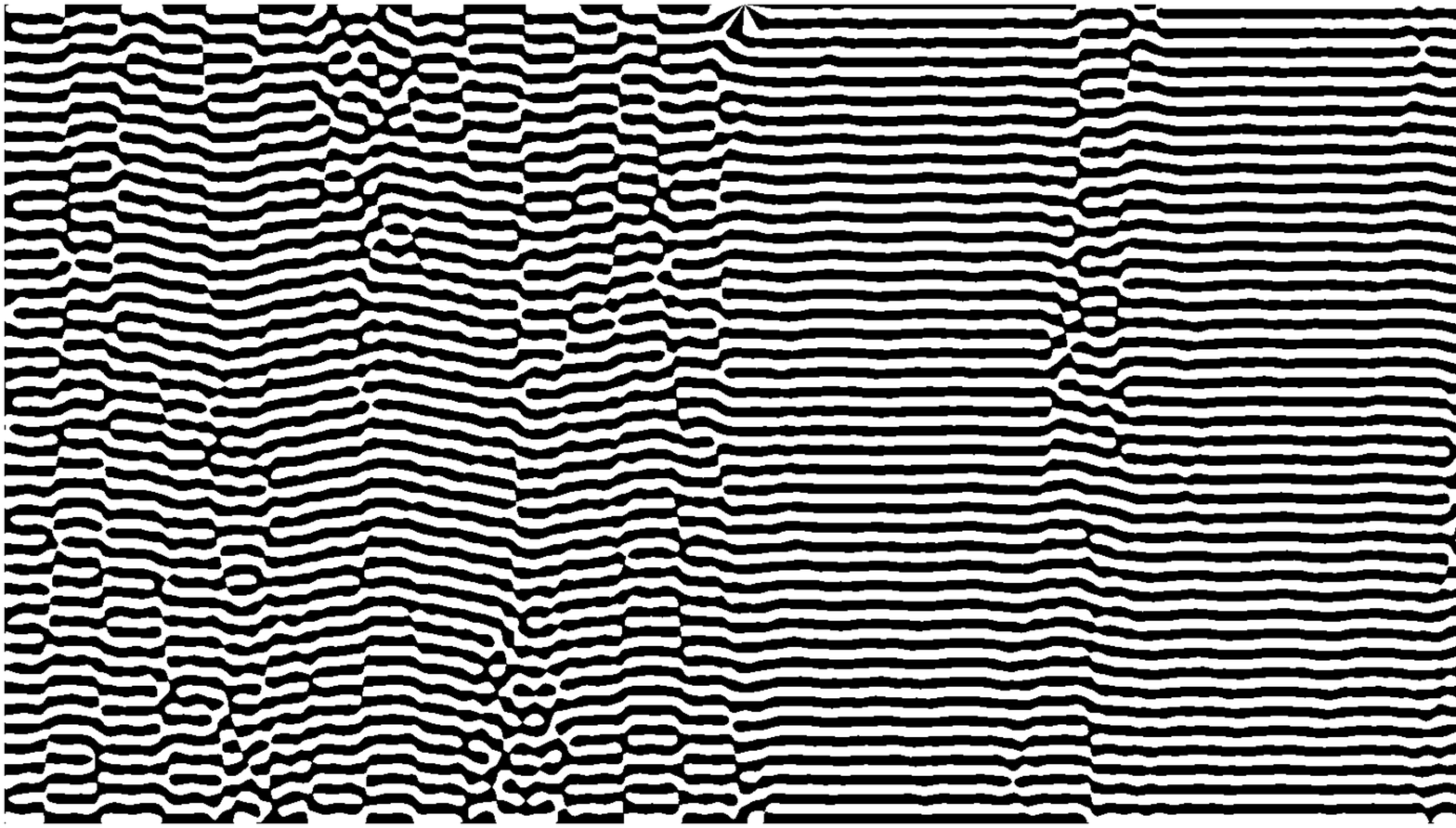} \\
b)
 \end{center}
 \end{minipage}
 \caption{Nodal domains of numerically calculated eigenfunctions of the barrier billiard. (a)  with $E_{\mathrm{exact}}=10089.70$. The superscar energy $\mathcal{E}_{85,1}^{\mathrm{DN}}=10089.91$. 
 (b)  with $E_{\mathrm{exact}}=10094.44$. The superscar energy $\mathcal{E}_{85,2}^{\mathrm{DN}}=10095.15$. }
 \label{vertical_DN}
 \end{figure}

\begin{figure}
\begin{minipage}{.49\linewidth}
\begin{center}
\includegraphics[angle=-90,width=.99\linewidth]{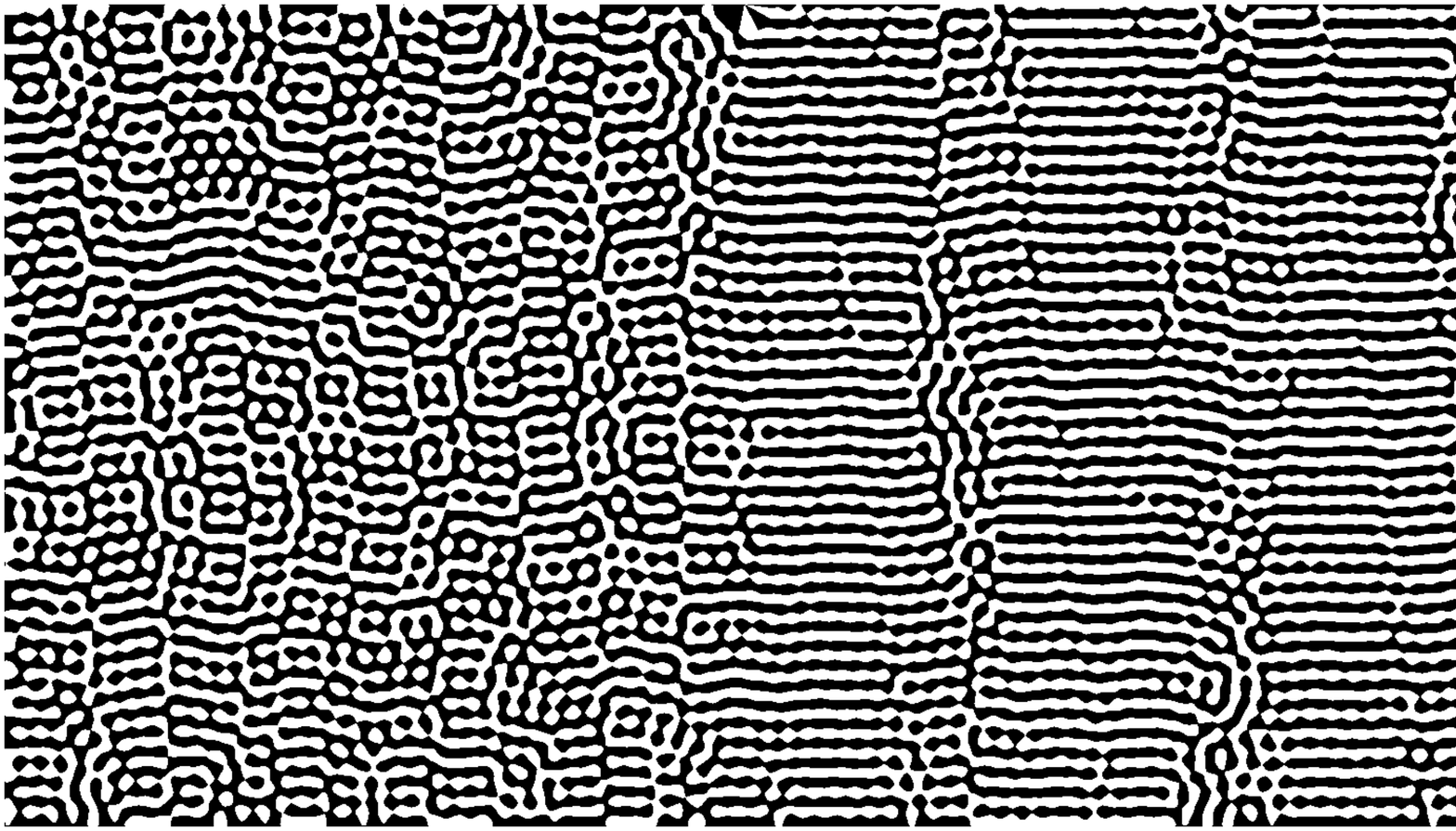} \\
a)
 \end{center}
 \end{minipage}
 \begin{minipage}{.49\linewidth}
\begin{center}
\includegraphics[angle=-90,width=.99\linewidth]{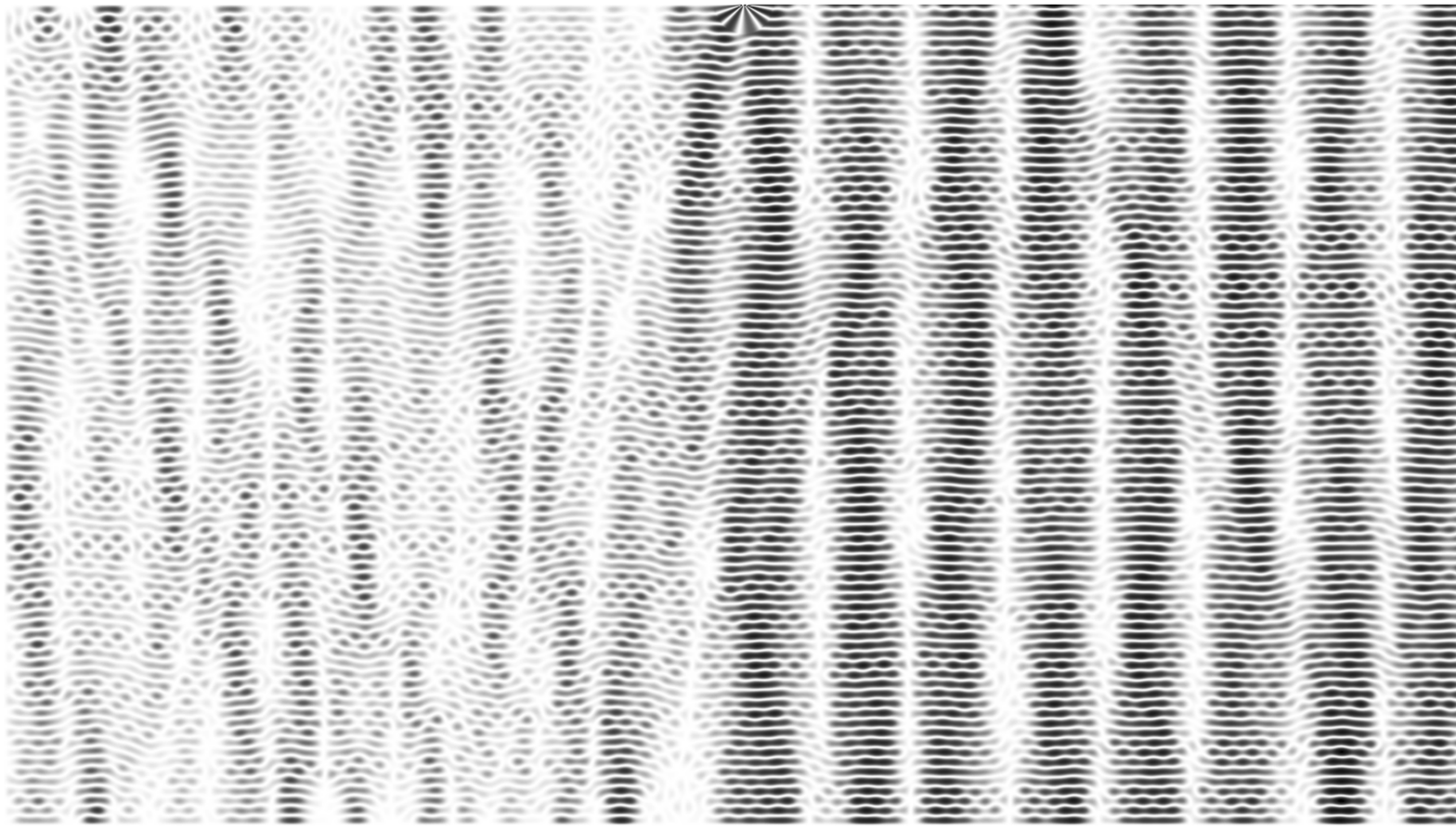} \\
b)
 \end{center}
 \end{minipage}
 \caption{ (a) The as at Fig.~\ref{vertical_DN} but  with $E_{\mathrm{exact}}=10102.03$. The superscar energy $\mathcal{E}_{85,3}^{\mathrm{DN}}=10103.88$. 
 (b)  Grey image of the modulus of the eigenfunction with $E_{\mathrm{exact}}=10192.82$. The superscar energy $\mathcal{E}_{85,8}^{\mathrm{DN}}=10199.93$. }
 \label{vertical_DN_2}
 \end{figure}

 At Figs.~\ref{vertical_DN} and \ref{vertical_DN_2} a few images of eigenfunctions with clear structure of the Dirichlet-Neumann vertical bouncing balls are presented.  The corresponding superscar waves correspond to $m=85$ and $n=1,2,3,8$ with energies (noted in figure captions)  very close to the exact energies. 

\begin{figure}
\begin{minipage}{.49\linewidth}
\begin{center}
\includegraphics[width=.99\linewidth]{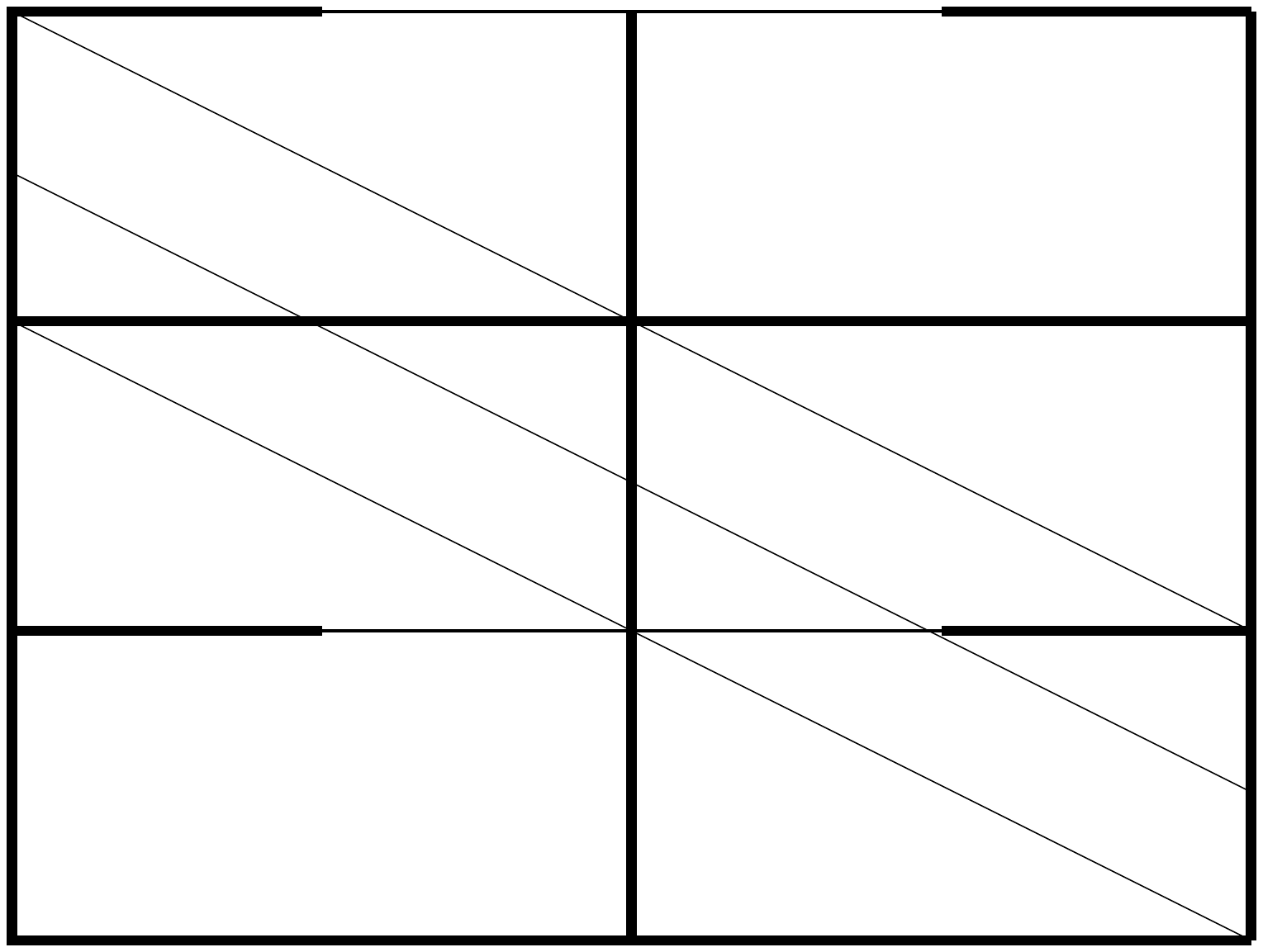} \\
a)
 \end{center}
 \end{minipage}
 \begin{minipage}{.49\linewidth}
\begin{center}
\includegraphics[width=.99\linewidth]{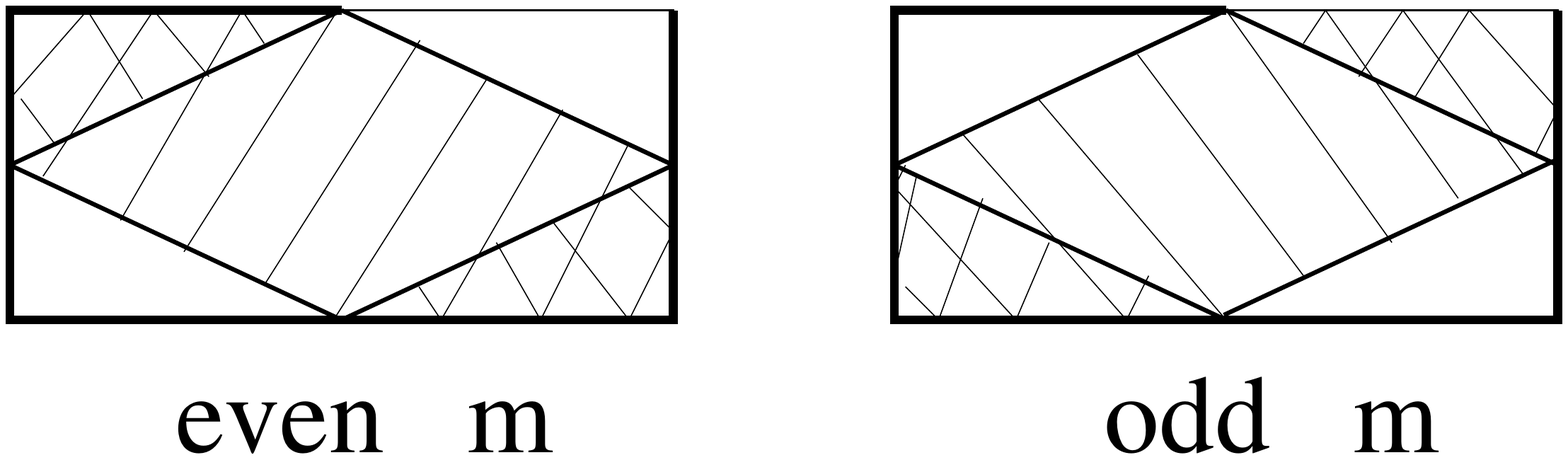} \\
b)
 \end{center}
 \end{minipage}
 \caption{ (a) The $(1-1)$ orbit unfolded. (b) Two possible pencils of the $(1-1)$ orbit and wavefronts of two corresponding superscar waves. }
 \label{1_1_orbit}
 \end{figure}

\begin{figure}
\begin{minipage}{.49\linewidth}
\begin{center}
\includegraphics[angle=-90,width=.99\linewidth]{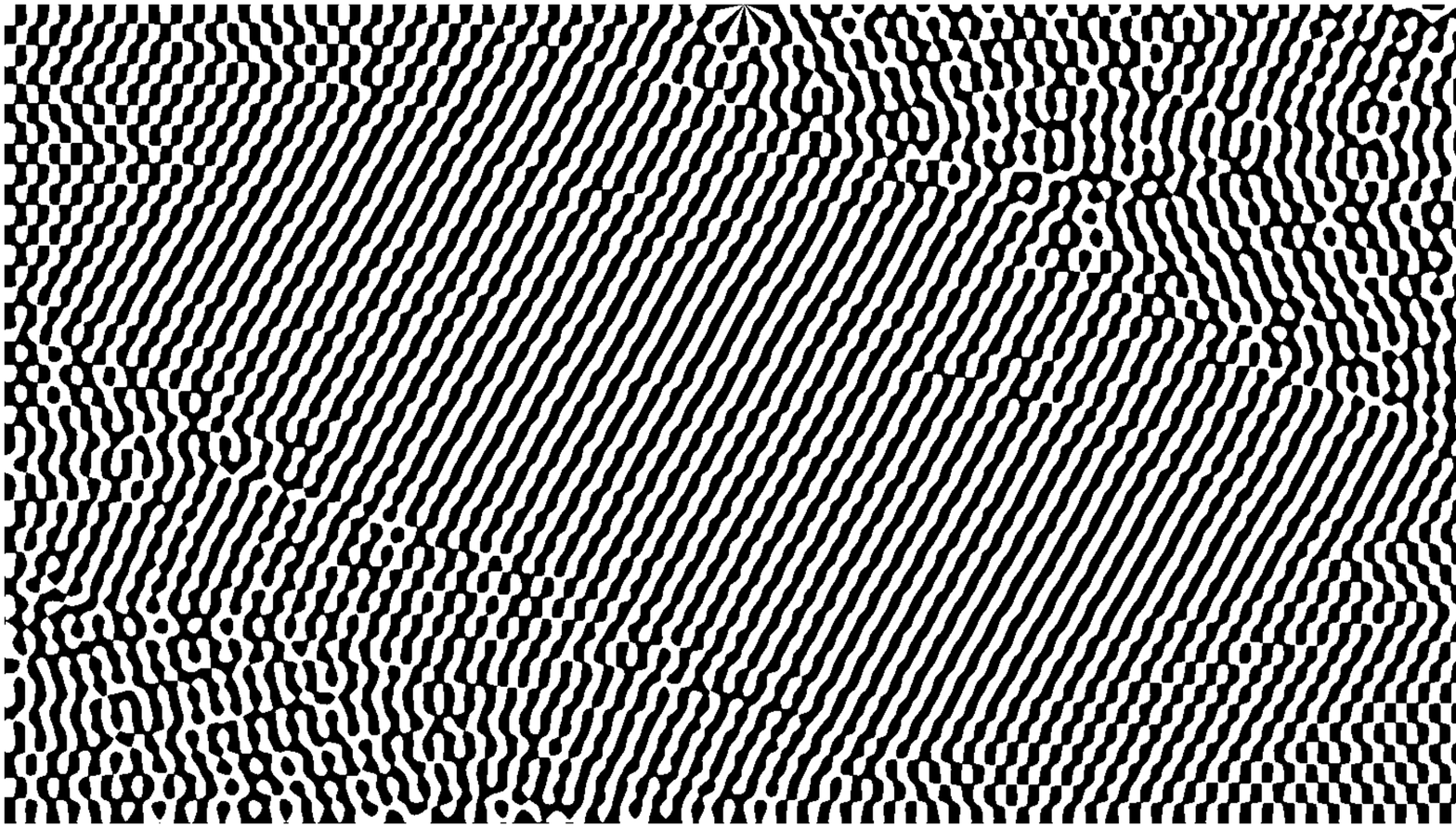} \\
a)
 \end{center}
 \end{minipage}
 \begin{minipage}{.49\linewidth}
\begin{center}
\includegraphics[angle=-90,width=.99\linewidth]{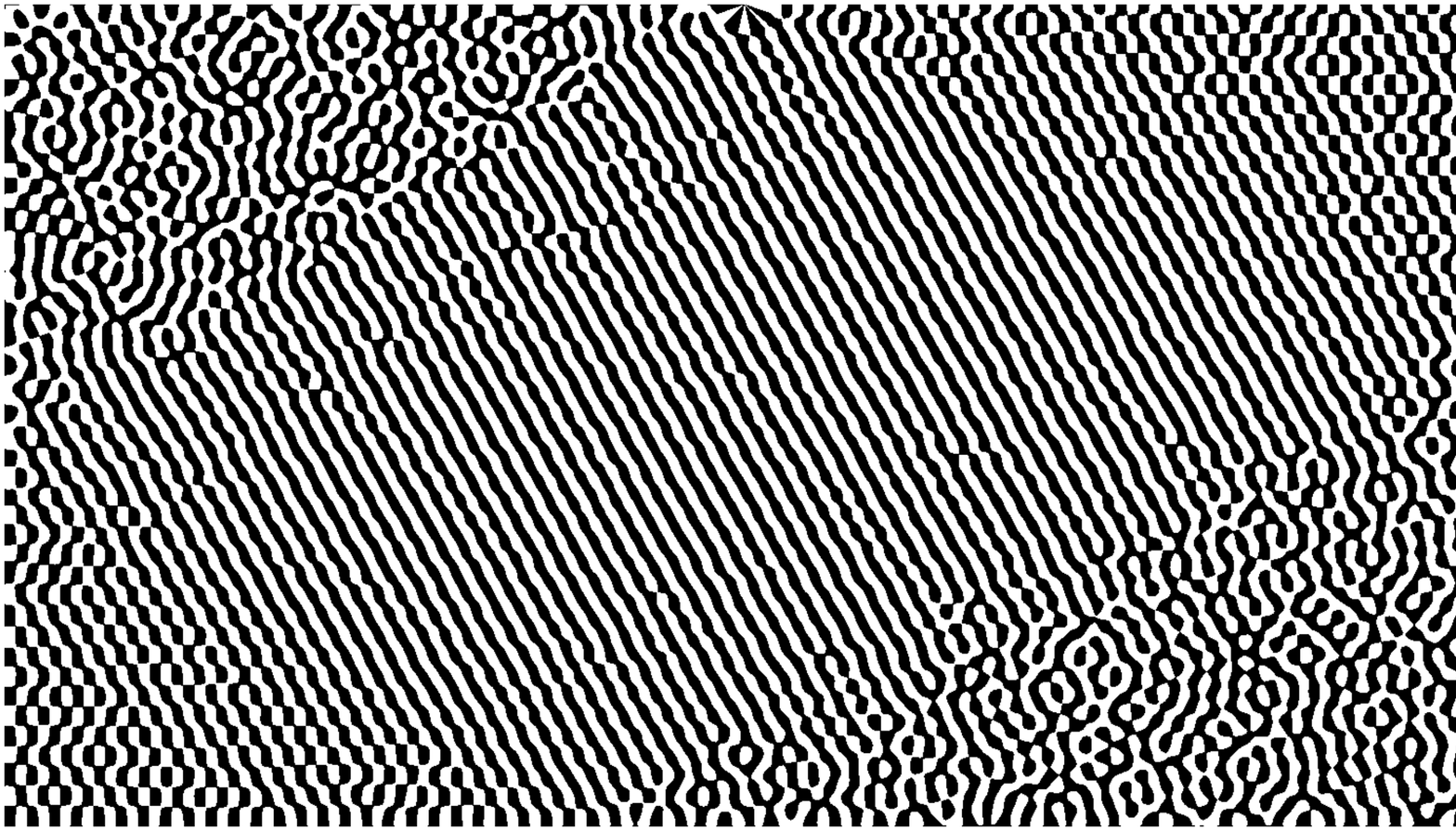} \\
b)
 \end{center}
 \end{minipage}
 \caption{ Nodal domains for numerically calculated eigenfunctions  with (a) $E_{\mathrm{exact}}=10099.58$  and (b) $\mathcal{E}_{\mathrm{exact}}=10041.41$ . The corresponding superscar energies are $E_{348,1}^{(\mathrm{1-1})}=10099.82$  and $\mathcal{E}_{347,1}^{(\mathrm{1-1})}=10041.87$. }
 \label{348_1_347_1}
 \end{figure}

 The horizontal and vertical boning balls are the only  periodic motions inside the barrier billiard which do not require the folding of periodic orbits pencils.  Other orbits are more complicated and should be folded inside the billiard.  The simplest of such orbits is the $(1-1)$ orbit indicated at Fig.~\ref{1_1_orbit}~a) with length  $ L_p=\sqrt{(2a)^2+(2b)^2}$.  When folding back to original barrier billiard this orbit gives rise to two periodic orbit pencils  as shown at  Fig.~\ref{1_1_orbit}~b).  In the usual rectangular billiard these two pencils may be continuously transformed one into another but in the barrier billiard they are restricted by singular vertices and constitute two different pencils which should be treated separately. The superscar waves propagating in the pencils have energies given by the expression
 \begin{equation}
 \mathcal{E}_{m,n}^{(\mathrm{1-1})}=\frac{\pi^2\, m^2}{L_p^2}+\frac{\pi^2\,n^2}{w^2}, \qquad L_p=\sqrt{(2a)^2+(2b)^2},\quad w=\frac{2ab}{L_p}\, .
 \end{equation}  
The existence of different pencils is manifest in different phase accumulated by a wave when propagating inside the pencils. It is plain that even and odd $m$ correspond to the pencils indicated at  Fig.~\ref{1_1_orbit}~b).   
 At Fig.~\ref{348_1_347_1} examples of superscar waves associated with the $(1-1)$ orbit are presented. The first  corresponds to $m=348$ and $n=1$ and the second to $m=347$ and $n=1$. The effect of switching from one pencil to another for even and odd $m$ is clear visible. 
 
 \begin{figure}
\begin{minipage}{.49\linewidth}
\begin{center}
\includegraphics[angle=-90,width=.99\linewidth]{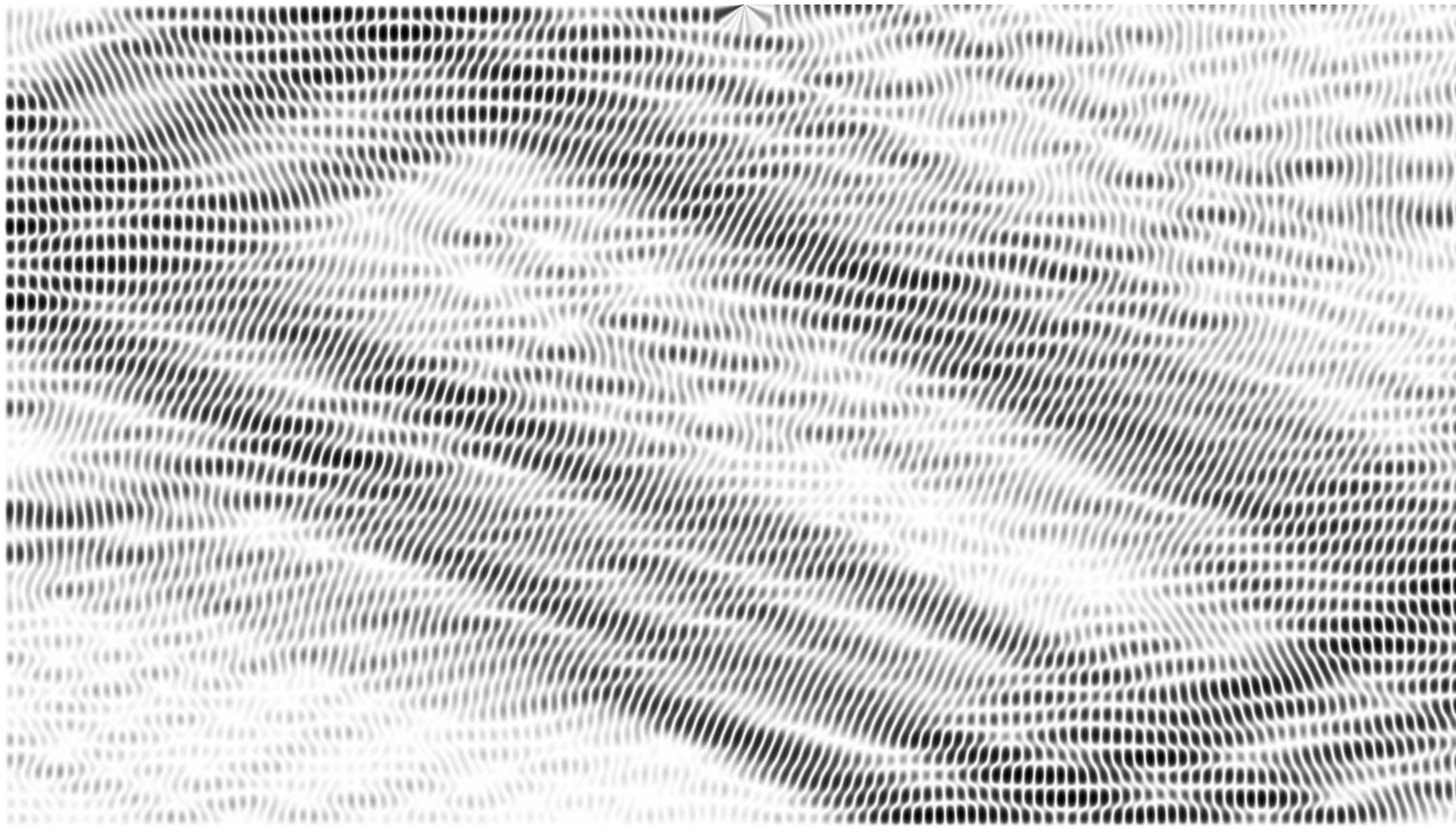} \\
a)
 \end{center}
 \end{minipage}
 \begin{minipage}{.49\linewidth}
\begin{center}
\includegraphics[angle=-90,width=.99\linewidth]{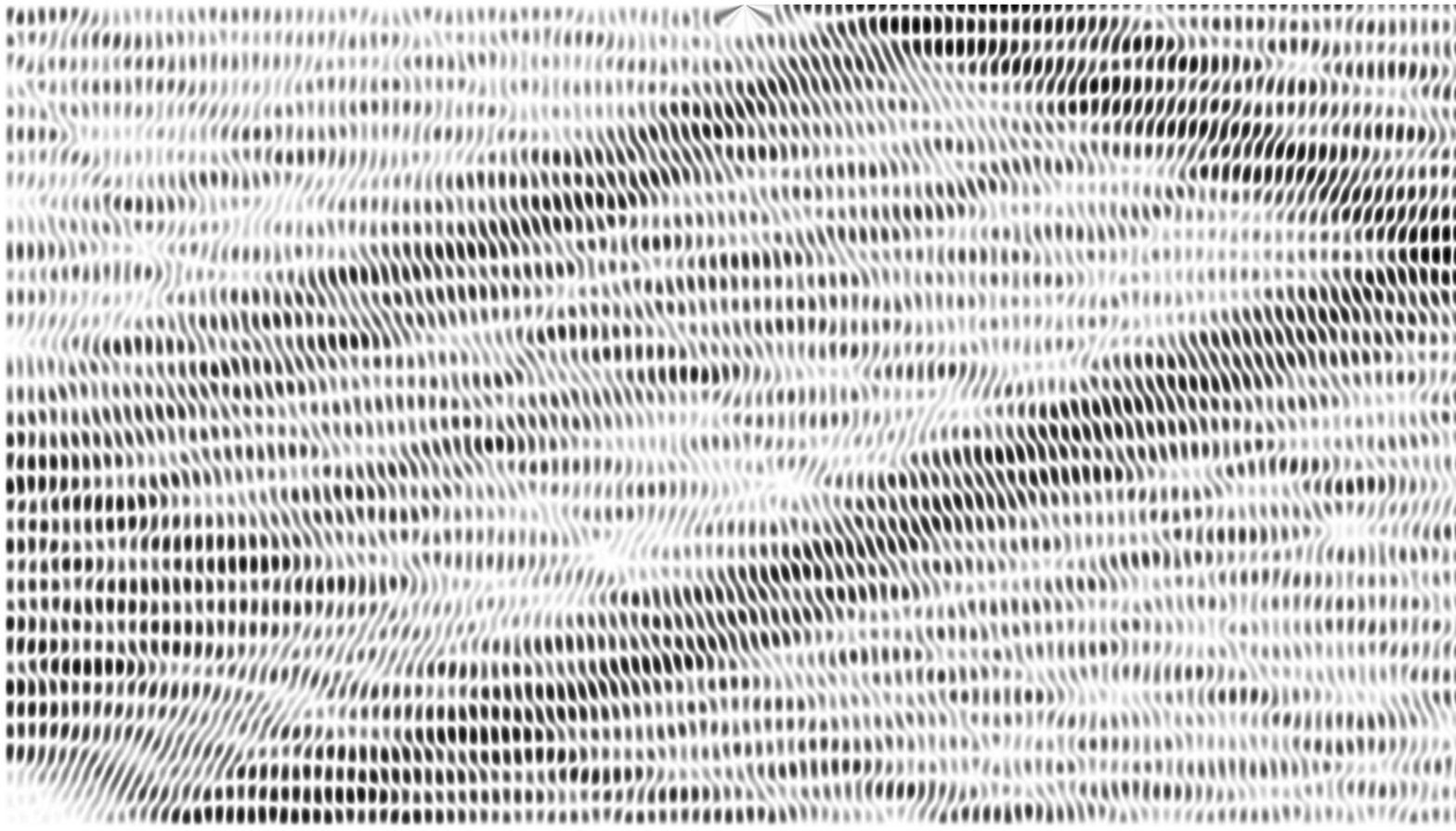} \\
b)
 \end{center}
 \end{minipage}
 \caption{ Grey images of the modulus of numerically calculated eigenfunctions with (a)  $E_{\mathrm{exact}}=10104.26$ and (b) $E_{\mathrm{exact}}=10045.51$. The corresponding superscar energies  $\mathcal{E}_{348,2}^{(\mathrm{1-1})}=10105.37$ and 
 $\mathcal{E}_{347,2}^{(\mathrm{1-1})}=10047.42$. }
 \label{347_1_348_1}
 \end{figure}

\begin{figure}
\begin{minipage}{.49\linewidth}
\begin{center}
\includegraphics[width=.99\linewidth]{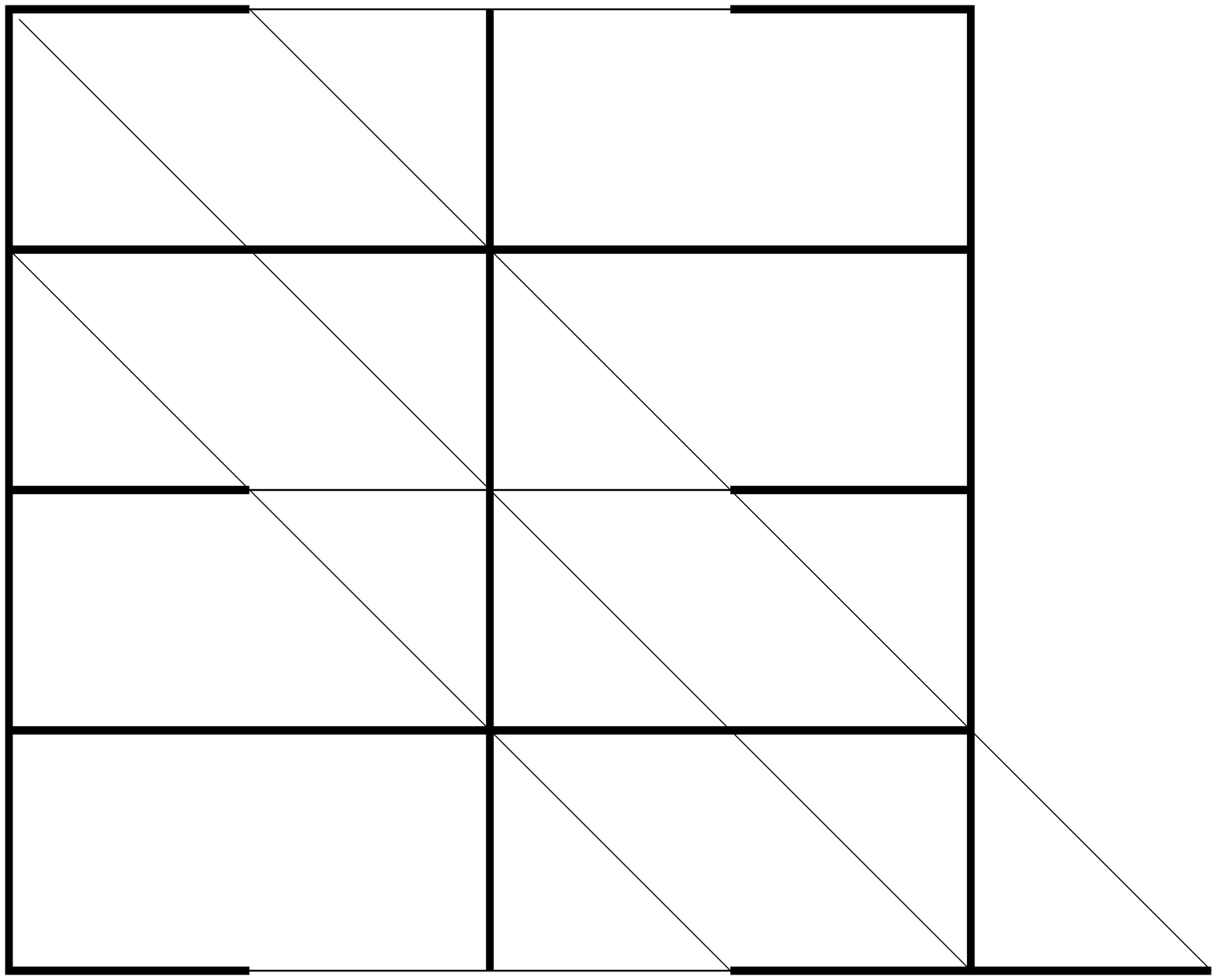} \\
a)
 \end{center}
 \end{minipage}
 \begin{minipage}{.49\linewidth}
\begin{center}
\includegraphics[width=1.1\linewidth]{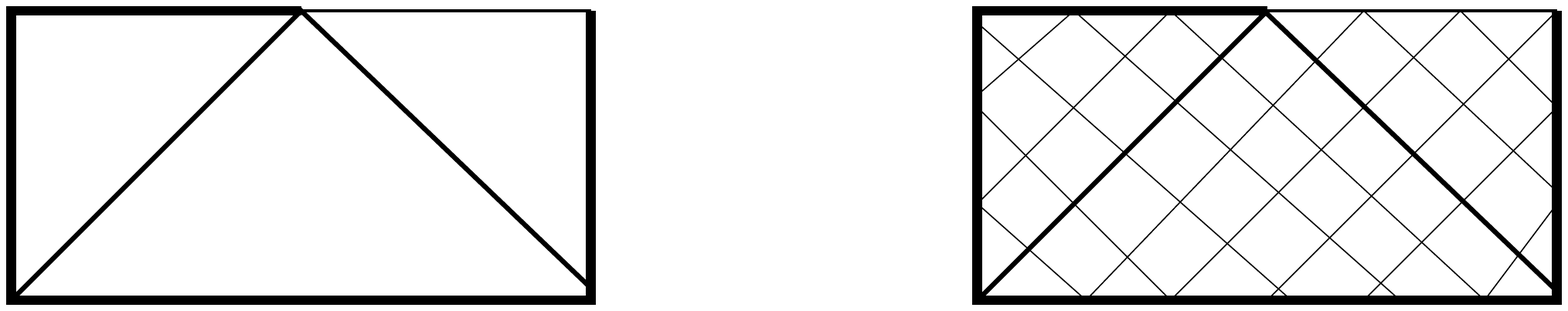} \\
b)
 \end{center}
 \end{minipage}
 \caption{ (a) Unfolding of the $(2-1)$ orbit. (b) Folded $(2-1)$ orbit and the wavefronts of the corresponding superscar wave. }
 \label{2_1_orbit}
 \end{figure}

\begin{figure}
\begin{minipage}{.49\linewidth}
\begin{center}
\includegraphics[angle=-90,width=.99\linewidth]{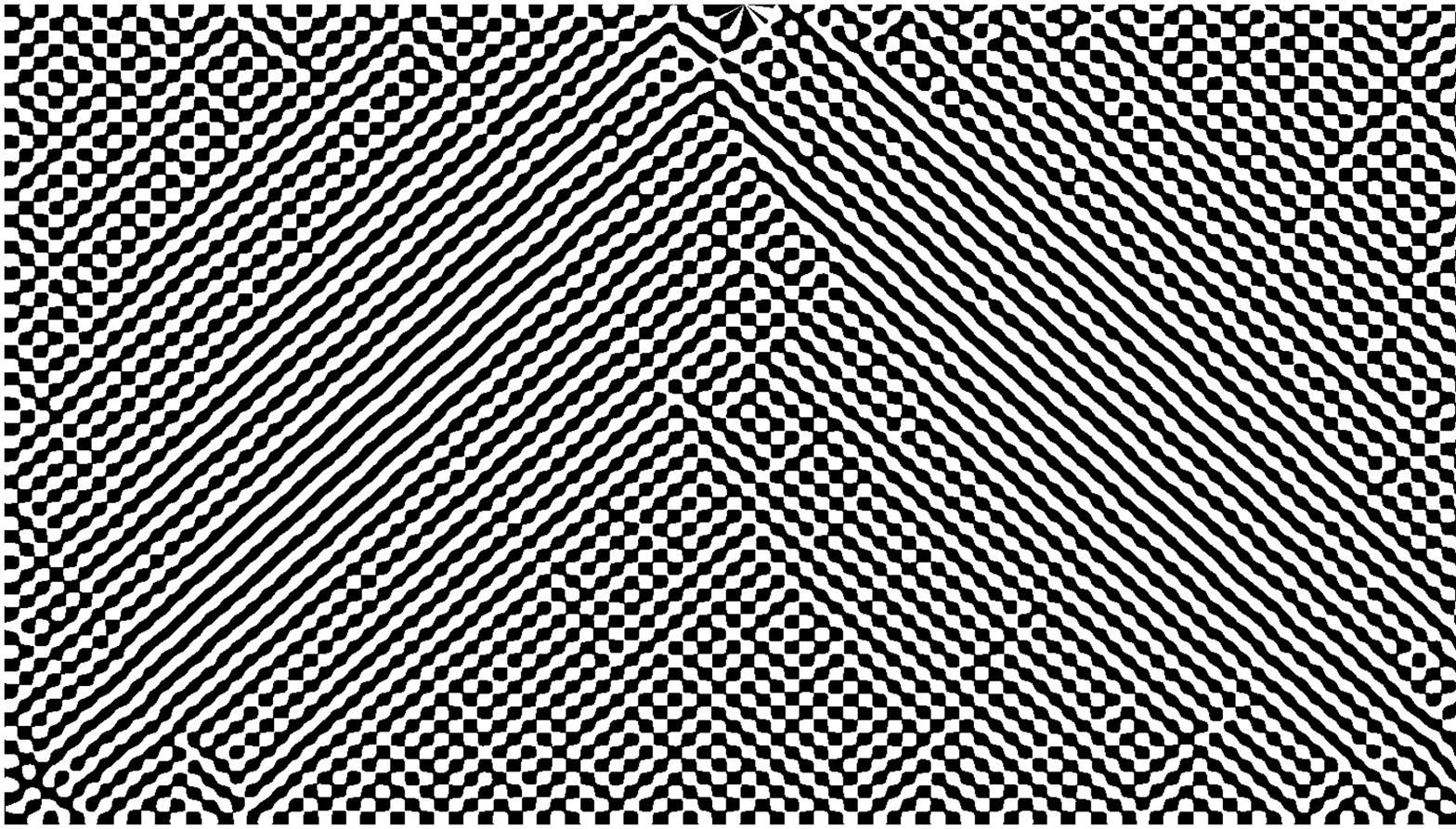} \\
a)
 \end{center}
 \end{minipage}
 \begin{minipage}{.49\linewidth}
\begin{center}
\includegraphics[angle=-90,width=.99\linewidth]{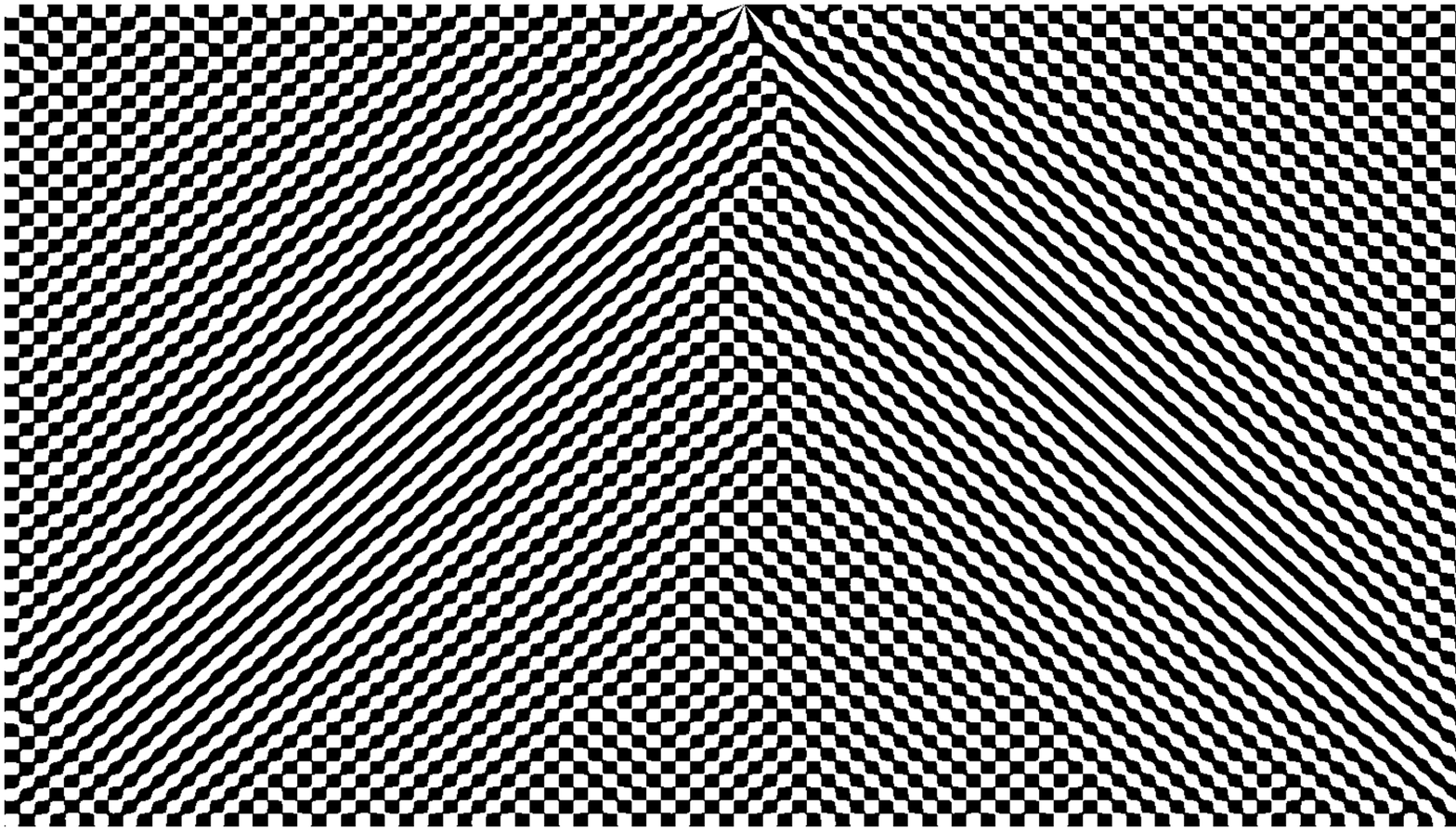} \\
b)
 \end{center}
 \end{minipage}
 \caption{ Nodal domains of numerically calculated eigenfunctions with (a)  $E_{\mathrm{exact}}=10017.57$ and (b)
  $E_{\mathrm{exact}}=10045.51$. The corresponding superscar energies  $\mathcal{E}_{227,1}^{(\mathrm{2-1})}=10017.67$ and 
 $\mathcal{E}_{228,1}^{(\mathrm{2-1})}=10106.31$. }
 \label{227_1_228_1}
 \end{figure}

\begin{figure}
\begin{minipage}{.49\linewidth}
\begin{center}
\includegraphics[angle=-90,width=.99\linewidth]{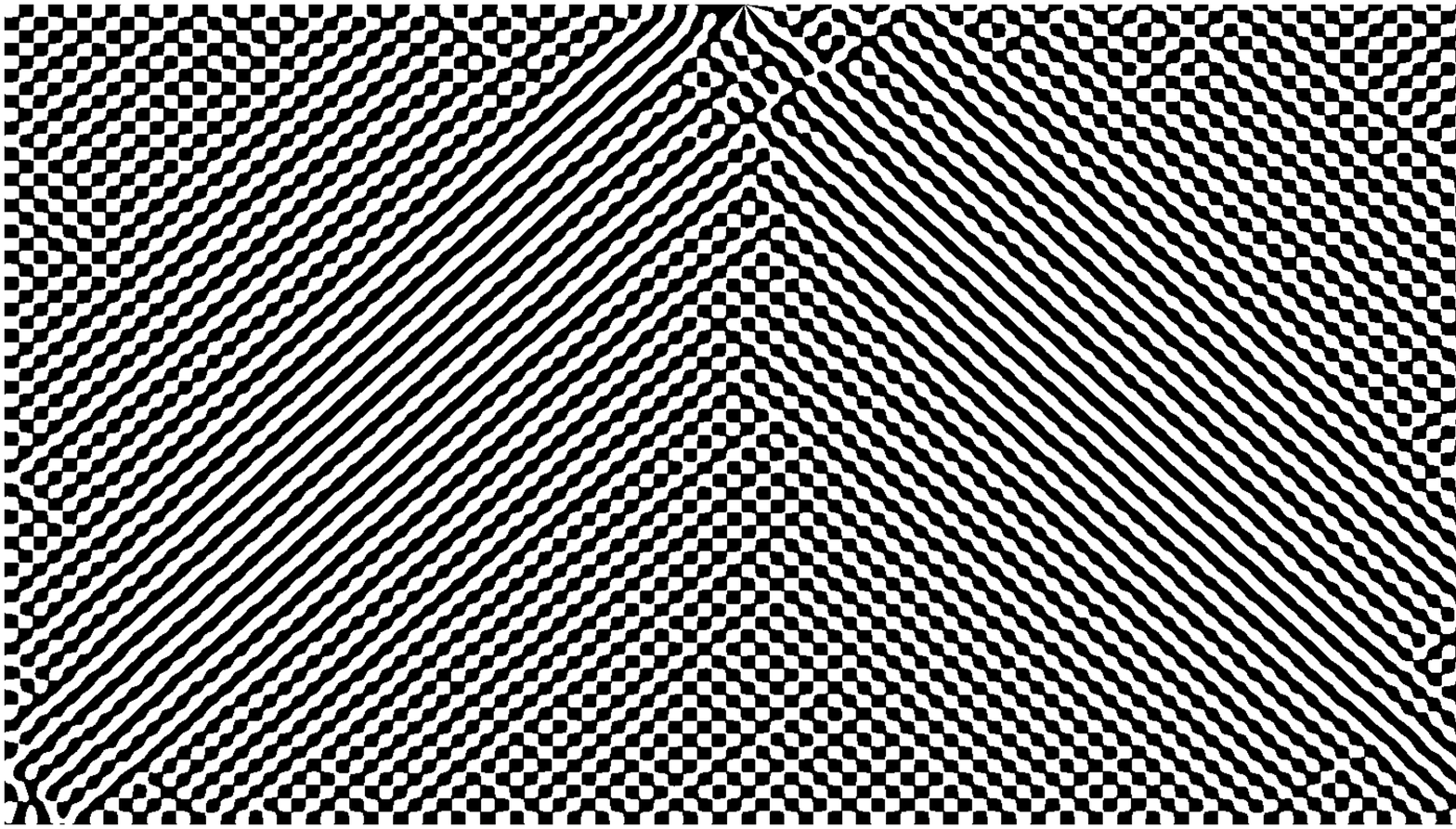} \\
a)
 \end{center}
 \end{minipage}
 \begin{minipage}{.49\linewidth}
\begin{center}
\includegraphics[angle=-90,width=.99\linewidth]{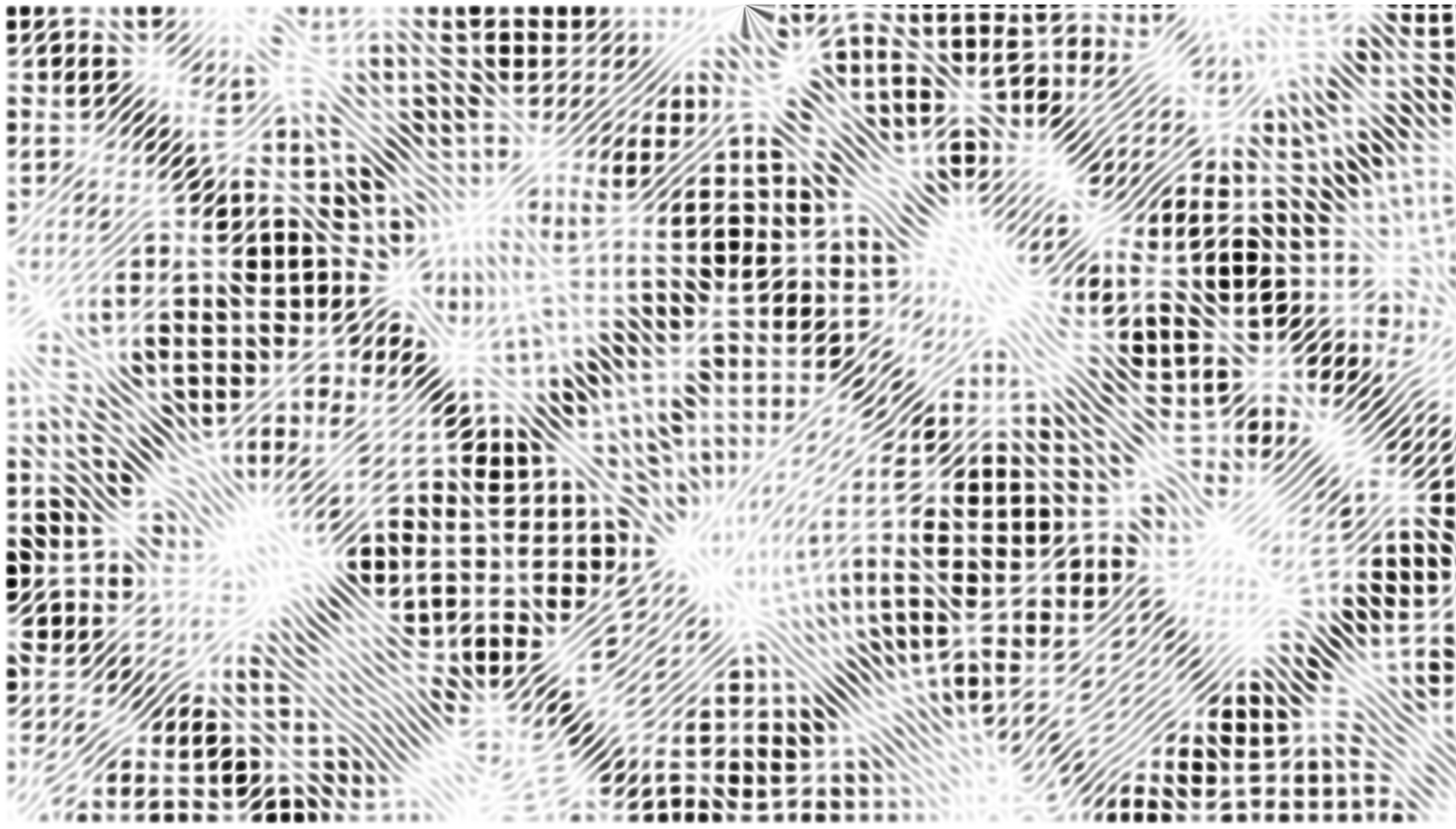} \\
b)
 \end{center}
 \end{minipage}
 \caption{ (a) Nodal domains of numerically calculated eigenfunctions with $E_{\mathrm{exact}}=10195.30$. (b) Grey image of the modulus of eigenfunction with $E_{\mathrm{exact}}=10201.28$. The corresponding superscar energies  $\mathcal{E}_{229,1}^{(\mathrm{2-1})}=10195.35$ and $\mathcal{E}_{229,3}^{(\mathrm{2-1})}=10201.67$. }
 \label{229_1_229_3}
 \end{figure}

At Fig~\ref{2_1_orbit} the unfolded and folded $(2-1)$ orbit is plotted. The pencil of this orbit has length  $L_p=\sqrt{(4a)^2+(2b)^2}$ and width  $w=4ab/L_p$. When folded it cover the whole barrier billiard surface. The superscar energies for such orbit are
\begin{equation}
\mathcal{E}_{m,n}^{(\mathrm{2-1})}=\frac{4\pi^2\, (m-1/2)^2}{L_p^2}+\frac{\pi^2\,n^2}{w^2}, \qquad L_p=\sqrt{(4a)^2+(2b)^2},\quad w=\frac{4ab}{L_p}\, .
\label{2_1_energy}
\end{equation}
At Figs.~\ref{227_1_228_1} and \ref{229_1_229_3} a)  eigenfunctions corresponding to superscar waves with $n=1$ and $m=227,\, 228,\, 229$ are represented. Notice exceptionally regular shape of nodal domains of these high-excited eigenfunctions and excellent  agreement of exact eigenvalues with superscar  energies \eqref{2_1_energy}.

\begin{figure}
\begin{center}
\includegraphics[width=.5\linewidth]{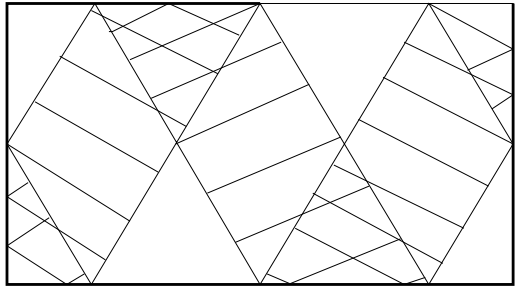}
\end{center}
\caption{One of periodic orbit pencils for the $(3-1)$ orbit corresponding to  a superscar wave with odd $m$ and its wavefronts. The second pencil with even $m$ occupies the complimentary part of the billiard.}
\label{3_1_orbit}
\end{figure}

\begin{figure}
\begin{minipage}{.49\linewidth}
\begin{center}
\includegraphics[angle=-90,width=.99\linewidth]{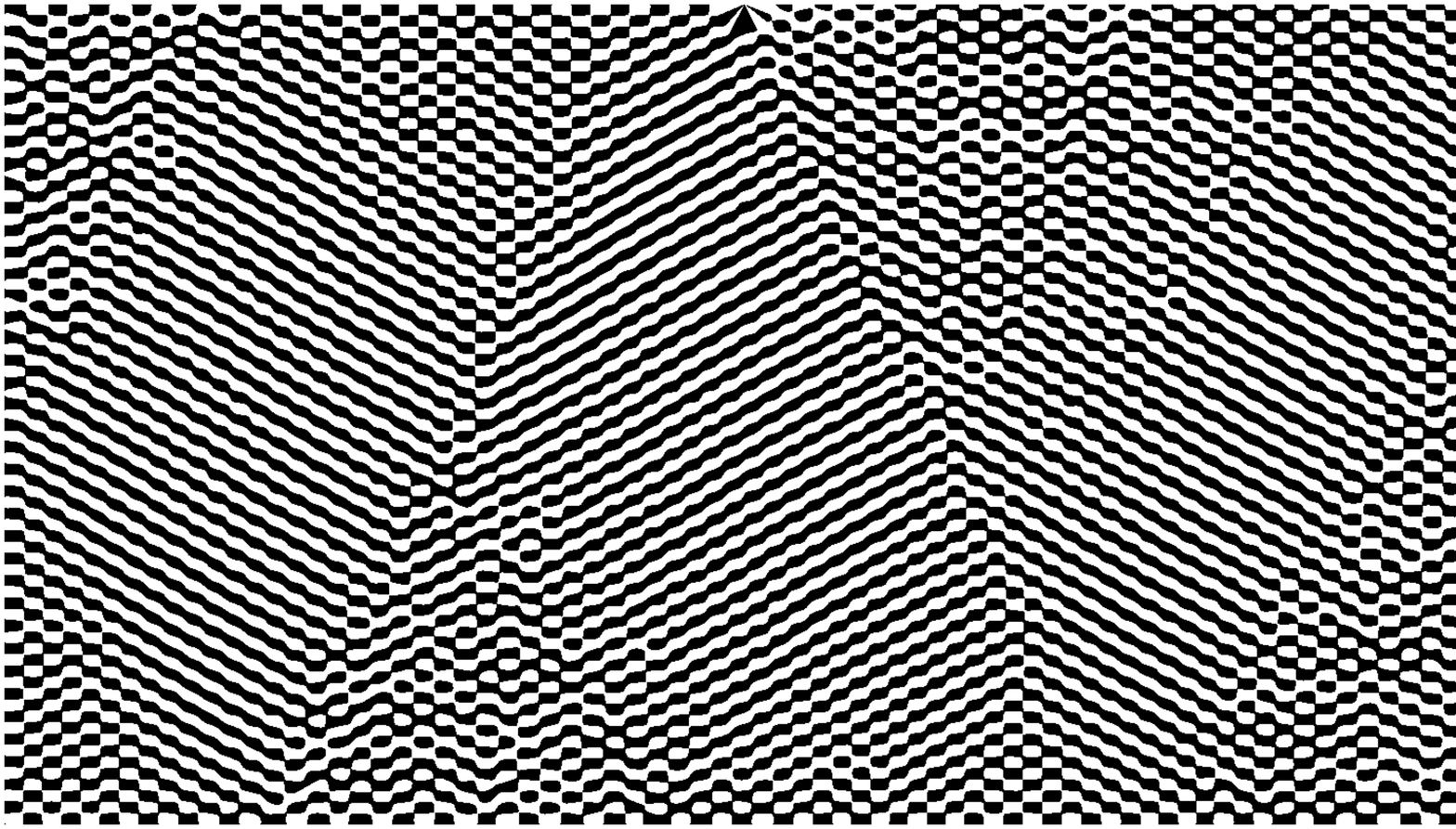} \\
a)
 \end{center}
 \end{minipage}
 \begin{minipage}{.49\linewidth}
\begin{center}
\includegraphics[angle=-90,width=.99\linewidth]{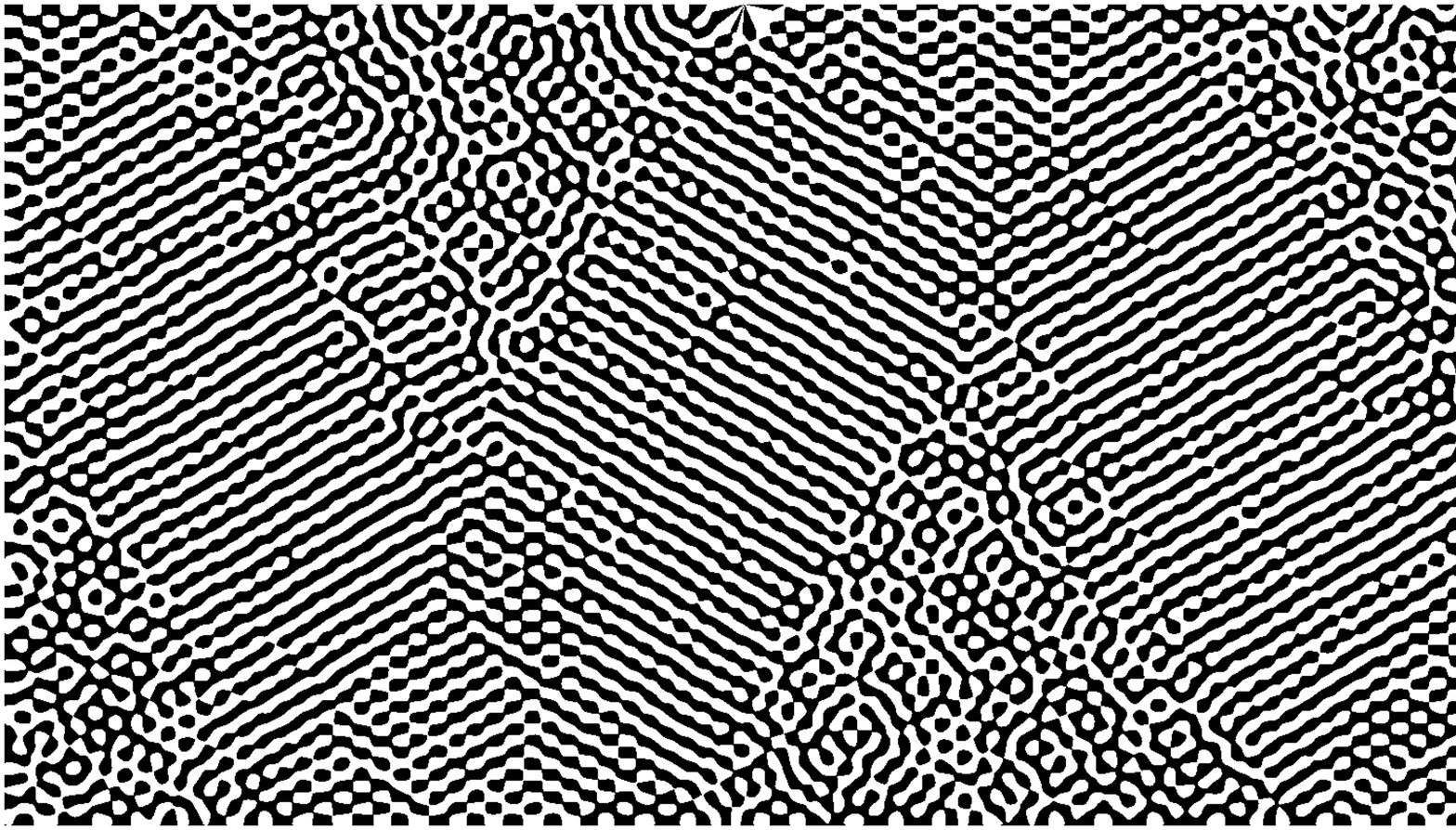} \\
b)
 \end{center}
 \end{minipage}
 \caption{ Nodal domains of numerically calculated eigenfunctions with (a)  $E_{\mathrm{exact}}=10018.13$ and (b)
  $E_{\mathrm{exact}}=10120.59$. The corresponding superscar energies  $\mathcal{E}_{589,1}^{(\mathrm{3-1})}=10019.80$ and 
 $\mathcal{E}_{592,1}^{(\mathrm{3-1})}=10122.07$. }
 \label{589_1_582_1}
 \end{figure}

The next example corresponds to the  $(3-1)$-orbit, see Fig.~\eqref{3_1_orbit}. It this case there are two symmetric channels with different parity of longitudinal quantum number $m$. The length of each of such channels is $L_p=\sqrt{(6a)^2+(2b)^2}$ and their width  is $w=2ab/L_p$. The superscar wave propagating in these channels have energy equal to
\begin{equation}
\mathcal{E}_{m,n}^{(\mathrm{3-1})}=\frac{\pi^2\, m^2}{L_p^2}+\frac{\pi^2\, n^2}{w^2},\qquad L_p=\sqrt{(6a)^2+(2b)^2},\quad 
w=\frac{2ab}{L_p}\, . 
\label{3_1_energy}
\end{equation}
At Fig.~\ref{589_1_582_1} two examples of eigenfunctions corresponding to odd and even $m$ are presented. The structure of propagating superscar waves is clearly visible and the exact energies are very close to the superscar ones.  
 
\begin{figure}
\begin{minipage}{.51\linewidth}
\begin{center}
\includegraphics[width=1.1\linewidth]{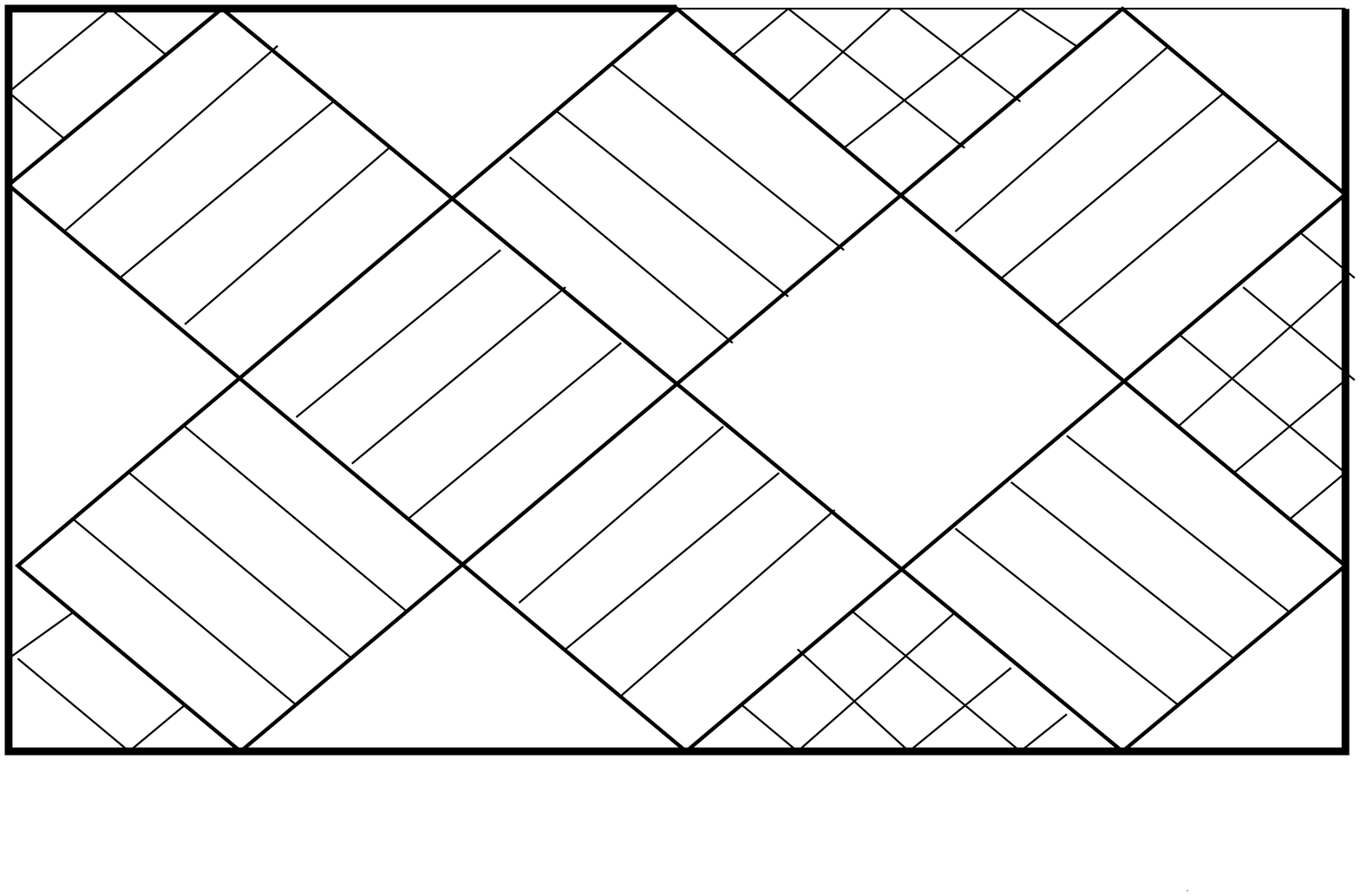} \\
a)
 \end{center}
 \end{minipage}
 \begin{minipage}{.48\linewidth}
\begin{center}
\includegraphics[angle=-90,width=.99\linewidth]{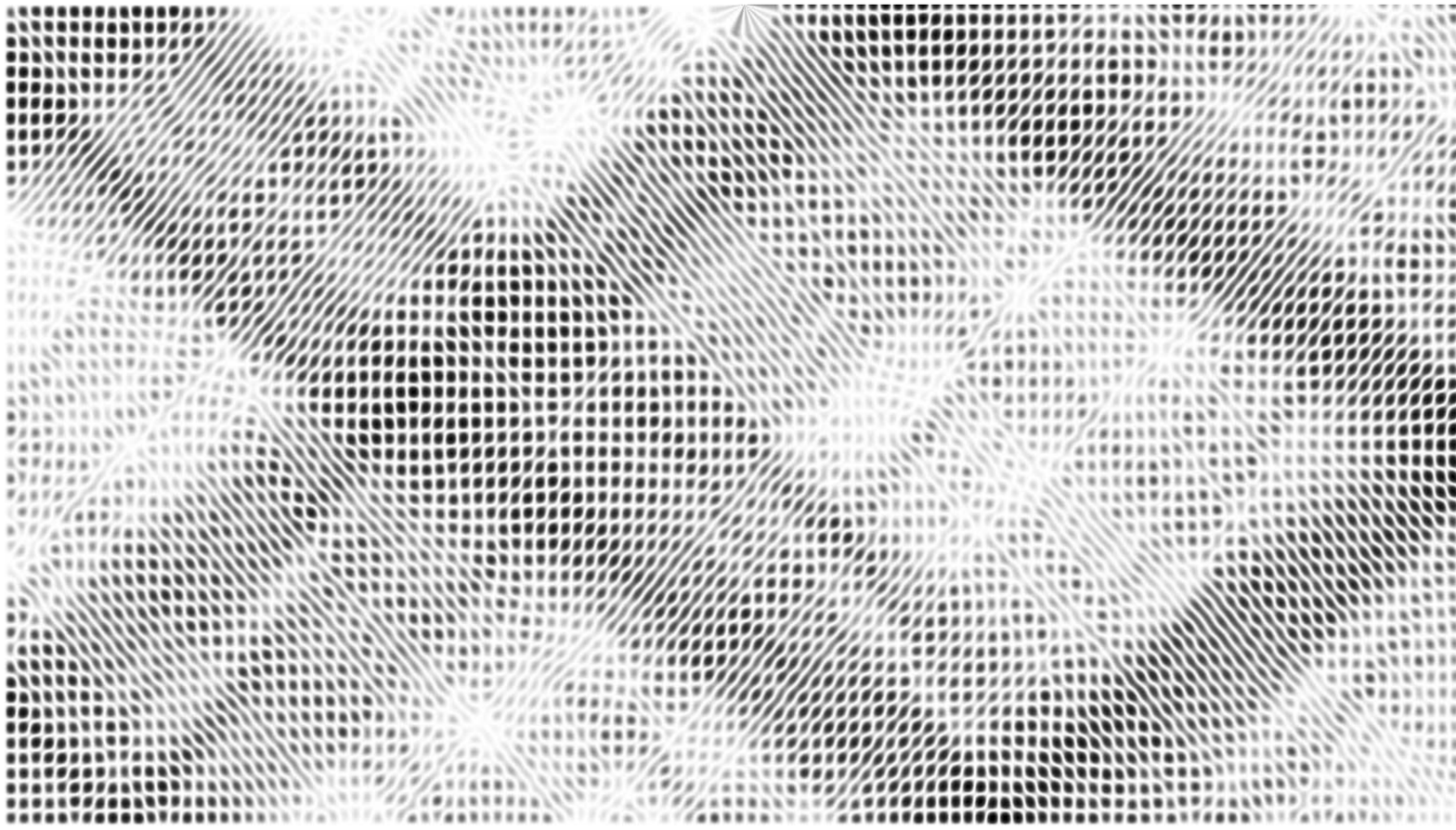} \\
b)
 \end{center}
 \end{minipage}
 \caption{ (a) One of periodic orbit channels for the $(3-2)$ orbit corresponding to  a superscar wave with even $m$ and its wavefronts. (b) Grey image of the modulus of eigenfunction with 
  $E_{\mathrm{exact}}=10152.12$. The corresponding superscar energies  $\mathcal{E}_{794,1}^{(\mathrm{3-2})}=10157.06$. }
 \label{3_2_orbit}
 \end{figure}

The last example is the $(3-2)$ orbit as at Fig.~\ref{3_2_orbit}~a) for which $L_p=\sqrt{(6a)^2+(4b)^2}$ and $w=2ab/L_p$. The corresponding eigenfunction with superscar structure is indicated at Fig.~\ref{3_2_orbit} b). 

\section{Quantitative characteristics of superscars}\label{Fourier_expansion} 

Numerous pictures  of superscar waves formation were presented in the previous Section. But such pictures are useful only to illustrate  a few superscar waves associated with short-period  orbit pencils. To get quantitative  information about the whole  structure of eigenfunctions in plane polygonal billiards it is convenient to calculate numerically  the overlap between an exact eigenfunction with energy $E_{\lambda}$ and a superscar wave propagated in a fixed periodic orbit pencil
 \begin{equation}
 C_{m,n}(E_{\lambda})=\int \Psi^{(\mathrm{superscar})}_{m,n}(x,y)\Psi_{E_{\lambda}}(x,y)\mathrm{d}x\,\mathrm{d}y\,. 
 \label{overlap}
 \end{equation}
Here $m$ and $n$ are integers corresponding to longitudinal and transverse quantum numbers of the superscar wave and the integration is performed over the whole billiard surface. The both functions in this equation are assumed to be  normalised so $0\leq |C_{m,n}(E_{\lambda})|\leq 1$. When the exact energy differs considerably from the superscar energy this overlap should be small. It means that for fixed $m$ only one peak appears when $E_{\lambda}\approx \mathcal{E}_{m,n}$. 

In calculations the transverse quantum number $n$ (which exists only due to singular diffraction) is kept  fixed but longitudinal quantum number $m$ (denoted below by $m(E)$) has been adjusted for different energies $E$ in such a way  that energy difference $|E-\mathcal{E}_{m,n}|$ is minimal 
\begin{equation}
m(E)=\left [\frac{L_p}{\pi}\sqrt{E-\frac{\pi^2n^2}{w^2}}\, \right ]
\end{equation}
where $[x]$ denotes the integer closest to $x$. 

A technical difficulty in this approach is the calculation of the folded superscar wave, $\Psi^{(\mathrm{superscar})}_{m,n}(x,y)$. The superscar wave is simple when it is unfolded. Due to folding back of periodic orbit pencils, superscar waves inside the original billiard  become complicated. For simple orbits the folded wave can be directly calculated as it has been done in Eq.~\eqref{exact_folding}. In Appendix \ref{appendix_A} it is shown how to calculate folded superscar function associated with an  arbitrary periodic orbit pencil.  

\begin{figure}
\begin{minipage}{.49\linewidth}
\begin{center}
\includegraphics[width=.9\linewidth]{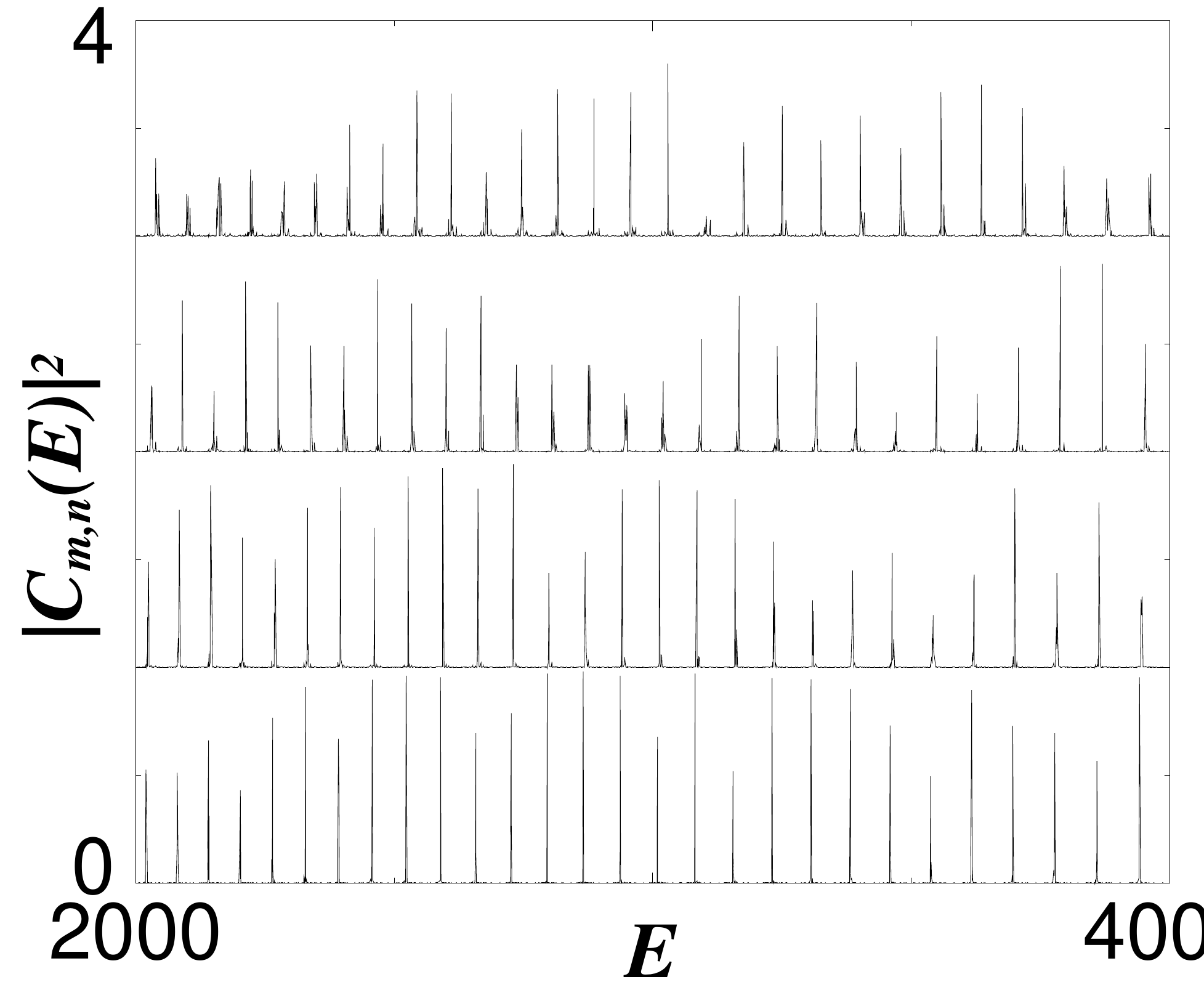} \\
a)
 \end{center}
 \end{minipage}
 \begin{minipage}{.49\linewidth}
\begin{center}
\includegraphics[width=.9\linewidth]{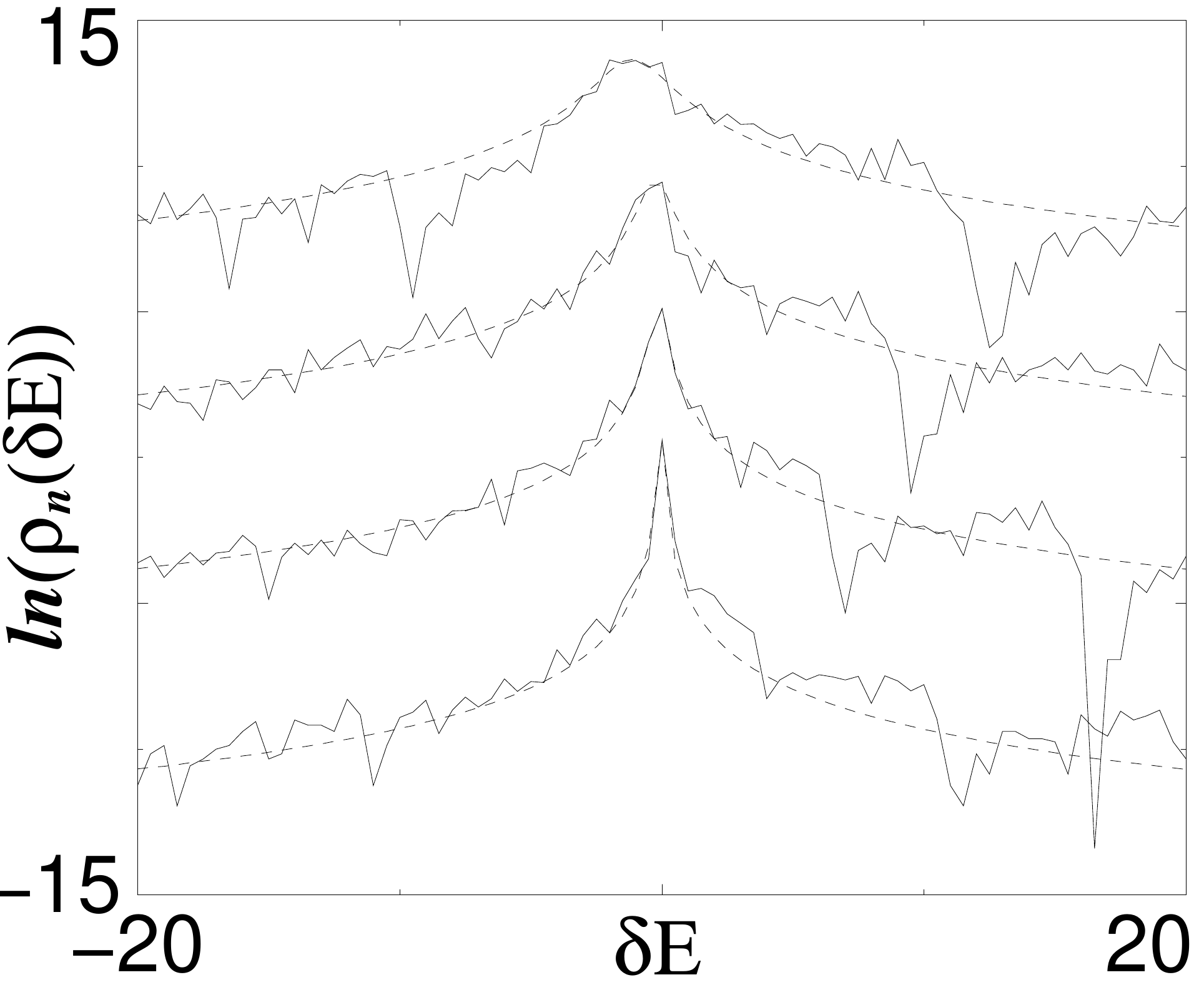} \\
b)
 \end{center}
 \end{minipage}
 \caption{(a) Overlap \eqref{overlap} for the superscar wave associated with the horizontal bouncing ball indicated at Fig.~\ref{horizontal}~a)   for different transverse quantum numbers: $n=1,\,2,\,3,\,4$. For clarity 4 plots were shifted vertically by 1. The lowest graph corresponds to $n=1$, the second to $n=2$ etc. (b) Local density given by Eq.~\eqref{local_density} calculated for the data of (a) for the same values of $n$. Different  curves are shifted vertically by 5 units. The abscissa axis is the shift of energy with respect to the difference $E_{\lambda}-\mathcal{E}_{m(E),n}$. The dashed lines are the Breit-Wigner fits \eqref{BW} to the local densities.}
 \label{overlap_0_1}
 \end{figure}

\begin{figure}
\begin{minipage}{.49\linewidth}
\begin{center}
\includegraphics[width=.99\linewidth]{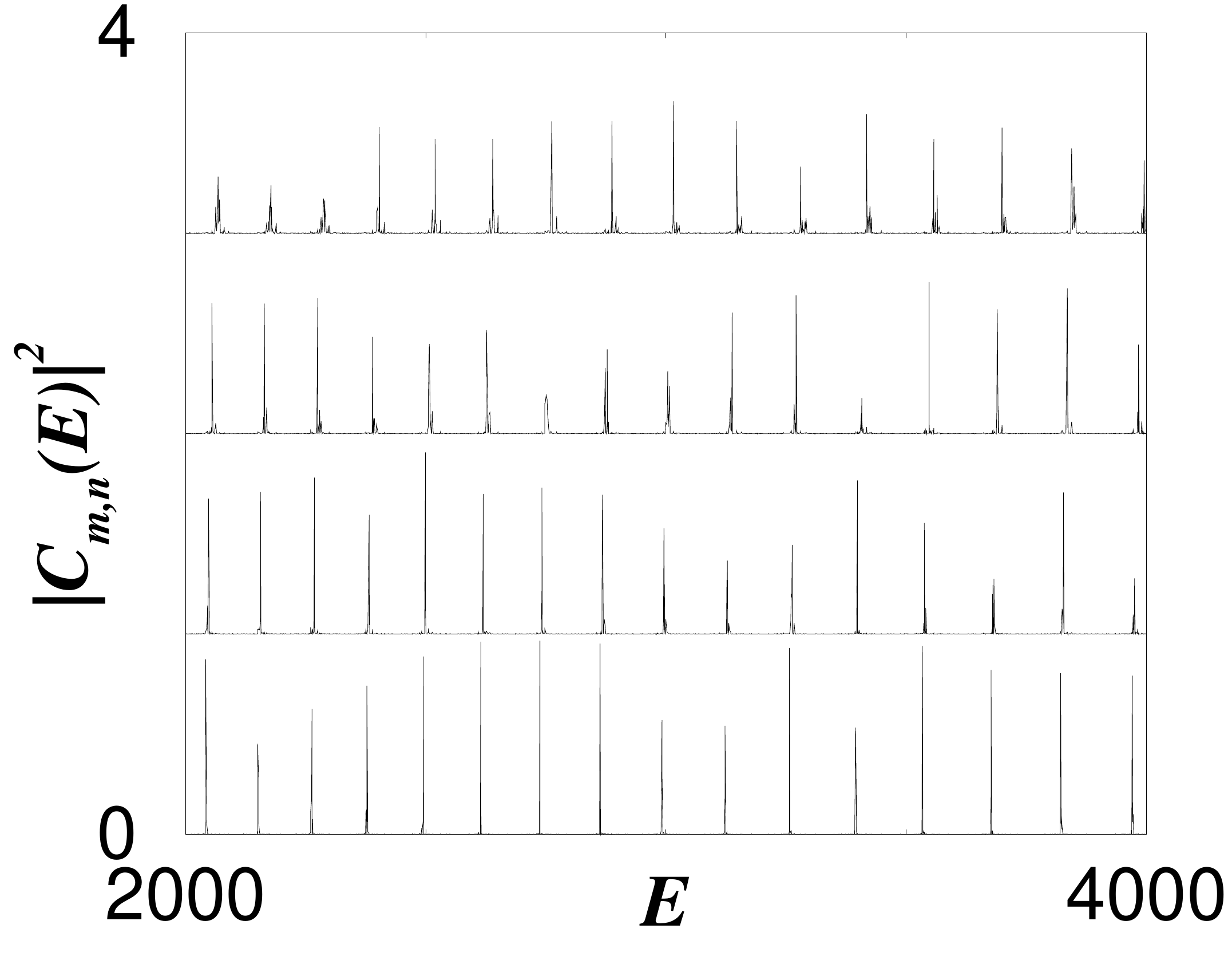} \\
a)
 \end{center}
 \end{minipage}
 \begin{minipage}{.49\linewidth}
\begin{center}
\includegraphics[width=.99\linewidth]{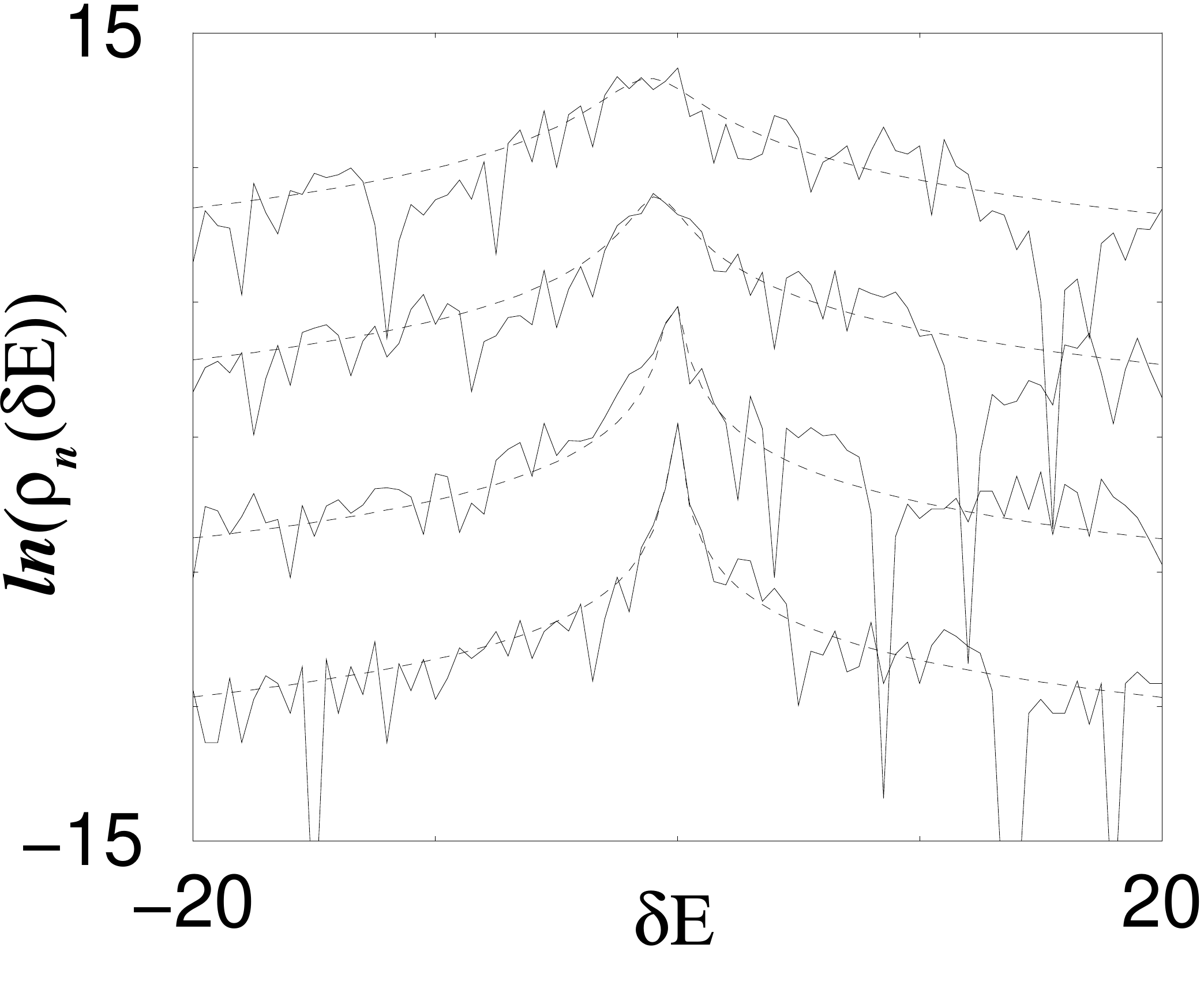} \\
b)
 \end{center}
 \end{minipage}
 \caption{ The same as at Fig.~\ref{overlap_0_1} but for left vertical bouncing ball indicated at Fig.~\ref{vertical}~b). }
 \label{overlap_1_0}
 \end{figure}

\begin{figure}
\begin{minipage}{.49\linewidth}
\begin{center}
\includegraphics[width=.99\linewidth]{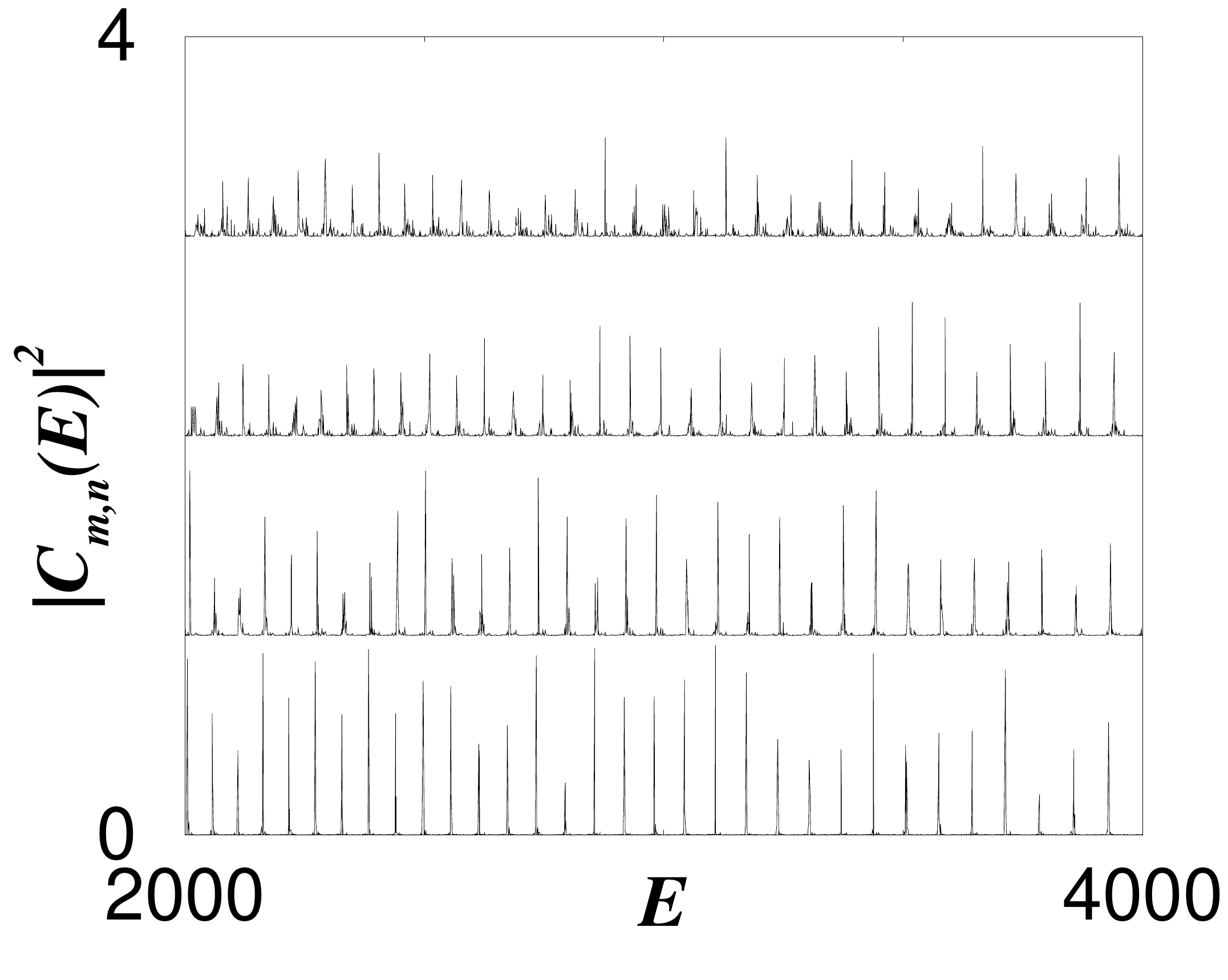} \\
a)
 \end{center}
 \end{minipage}
 \begin{minipage}{.49\linewidth}
\begin{center}
\includegraphics[width=.99\linewidth]{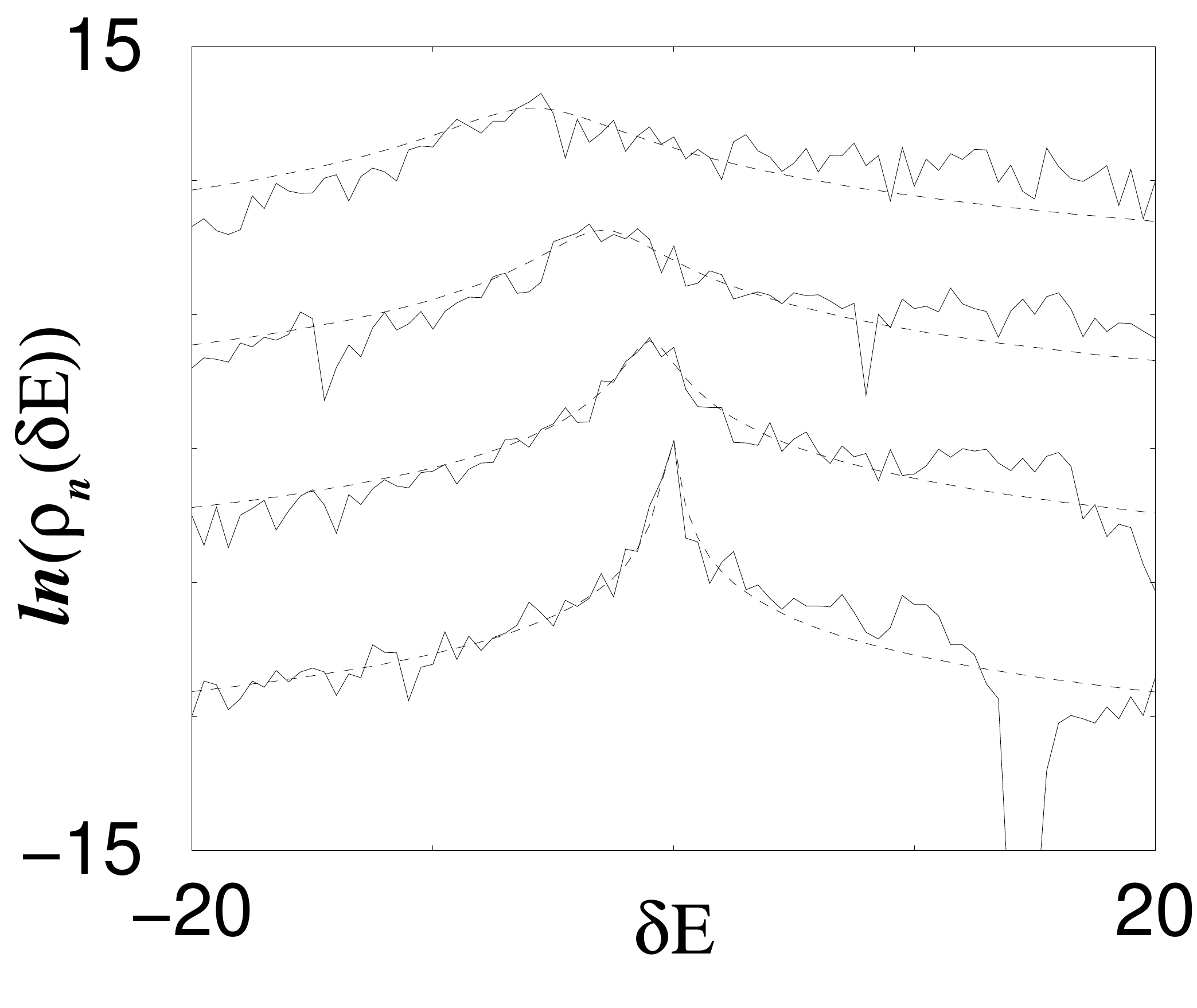} \\
b)
 \end{center}
 \end{minipage}
 \caption{The same as at Fig.~\ref{overlap_0_1} but for the $(1-1)$ periodic orbit  indicated at Fig.~\ref{1_1_orbit}~a).}
 \label{overlap_1_1}
 \end{figure}

The overlaps between all eigenstates in the interval $2000<E_{\lambda}<4000$ and all superscar waves propagating inside the $(0-1)$ pencil (horizontal bouncing ball),  the $(1-0)$ pencil (left vertical bouncing ball), and the $(1-1)$ pencil  are presented  at Figs.~\ref{overlap_0_1}~a) --\ref{overlap_1_1}~a).  Each of these figures shows the overlap for the four lowest transverse quantum numbers, $n=1-4$. 

Every time when the energy of true eigenstate is close to the superscar energy the corresponding eigenfunction has a considerable  overlap  with the superscar wave. As expected, small $n$ leads to higher values of the overlap. To analyse quantitatively the structure of overlap peaks it is instructive to calculate their  local density for each fixed $n$ defined as follows ($\delta E$ is the difference between the true energy  $E_{\lambda}$ and the best superscar energy $\mathcal{E}_{m(E_{\lambda}),n}$)
\begin{equation}
\rho_n(\delta E)=\left \langle \sum_{\lambda} |C_{m,n}(E_{\lambda})|^2 \delta(\delta E-E_{\lambda}+\mathcal{E}_{m,n}) \Big |_{m=m(E_{\lambda})}\right \rangle
\label{local_density}
\end{equation}
where the averaging is taken over all peaks in a given energy interval $[E-e,E+e]$ where $e\ll E$. For $n=1,\,2,\,3,\, 4$ this local density is plotted at  Figs.~\ref{overlap_0_1}~b) --\ref{overlap_1_1}~b).  As has been discussed above superscar waves can be considered as long-lived states which interact weekly due to residual interactions governed by parameter \eqref{parameter}. From general considerations \cite{Wigner}-\cite{Shepelyansky}, it is expected that in such situation the local density should be well approximated by  the  Breit-Wigner distribution
\begin{equation}
\rho_n(\delta E)\approx\frac{\Gamma_{n}(E)}{2\pi[(\delta E-\epsilon_{n}(E))^2+\Gamma_{n}^2(E)/4]}
\label{BW}
\end{equation}
where $\epsilon_n(E)$ and $\Gamma_{n}(E)$  are certain parameters (depending on the energy interval) which, in principle, could be calculated from perturbation series.  In Figs.~\ref{overlap_0_1}~b)--\ref{overlap_1_1}~b) it is demonstrated that such fits indeed approximate well the local densities for all considered cases. In the previous Section it was argued that the width $\Gamma_{n}(E)$ asymptotically should have the form indicated in Eq.~\eqref{imaginary_part}
\begin{equation}
\Gamma(E)=C\frac{\pi n^2}{w^2}\sqrt{\frac{d}{kw^2}},\qquad C=-\frac{2\zeta(1/2)}{\sqrt{\pi}}\approx 1.65\, .
\label{Gamma} 
\end{equation}
Here $w$ is the width of a periodic orbit pencil and  $d$ is the distance between singular vertices  along the pencil boundary. 

Numerical fits confirm this estimation. For example, for the $(1-1)$ orbit the data on Fig.~\ref{overlap_1_1}~b) are fitted  well by expression $\Gamma_{n}(E)\approx 3.5 n^2/\sqrt{k}$. When calculating analytically from  Eq.~\eqref{Gamma} one gets 
$\Gamma_{n}(E)=3.52n^2/\sqrt{k}$. 
 
 Another method to obtain quantitative measure of the superscar phenomenon  in plane polygonal billiards  consists in  the Fourier-type expansion of eigenfunctions.  For the full barrier billiard as at Fig.~\ref{examples}~b) with the Dirichlet boundary conditions along the rectangle $(2a,b)$ it is natural to represent eigenfunctions in the following basis
\begin{equation}
	\Psi(x,y)=\sum_{k ,p=1}^{\infty}  F_{p\, k}\sin\Big ( \frac{\pi p (x+a)}{2a} \Big )\sin \Big (\frac{\pi k y}{b}\Big ). 
\label{big_rectangle_expansion}
\end{equation}
For desymmetrised barrier billiard as at Fig.~\ref{examples}~c) no preferential system of expansion exists. Expansion \eqref{big_rectangle_expansion} inside the desymmetrised barrier billiard gives rise to two different series depended on the parity of $k$. For odd $p=2q-1$ function $\Psi(x,y)$ is even, $\Psi(-x,y)=\Psi(x,y)$
\begin{equation}
	\Psi_{\mathrm{even}}(x,y)=\sum_{q,k=1}^{\infty} f_{q\, k}\cos \Big (\frac{\pi(q-1/2)x}{a} \Big ) \sin \Big (\frac{\pi k y}{b} \Big ) 
 \label{even_function}
\end{equation}
and for even $p=2q $ function $\Psi(x,y)$ is odd, $\Psi(-x,y)=-\Psi(x,y)$
\begin{equation}
\Psi_{\mathrm{odd}}(x,y)=\sum_{q,k=1}^{\infty}  g_{q\,k}\sin \Big (\frac{\pi q x }{a}\Big ) \sin \Big (\frac{\pi k y}{b}\Big ).
\label{odd_function}
\end{equation} 
Formally the both series can be used on equal footing as  inside the desymmetrised barrier billiard these two expansions are equivalent because   
\begin{equation}
\sin \frac{\pi}{a}mx=\sum_{n=1}^{\infty} A_{mn} \cos \frac{\pi}{a}(n-\frac{1}{2})x
\label{Gibbs}
\end{equation}
where matrix $A_{mn}$ is an orthogonal matrix
\begin{equation}
A_{mn} = \frac{1}{\pi}\left ( \frac{1}{m+n-1/2}+\frac{1}{m-n+1/2}\right ).
\end{equation}
This possibility of re-expansion constitutes is a kind of the Gibbs phenomenon as  series \eqref{Gibbs} is only conditionally converges.

\begin{figure}
\begin{minipage}{.49\linewidth}
\begin{center}
\includegraphics[width=1.1\linewidth]{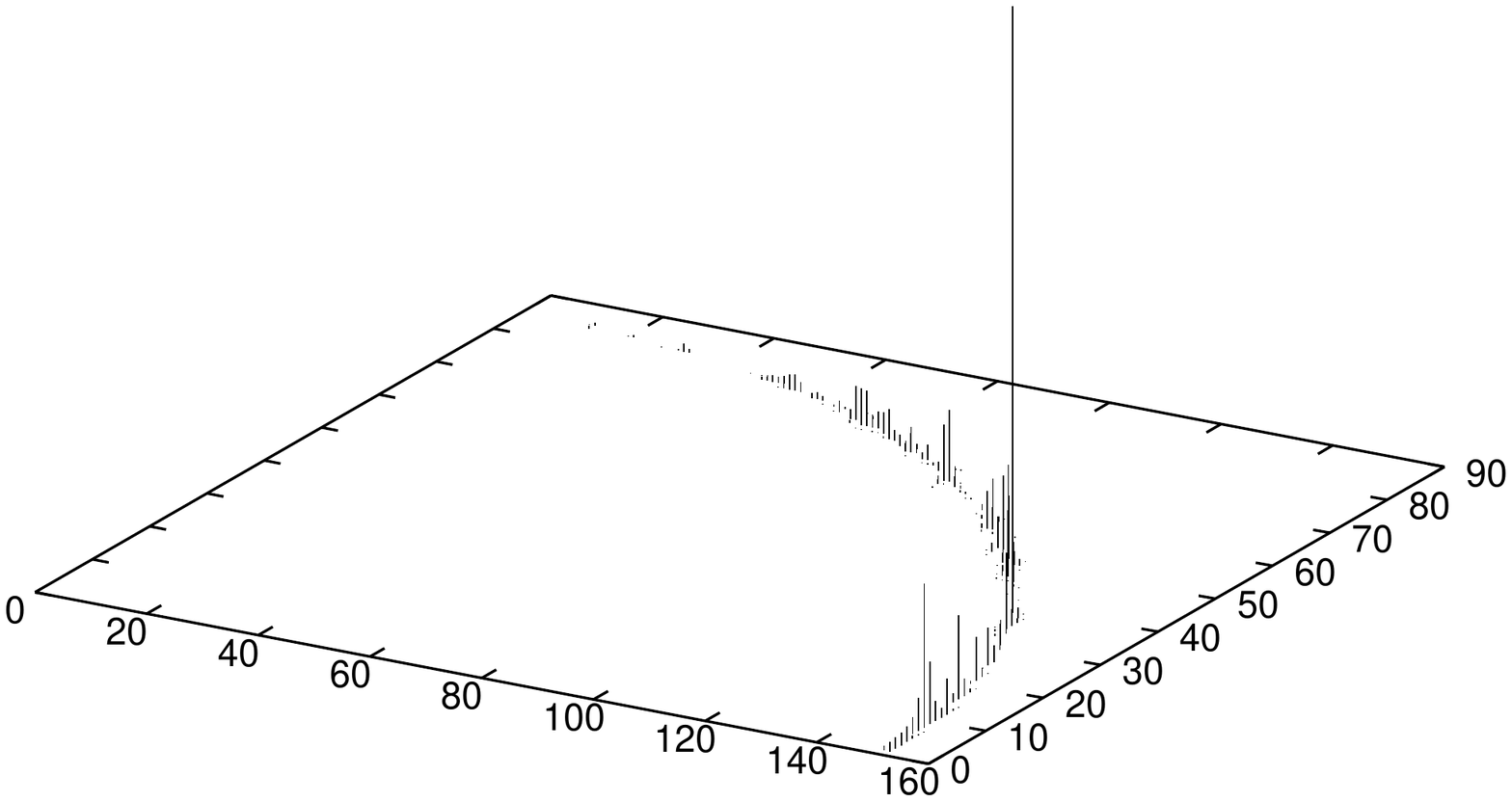} \\
a)
 \end{center}
 \end{minipage}
 \begin{minipage}{.49\linewidth}
\begin{center}
\includegraphics[width=1.1\linewidth]{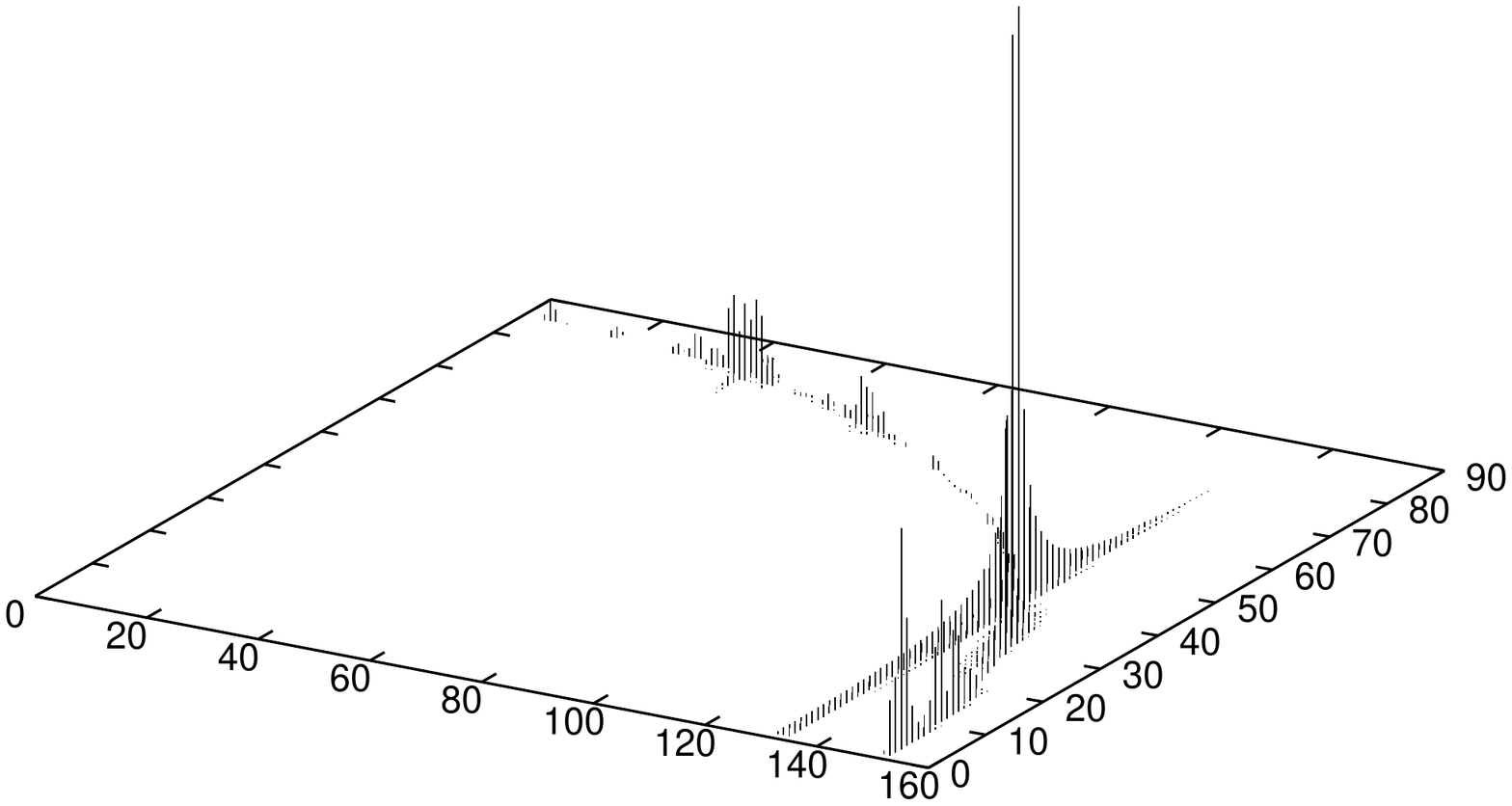} \\
b)
 \end{center}
 \end{minipage}
 \caption{Modulus of Fourier expansion coefficients for even series \eqref{even_function} for two eigenfunctions as at Fig.~\ref{348_1_347_1} corresponded to the $(1-1)$ orbit with energies:   (a) $E_{\mathrm{exact}}=10041.41$ and (b)   $E_{\mathrm{exact}}=10099.58$. }
 \label{coefficients_1_1}
 \end{figure}

\begin{figure}
\begin{minipage}{.49\linewidth}
\begin{center}
\includegraphics[width=1.1\linewidth]{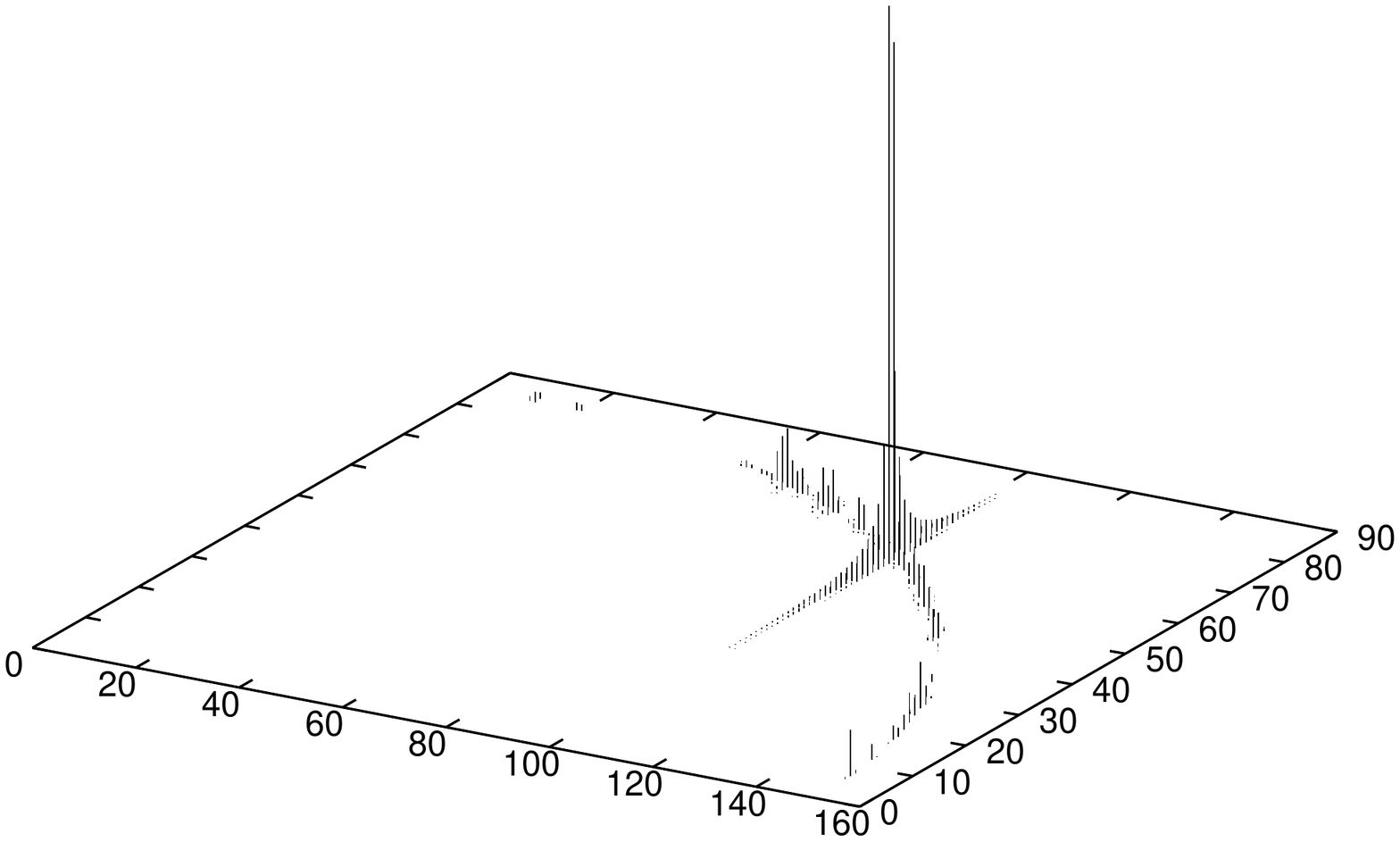} \\
a)
 \end{center}
 \end{minipage}
 \begin{minipage}{.49\linewidth}
\begin{center}
\includegraphics[width=1.1\linewidth]{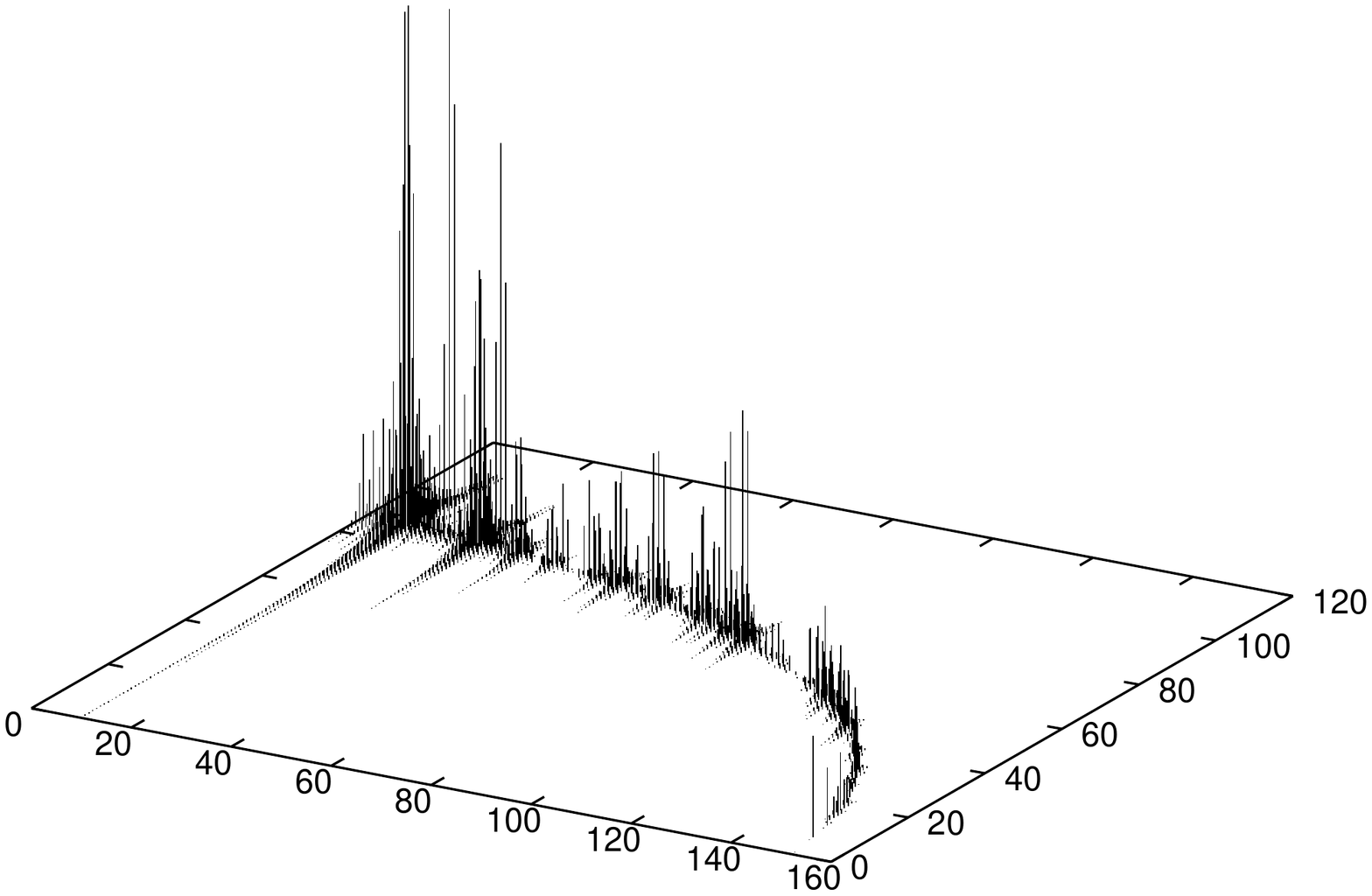} \\
b)
 \end{center}
 \end{minipage}
 \caption{Modulus of Fourier expansion coefficients for even series \eqref{even_function}:   (a) for eigenfunction indicated at Fig.~\ref{227_1_228_1}~a) corresponded to the $(2-1)$ orbit with $E_{\mathrm{exact}}=10017.57$ and  (b) for  an arbitrary chosen state with  $E_{\mathrm{exact}}=10013.57$. }
 \label{coefficients_2_1}
 \end{figure}

The existence of  superscars  manifests itself in the appearance of large coefficients in such Fourier-type expansions (see Appendix~\ref{appendix_A}).  A few examples of  these expansions for the barrier billiard are presented at Figs.~\ref{coefficients_1_1} and \ref{coefficients_2_1}. Energy conservation forces coefficients to be  close to a quarter to an ellipse curve in these figures. Noticeable exception is seen at Fig.~\ref{coefficients_1_1} b) where certain  coefficients deviate considerably from  the constant energy curve. It is plain that it corresponds to the above mentioned Gibbs phenomenon.  If this eigenfunction is expanded into odd series \eqref{odd_function} such  large deviations would disappear (cf. Appendix~\ref{appendix_A}). But for orbits with even $M$ like the $(2-1)$ orbit indicated at Fig.~\ref{2_1_orbit}~a) expansion coefficients always have such Gibbs tail which could not be removed by simple change of the basis  (cf. Fig.~\eqref{coefficients_2_1} a)).

It is clear that well isolated  superscar states associated with short-period orbits  are rare. Typical eigenfunctions may contain a certain number of large coefficients corresponded to a kind of superposition of many different superscar waves (see Fig.~\ref{coefficients_2_1}~b)). 
  
\section{Fractal dimensions}\label{fractal_dimensions}

Everything discussed in the previous Sections about  a superscar wave propagating inside a periodic orbit pencil could be applied  to an arbitrary  polygonal billiards where there exists  at least one  periodic orbit family. Unfortunately periodic orbits in generic polygonal billiards is an elusive subject and even the existence of one periodic orbit is not guaranteed.  Only for a special sub-class of rational polygonal billiards called Veech polygons \cite{Veech} where there exits a hidden group structure one can control all classical periodic orbits.  

The knowledge of periodic orbits in such models permits to calculate analytically an important characteristic of their spectral statistics, namely the spectral compressibility, $\chi$, \cite{Giraud}, \cite{Wiersig}   which determines the linear growth of the number variance with the length of the interval
\begin{equation}
\Sigma^2(L)\equiv \langle (n(L)-\bar{n}(L))^2\rangle =\chi L .
\end{equation}   
Here $n(L)$ is the number of energy levels in an interval $L$, $\bar{n}(L)$ is the mean number of levels in this interval normalised to unit density, $\bar{n}(L)=L$, and the average is taken over different intervals of length $L$ in a small energy window.  For the Poisson distribution typical for spectral statistics of integrable systems $\chi=1$ and for standard random matrix ensembles which describe spectral statistics of chaotic systems $\chi=0$. The right triangular billiard with $\pi/8$ angle  has $\chi=\frac{5}{9}$ \cite{Giraud} and the barrier billiard  considered in the paper has $\chi=\frac{1}{2}$ \cite{Wiersig}.
 
Spectral statistics of models with non-trivial compressibility, $0<\chi <1$, are called intermediate statistics. Many pseudo-integrable billiards belongs to this class \cite{Richens}, \cite{Cheon}--\cite{Wiersig}. 
The characteristic features of  intermediate statistics are  (i) level repulsion on small distances as for usual random matrix ensembles, (ii) exponential decrease of the nearest-neighbour distribution on large distances similar to the Poisson distribution.  These properties have been observed in numerical calculations but have not been fully proved mathematically.  Numerics (and certain analytical arguments \cite{entropy}) also suggest that for models with intermediate spectral statistics eigenfunctions are fractal (in general, even multifractal).
 
The notion of multifractality (see e.g. \cite{Mirlin}, \cite{Evers} and references therein) is related with a natural question about the number of important components in eigenfunctions. Let an eigenfunction with eigenvalue $E$ be written as an expansion in a certain  basis
\begin{equation}
\Psi_{E}(x,y)=\sum_{\nu=1}^{\mathcal{N}}A_{\nu}(E) \phi_{\nu}(x,y), \qquad A_{\nu} =\int \Psi(x,y)\phi_{\nu}(x,y)\mathrm{d}x\mathrm{d}y,  \qquad  \qquad \sum_{\nu=1}^{\mathcal{N}}A_{\nu}^2=1.
\label{wave_expansion}
\end{equation} 
Here $\mathcal{N}$ is the total  number of components. 

The central question in the multifractal formalism is the scaling of the moments of expansion coefficients with $\mathcal{N}$. Let define the moments with arbitrary $q$ as follows 
\begin{equation}
M_q(E)=\sum_{\nu=1}^{\mathcal{N}} |A_{\nu}(E)|^{2q}.
\label{moments}
\end{equation}
The inverse of these moments, $R_q=M_q^{-1}$, is called the participation ratios.

Multifractality means that moments of eigenfunctions (or their inverse)  scale as a certain power of total number of wave function components
\begin{equation}
M_q\underset{\mathcal{N}\to\infty}{\longrightarrow} \mathcal{N}^{-\tau(q)}, \tau(q)=(q-1)D_q
\end{equation}  
where $D_q$ are called generalised fractal dimensions. 

If only a finite number of components gives contribution to an eigenfunction \eqref{wave_expansion} (e. g.  for localised states) then $D_q=0$. In the opposite case of completely extended states when all components are of the same order then from normalisation 
$A_{\nu}\sim \mathcal{N}^{-1/2}$ and consequently $D_q=1$. 

In Ref.~\cite{Shklovskii} the multifractality was observed in the 3-dimensional Anderson model at the metal-insulator transition and later it has been investigated in different matrix models \cite{Evers}.  
 
For billiards the sum in \eqref{wave_expansion} includes formally an  infinite number of summands. For 2-dimensional billiards a natural basis consists of elementary trigonometric functions with fixed momentum. Physically it is clear that in the semiclassical limit $k\to\infty$ the number of important (large) components should be of the order of the number of possible quantum cells on the constant momentum  surface. For 2-dimensional billiards this surface is a circle of radius $k$ and therefore $\mathcal{N}\sim k $ (as we are interested only in powers of $\mathcal{N}$ precise pre-factor is irrelevant). Consequently, for billiards fractal dimensions determine the behaviour of the moments as function of the momentum
\begin{equation}
M_q(E)=\sum_{\nu=1}^{\infty} |A_{\nu}(E)|^{2q} \underset{k\to\infty}{\longrightarrow} k^{-\tau(q)}, \qquad k=\sqrt{E}.
\end{equation}  
For systems with non-trivial spectral compressibility, $0<\chi<1$, numerical  and partly analytical calculations \cite{entropy} suggest that fractal dimensions should be  also non-trivial, $0<D_q<1$. To check these predictions numerical calculations of fractal dimensions were performed for high-excited states in  the barrier billiard. Each eigenfunction has been expanded into series \eqref{even_function} 
\begin{equation}
	\Psi_{E}(x,y)=\sum_{q,k=1}^{\infty} f_{q, k}(E)\cos \Big (\frac{\pi(q-1/2)x}{a} \Big ) \sin \Big (\frac{\pi k y}{b} \Big ) 
 \label{expansion_fractal}
\end{equation}
and  expansion coefficients $A_{\nu}(E)\equiv   f_{q,k}(E)$ were computed.  

\begin{figure}
\begin{center}
\includegraphics[width=.99\linewidth]{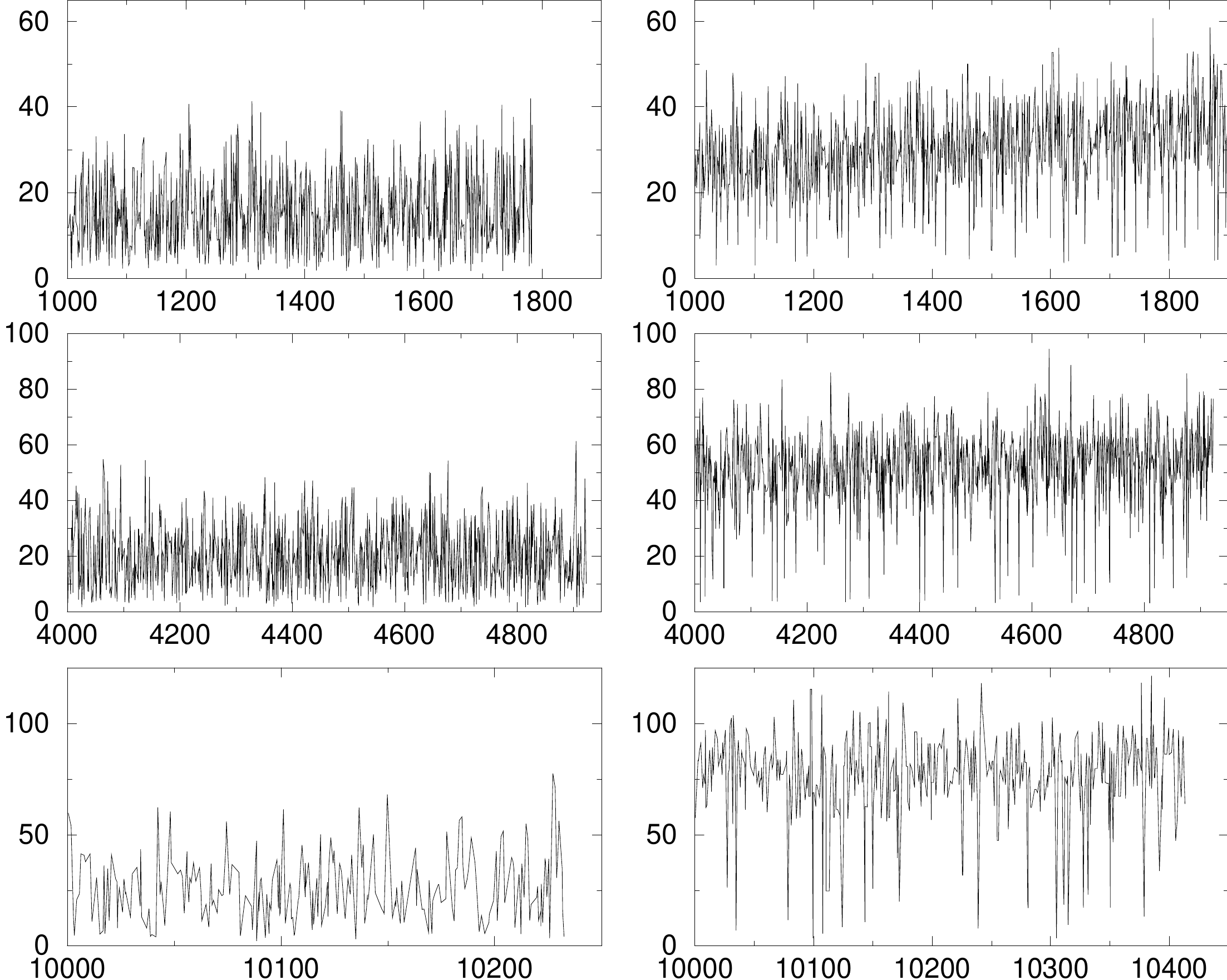} 
\end{center}
\caption{ Participation ratio $R_2(E)\equiv M_2^{-1}$ in different energy intervals  for barrier billiard (left) and the stadium billiard (right). }
\label{barrier_stadium}
\end{figure}

\begin{figure}
 \begin{minipage}{.49\linewidth}
\begin{center}
\includegraphics[width=.99\linewidth]{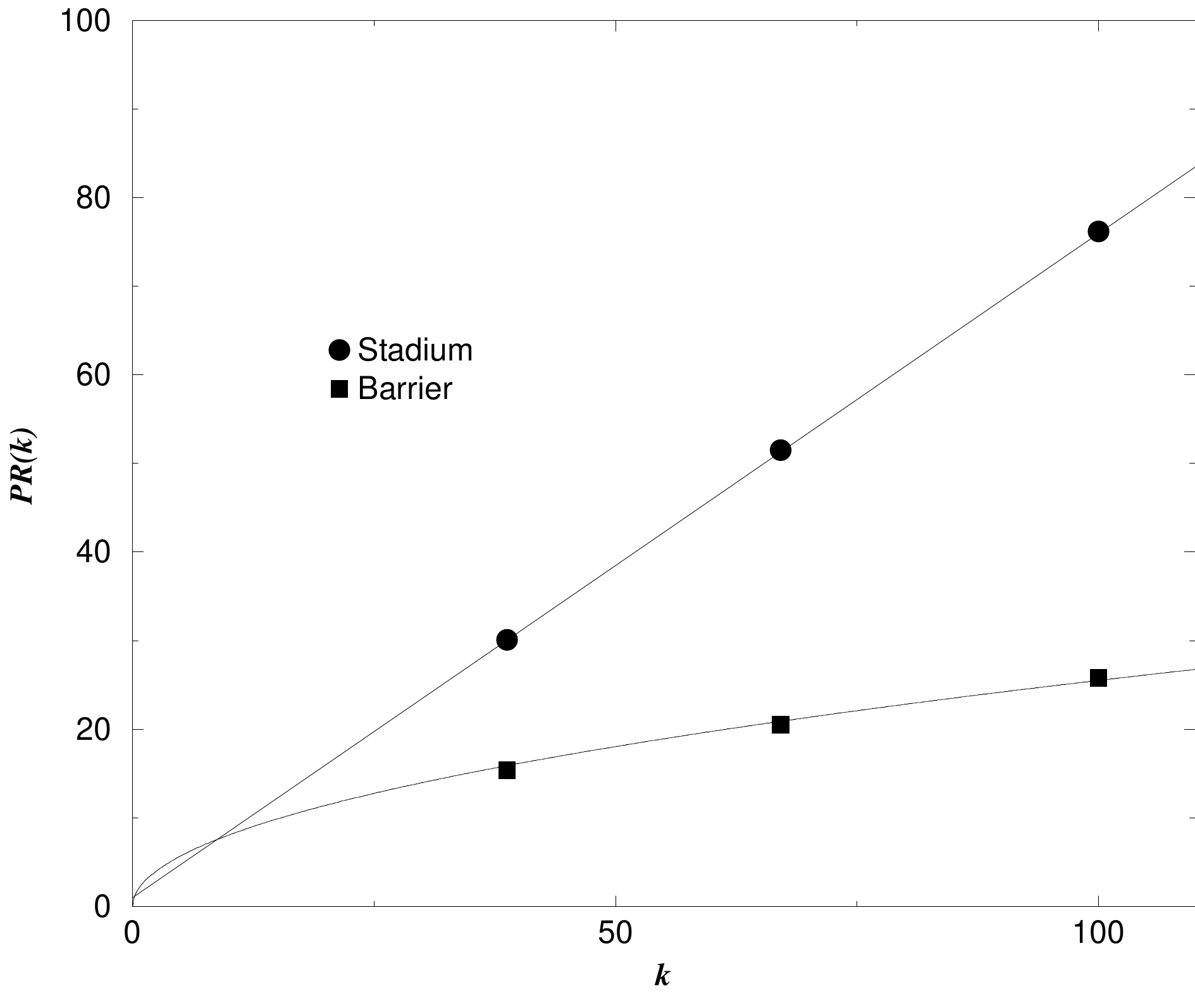} \\
a)
 \end{center}
 \end{minipage}
 \begin{minipage}{.49\linewidth}
 \begin{center}
\includegraphics[width=.99\linewidth]{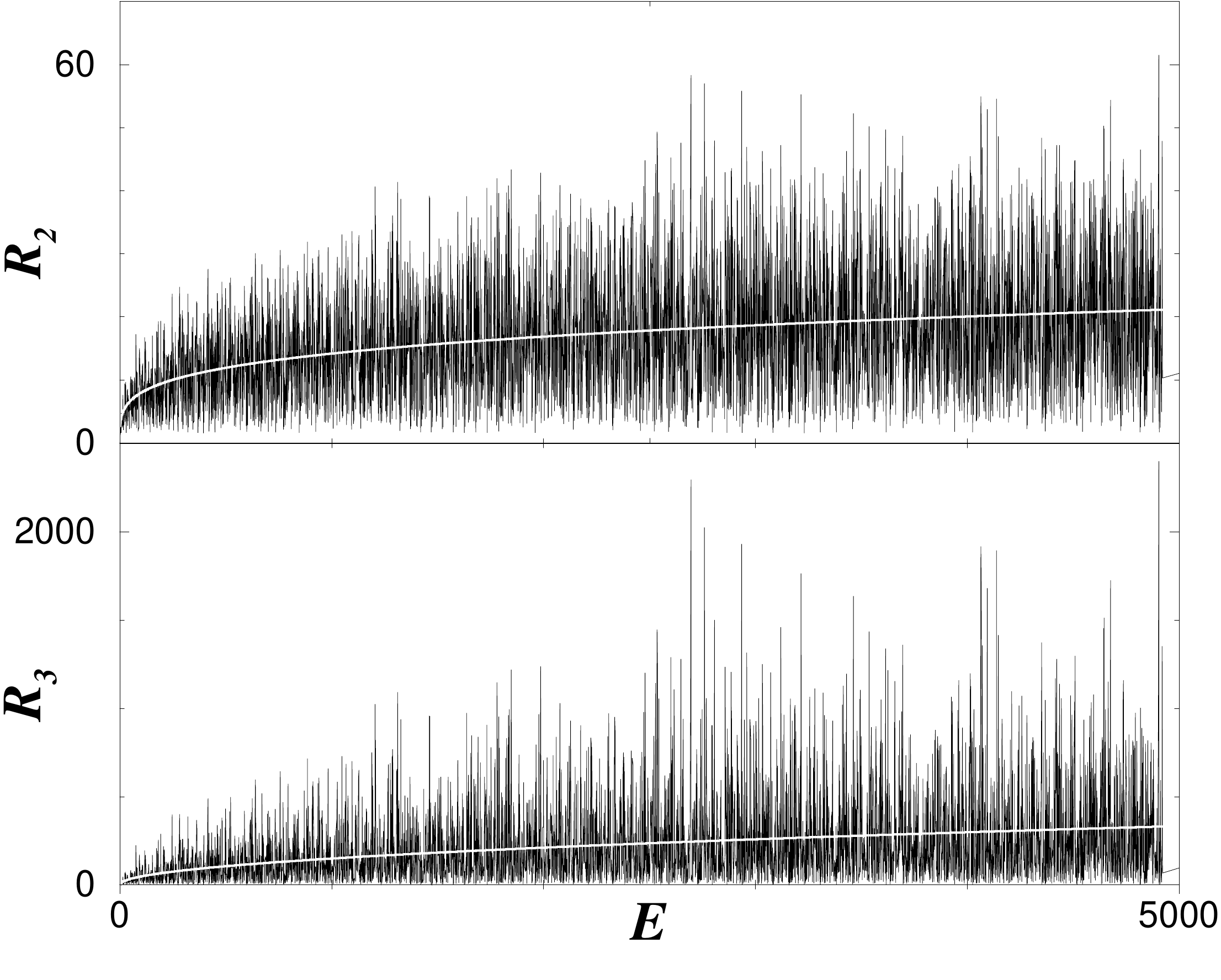} \\
b)
 \end{center}
 \end{minipage}
 \caption{ (a) Scaling with momentum of participation ratio $R_2$ for the barrier billiard (square) and the stadium billiard (circle). Solid lines indicate the fits to the data.  Stadium fit: $R_2(k)\approx 0.75 k+1$. Barrier billiard fit:  $R_2(k)\approx 2.55\sqrt{k}$. 
 (b) Participation ratios $R_2$ (top) and $R_3$ (bottom) versus energy for the barrier billiard. White lines indicate the fits  $R_2\approx 2.52\sqrt{k}$ and $R_3\approx 4.7k$.  }
\label{fits_barrier_stadium}
 \end{figure}

At Fig.~\ref{barrier_stadium}~left)   the participation ratio $R_2$ for 3 energy intervals close to the $1000^{\mathrm{th}}$, the $4000^{\mathrm{th}}$, and the $10000^{\mathrm{th}}$ levels for the barrier billiard  are presented. For comparison at this figure the same quantity but for the quarter of the (chaotic) stadium billiard with the same aspect ratio   are shown for comparison. The area of the both billiards is $4\pi$ and the energies approximately equal the level numbers.  
At Fig.~\ref{fits_barrier_stadium}~a)  these data were used to calculate average values of the participation ratios for the barrier billiard and the stadium billiard. As expected, for the chaotic stadium billiard participation ratio scales linear with the momentum. The best fit gives  $R_2(E)=0.75 k$. But for the barrier billiard the best  fit suggests that  $R_2(E)=2.55\sqrt{k}$ which means that fractal dimension  $D_2$ is non-trivial, $D_2=0.5$. At Fig.~\ref{fits_barrier_stadium} the participation ratios $R_2(E)$ and $R_3(E)$ for all states till $E=5000$ are plotted for the barrier billiard. The best fits (indicated by white lines at this figure) give $R_2(E)\approx 2.52\sqrt{k}$ and  $R_3(E)\approx 4.7k$. Therefore, these results suggest that $D_2\approx D_3\approx 0.5$.  

 Of course, much more calculations should be done to establish correct values of fractal dimensions for pseudo-integrable billiards.
 It has been briefly mentioned  in Ref.~\cite{main} that such fractal dimensions could  be obtained analytically  in an oversimplified model of the barrier  billiard eigenfunctions based on  the assumption that mean variance of expansion coefficients has the Breit-Wigner form similar as for the overlap coefficients \eqref{BW}.  In a sense, the model is resemble  the Rosenzweig-Porter model \cite{Kravtsov}, \cite{RP} which has been investigated later. The detailed discussion of that model is beyond the scope of this paper and will be performed  somewhere.


\section{Summary}\label{summary}

Wave functions are on the very basis of quantum mechanical calculations and the investigation of their properties are important for many applications. For generic classically chaotic systems Berry's conjecture \cite{Berry_2}, \cite{Berry_3} stipules that  in the semiclassical limit typical quantum eigenfunctions are random superpositions of elementary functions with fixed momentum whose coefficients are independent Gaussians with zero mean and variance determined from the normalisation. Nevertheless, it does not signify  that all chaotic eigenfunctions are completely structureless (cf. \cite{Shnirelman}). It is well known that eigenfunctions in certain models 
may have structures (called scars) in a vicinity of unstable periodic orbits \cite{scar_heller}-\cite{scar_berry}. Contributions of individual unstable periodic orbits decrease with increasing of energy and a general belief is that scar phenomenon in chaotic systems will be strongly suppressed (or even disappear) in the semiclassical limit though a certain increase of amplitudes could still be detected \cite{kaplan}. 

The main message of this paper is that  eigenfunctions of plane polygonal billiards have clear structures associated with  periodic orbit families. The mechanism of formation of such structures is not  periodic orbits themselves but singular  scattering on billiards corners whose angles $\neq \pi/n$ with integer $n$. Classical ray scattering on such scatters are discontinuous but quantum mechanics substitutes  singularities by smooth strong filed regions (called optical boundaries). When optical boundaries of many scatters overlap  the result in the semiclassical limit corresponds to vanishing of eigenfunctions along straight lines formed by singular scatters. As periodic orbits in polygonal billiards form parallel families  (pencils) restricted by singular scatters such pencils can support propagating waves reflected from pencil boundaries  passing through singular vertices as from mirrors. The validity of such specular reflection from fictitious  mirrors becomes exact in strict semiclassical limit $k\to\infty$ which explains the existence of such structures at very high energies contrary to scars in chaotic systems. To stress this fact we propose to call them 'superscars'. Many pictures of superscars in simple billiards are presented in the main part of the paper.  For the barrier billiard certain superscar waves were observed in microwave  experiments \cite{Dietz}.

In principle, superscar waves may exist in any polygonal billiards.  But there is no theorem that guarantees the existence of even one classical periodic orbit  for generic polygonal billiards. Consequently the superscar construction could be applied for polygonal billiards with at least one periodic orbit family. To construct a superscar wave associated with a periodic orbit pencil it is necessary to know the periodic orbit length, $L_p$  and the width of the pencil, $w$,  restricted by singular billiard corners. The unfolded superscar wave has the form of the plane wave propagating inside the pencil as in Eq.~\eqref{main} and its energy is the same as for a wave propagating inside rectangular slab with the Dirichlet boundary conditions: $\mathcal{E}_{m,n}=\pi^2 m^2/L_p^2+\pi^2 n^2/w^2$. The parity of longitudinal quantum number $m$ is determined by the total phase accumulated by the periodic orbit. The transverse quantum number $n$ can be arbitrary but $n\ll m$. The simplest verification of this construction  consists in computing a few states with energies  in a vicinity of the superscar energy with different $m$ and a fixed $n$. Numerical calculations performed in the paper show  that  in a small vicinity of superscar energies there always exist true eigenstates which have clear structure related with the folded superscar wave.  To get quantitative view of such phenomenon it is useful to calculate numerically the overlap between true eigenfunctions and the folded superscar wave. At least for the barrier billiard such overlap has the expected  Breit-Wigner form \eqref{BW} whose parameters agree with analytical estimates.  

Further progress depends on possibility to control periodic orbits in polygonal billiards which is a complicated problem.
Only for a special sub-class of polygonal billiards called Veech polygons \cite{Veech}, \cite{Giraud} one can find analytically all periodic orbits and their parameters.   Right triangular billiard with one angle $\pi/n$ and the barrier billiard considered in the paper belong to this class. For Veech polygons one can argue that eigenfunctions in the momentum representation are fractal with non-trivial fractal dimensions which is confirmed  by numerical calculations for the barrier billiard. 

The investigations presented in the paper clearly demonstrate that eigenfunctions of polygonal billiards and especially of pseudo-integrable ones have  interesting and unusual properties different from the both integrable and chaotic systems. In the absence of mathematical theorems more detailed  examinations of such phenomena are highly desirable.


\appendix
\section{Expansion of a superscar wave into Fourier-type  series}\label{appendix_A}
   
The purpose of this Appendix is to calculate analytically folded superscar waves associated with a given periodic orbit pencil in the barrier billiard. For simple periodic orbits the folding can be done by inspection as in \eqref{exact_folding}.   For more complicated orbits it is convenient to find  the expansion of superscar waves  in   Fourier-type series \eqref{big_rectangle_expansion}, \eqref{even_function}, \eqref{odd_function}. 

An unfolded  periodic orbit in a 2 dimensional rectangle with sides $a$ and $b$ can be represented by a line which connects the origin (the point with coordinates $(0,0)$) with a point with coordinates $2Ma$ and $2Nb$ (cf. Fig.~\ref{examples}~c)).  Its length is 
\begin{equation}
L_p=\sqrt{(2Ma)^2+(2Nb)^2}.
\label{length}
\end{equation}
Primitive periodic orbit corresponds to co-prime integers, $(M,N)=1$. Periodic orbits in the rectangular billiard form families whose total width is $2w$ where 
\begin{equation}
w=\frac{2ab}{L_p}.
\label{width}
\end{equation}
Introduce a new coordinate system $(\xi,\eta)$ with coordinate $\xi$ along the orbit and $\eta$ perpendicular to it
\begin{equation}
\xi=x\cos \theta+y\sin \theta,\qquad  \eta=-x\sin \theta +y\cos \theta, \qquad \cos \theta=\frac{2Ma}{L_p},\qquad \sin \theta =\frac{2Nb}{L_p}
\label{new_coordinates}
\end{equation}
and $L_p$ is the periodic orbit length \eqref{length}. 

Periodic orbit pencils are determined by periodic orbits shifted parallel to one of them till touching singular points whose coordinates are
\begin{equation}
x_s=2a \alpha \quad y_s=b/2+b\beta , \qquad \alpha,\beta \in \mathbb{Z}. 
\end{equation}
Values of $\eta$ when periodic orbit channel hits such points follow from Eq.~\eqref{new_coordinates}
\begin{equation}
\eta_s=w(-2\alpha N+\beta M +\frac{1}{2}M). 
\end{equation} 
When $M$ is odd, $\eta_s=w(r+\frac{1}{2})$  with integer $r$. Therefore minimal distance of the chosen periodic orbit passing through the origin from singular points is $\pm w/2$ which corresponds to periodic orbit pencil width equals $w$ \eqref{width}.   For even $M=2M^{\prime}$, $\eta_s=w r $ with integer $r=-2\alpha N+(2\beta+1)M^{\prime}$. If $M^{\prime}$ is odd, $r\neq 0$ as $(N,M^{\prime})=1$ and minimal $\eta_s=\pm w$ which leads to the pencil width equals $2w$. If $M^{\prime}$ is even, periodic orbit passing through the origin hits singular points but $r$ is even. Therefore in this case the pencil width is also $2w$. 

To construct superscar waves it is necessary to know the total phase accumulated by periodic orbits inside a given periodic orbit pencil. Let us calculate this phase for the  periodic orbit passing through the origin. Its equation is
\begin{equation}
y=\frac{Nb}{Ma}x.
\end{equation}
The Neumann part  of the boundary after unfolding will be situated  at $x_j=2aj$, $j=0..M-1$ and will  occupy the segments 
\begin{equation}
I_j=[b/2+2rb,3b/2+2rb],\qquad r=0..N-1.
\label{intervals}
\end{equation}
The number of crossing of these boundaries is 
\begin{equation}
\mathcal{N}(M)=\sum_{j=0}^{M-1}\Theta\left (\Big \{\frac{N}{M}j\Big \}-\frac{1}{4}\right )\Theta\left (\frac{3}{4}-\Big \{\frac{N}{M}j\Big \}\right )
\end{equation}
where $\{x\}$ denoted the fractional part of $x$ and $\Theta(x)$ is the Heaviside function.  As $N,M$ are co-prime integers the series $\{\frac{N}{M}j\}$ can be substituted by $\{\frac{1}{M}j\}$ and simple algebra shows that  $\mathcal{N}$ is even for odd $M$ and   $\mathcal{N}$ is odd for even $M$ (when $M\equiv 0 \mod 4$ it is necessary to shift the intervals \eqref{intervals} by a small amount). 

As the total number of reflections from the boundaries is even $(2M+2N)$  the pencil which includes the orbit passing through the origin for odd $M$  has even phase but for even $M$ the total accumulated phase is always odd. 

Each periodic orbit pencil fixed by 2 integers, $M$ and $N$ can support superscar waves propagating inside the pencil.  In the direction along the orbit such wave is simply a plane wave and in perpendicular direction it has to vanish at  (effective) boundaries  of periodic orbit pencil.
\begin{equation}
\Psi^{(\mathrm{superscar})}_{mn}(\xi,\eta)=\mathrm{e}^{\mathrm{i}\pi m \xi /L_p}\varphi_n(\eta)
\label{main}
\end{equation}
 Different cases differ by function $\varphi_n(\eta)$. Combining the above expressions one gets the following formulae.
\begin{itemize}
\item For pencils with odd $M$ and even longitudinal quantum number $m$
\begin{equation}
\varphi_n(\eta) =\left \{\begin{array}{cc} \sin\left (\frac{\pi n}{w}(\eta+\frac{w}{2}) \right ),  & -\frac{w}{2}<\eta<\frac{w}{2}\\ 
0 ,& \frac{w}{2}<\eta<\frac{3w}{2}\end{array}\right . .
\label{odd_M_even_m}
\end{equation}
\item For odd $M$ and odd $m$ 
\begin{equation}
\varphi_n(\eta)=\left \{\begin{array}{cc}   0, & -\frac{w}{2}<\eta<\frac{w}{2}\\ \sin\left (\frac{\pi n}{w}(\eta+\frac{3w}{2}) \right )
& \frac{w}{2}<\eta<\frac{3w}{2}\end{array}\right . .
\label{odd_M_odd_m}
\end{equation}
\item For even $M$  $m$ is always odd and for $M\not\equiv 0\mod 4$, 
 \begin{equation}
\varphi_n(\eta)=\sin\left (\frac{\pi n}{2w}(\eta+w) \right ) ,\qquad -w<\eta<w .
\label{even_M_2}
\end{equation}
\item For  $M\equiv 0\mod 4$ 
 \begin{equation}
\varphi_n(\eta)=\sin\left (\frac{\pi n}{2w}\eta \right ) ,\qquad 0<\eta<2 w.
\label{even_M_4}
\end{equation}
\end{itemize} 
The superscar energy is 
\begin{equation}
\mathcal{E}_{mn}=\frac{\pi^2}{L_p^2} m^2+\left \{\begin{array}{cc}\frac{\pi^2}{w^2}n^2,&\mathrm{odd}\;M\\ \frac{\pi^2}{4w^2}n^2,&\mathrm{even}\;M
\end{array}\right . .
\label{superscar_energy}
\end{equation} 
Expression \eqref{main} with  \eqref{odd_M_even_m}-\eqref{even_M_4} determines the unfolded superscar waves. To find folded back waves it is convenient to calculate the corresponding  series \eqref{big_rectangle_expansion}. To do it one has to find the overlap of  $\sin\Big ( \frac{\pi p (x+a)}{2a} \Big )\sin \Big (\frac{\pi k y}{b}\Big ) $ with the folded wave. As unfolding of product of trigonometric functions is obvious such overlap can be calculated as follows  
\begin{equation}
 F_{p\, k}=\frac{2}{ab} \int_{\mathrm{POP}}  \sin\Big ( \frac{\pi p (x(\xi,\eta)+a)}{2a} \Big )\sin \Big (\frac{\pi k y(\xi,\eta) }{b}\Big ) \Psi^{(\mathrm{superscar})}_{mn}(\xi,\eta)\, \mathrm{d}\xi\, \mathrm{d}\eta  
\end{equation}
where the integration is performed over the given periodic orbit pencil.

 One has
\begin{equation}
\sin\Big (\frac{\pi p(x+a)}{2a}\Big )\sin \Big (\frac{\pi k y}{b}\Big ) =-\frac{1}{4}\sum_{p=\pm |p|,\, k=\pm |k|}\epsilon_{kp} \mathrm{e}^{\mathrm{i} \pi (p x/2a+k y/b)}
\label{four_exponentials}
\end{equation}
where $\epsilon_{kp}=\sgn(k)\sgn(p)\mathrm{e}^{\mathrm{i}\pi p/2}$.

From Eq.~\eqref{new_coordinates} it follows that 
\begin{equation}
\frac{p x}{2a}+\frac{k y}{b}=\xi \frac{m}{L_p}+ \frac{\eta}{abL_p}(-p Nb^2+2k Ma^2),\qquad m=kM+2pN.
\label{exponent_transform}
\end{equation}
Further transformations depend on parity of $M$.  For odd $M$, the integers  $M$ and $2N$ by construction have no common divisors, $(M,2N)=1$. According to well known theorem in this case  there exist two integers  $\mu$ and $\nu$ such that
\begin{equation}
 M\nu -2N\mu=1.
 \end{equation}
 Using this relation one can introduce instead of two integers $k$ and $p$ another two integers $m$ and $q$ such that   
\begin{equation}
k=qM-m \mu,\qquad p=-2qN+m \nu
\label{change_integers}
\end{equation}
whose inverse are $m=2kN+pM$, $q=k\nu+p\mu$.

Then 
\begin{equation}
\frac{1}{abL_p}(-p Nb^2+2k Ma^2)=\frac{1}{w} (q-mQ),\qquad Q=\frac{\nu b^2 N+2\mu Ma^2}{2M^2 a^2+2N^2 b^2)}, \quad w=\frac{2ab}{L_p}.
\end{equation} 
Notice that 
\begin{equation}
\frac{m^2}{L_p^2}+\frac{(q-mQ)^2}{w^2}=\frac{k^2}{(2a)^2}+\frac{p^2}{b^2}
\label{energies}
\end{equation}
as it should be.

The integration over $\xi$ is simple. If $m$ and $m_1$ are of the same parity then
\begin{equation}
\int_0^{L_p}\mathrm{e}^{\mathrm{i}(m-m_1)\xi/L_p}\mathrm{d}\xi=L_p\delta_{m\,m_1}
\label{integral_xi_odd_M}
\end{equation}
Therefore the first term in Eq.~\eqref{exponent_transform} for odd $M$  can be identify with the first factor in Eqs.~\eqref{main} which means that $m$ is fixed by the longitudinal quantum number of the superscar wave.

The necessary series is reduced to the expansion
\begin{equation}
\varphi_n(\eta)=\sum_q C_{q\,m} \mathrm{e}^{\mathrm{i} \frac{\pi \eta}{w} (q-mQ) }, \qquad C_{q\,m}=\frac{1}{2w} \int_{-w/2}^{3w/2} \varphi_n(\eta)
 \mathrm{e}^{-\mathrm{i} \frac{\pi \eta}{w} (q-mQ) }. 
\end{equation}  
Explicit formulae for odd $M$, Eqs.~\eqref{odd_M_even_m}, \eqref{odd_M_odd_m}, are cumbersome but easy to get. It is plain that the denominators of these expressions are  proportional to $n^2/w^2-(q-mQ)^2/w^2$. From \eqref{energies} it follows that this difference equals $E_{kp}-\mathcal{E}_{mn}$ where
$\mathcal{E}_{mn}$ is the superscar energy \eqref{superscar_energy}.  Therefore the largest expansion coefficient corresponds to the minimum of this difference.    Notice that from 4 terms in \eqref{four_exponentials}  only one corresponding to \eqref{change_integers} gives nonzero contributions. Other 3 terms will give values of $m_1= \pm 2kN \pm pM$ which are of the same parity as $m=2kN+pM$ (when $N,M\neq 0$) and they vanish after integration over $\xi$ in \eqref{integral_xi_odd_M}. It means that $F_{p,k}\sim C_{q\, m}$ when $k=qM-m \mu$ and $p=-2qN+m \nu$. 

For even $M$, $M=2M^{\prime}$ the argumentation is similar. In this case $N$ should be odd and $(M^{\prime},N)=1$. Therefore  there exist two integers $\nu^{\prime}$ and $\mu^{\prime}$ such that  
\begin{equation}
M^\prime \nu^{\prime}-N\mu^{\prime}=1.
\end{equation}
Instead of \eqref{change_integers} one gets 
\begin{equation}
k=qM^{\prime}-m^{\prime} \mu^{\prime},\qquad p=-qN+m^{\prime} \nu^{\prime}
\end{equation}
and 
\begin{equation}
\frac{p x}{2a}+\frac{k y}{b}=\xi \frac{2m^{\prime}}{L_p}+ \frac{\eta}{w}(q-m^{\prime} Q^{\prime}),\qquad Q^{\prime}=\frac{2 M a^2\mu^{\prime}+Nb^2\nu^{\prime}}{2M^2 a^2+2N^2b^2}.
\label{total_even_M}
\end{equation}
But for even $M$ the superscar wave should have the factor $\mathrm{e}^{\pi m/L_p}$ with odd $m$ but all terms in the expansion \eqref{big_rectangle_expansion} have even $m$ (cf. \eqref{total_even_M}). It is clear that this is the manifestation of the fact that periodic orbits with even $M$ for desymmetrised barrier billiard are a half of periodic orbits for the full barrier billiard. As above the solution is to use an analog of the Gibbs phenomenon  because 
\begin{equation}
B_{m\,m^{\prime}}=\int_0^{L_p} \mathrm{e}^{\mathrm{i} \pi (2m-1-2m^{\prime})\xi/L_p}=\frac{L_p}{\pi(m^{\prime}+1/2-m)}.
\end{equation}
Therefore the necessary expansion takes the form 
\begin{equation}
\mathrm{e}^{\mathrm{i}\pi(2 m-1) \xi /L_p}\varphi_n(\eta)=\sum_{m^{\prime},q} B_{m\,m^{\prime}} \mathrm{e}^{\mathrm{i}\pi m^{\prime} \xi /L_p}
C_{q\,m^{\prime}}\mathrm{e}^{\mathrm{i} \frac{\pi \eta}{w} (q-m^{\prime}Q^{\prime})} . 
\end{equation}
Functions $\varphi_n(\eta)$ for even $M$ are given by Eqs.~\eqref{even_M_2} and \eqref{even_M_4}.  As the result includes the summation over all $m^{\prime}$ the 4 terms in \eqref{four_exponentials} give the same contribution and for even $M$ $F_{p\, k}\sim B_{m\,m^{\prime}} C_{q\, m^{\prime}}$ when $k=qM^{\prime}-m^{\prime} \mu^{\prime}$ and $p=-qN+m^{\prime} \nu^{\prime}$.


\end{document}